\documentclass [12pt]{article}
\def\maj#1{\ifmmode\mbox{\usefont{U}{msb}{m}{n}#1}\else{\usefont{U}{msb}{m}{n}#1}\fi}
\def\v#1{\mathbf{#1}}
\makeatletter \@addtoreset{equation}{section} \makeatother

\usepackage{graphics,epsfig,graphicx}
\usepackage{color}

\begin{document}

\title{\textbf{The trion: two electrons plus one
hole
\\ versus one electron plus one exciton}}
\author{M. Combescot, O. Betbeder-Matibet and F.
Dubin
 \\ \small{\textit{GPS, Universit\'e Pierre et Marie
Curie and Universit\'e Denis Diderot,
CNRS,}}\\ \small{\textit{Campus Boucicaut, 140 rue de
Lourmel, 75015 Paris, France}}}
\date{}
\maketitle

\begin{abstract}
We first show that, for problems dealing with
trions, it is totally hopeless to use the
standard many-body description in terms of
electrons and holes and its associated Feynman
diagrams.
We then show how, by using
the description of a trion as an electron
interacting with an exciton, we can obtain the
trion absorption through far simpler diagrams,
written with electrons and
\emph{excitons}. These diagrams are quite
novel because, for excitons being not exact
bosons, we cannot use standard procedures designed
to deal with interacting true fermions or true
bosons. A new many-body formalism is necessary to
establish the validity of these
electron-exciton diagrams and to derive
their specific rules. It relies on the
``commutation technique'' we recently developed to
treat interacting close-to-bosons. This
technique generates a  scattering
associated to direct Coulomb processes between
electrons and excitons and a dimensionless
``scattering'' associated to electron exchange
inside the electron-exciton pairs --- this
``scattering'' being the original part of our
many-body theory. It turns out that, although
exchange is crucial to differentiate singlet from
triplet trions, this ``scattering'' enters the
absorption explicitly when the photocreated
electron and the initial electron have the same
spin ---
\emph{i}.\
\emph{e}., when triplet trions are the only ones
created --- \emph{but not} when the two spins are
different, although triplet trions are also
created in this case. The physical reason
for this rather surprising result will be given.
\end{abstract}

PACS.: 71.35.-y Excitons and related phenomena

\newpage

\section{Introduction}

While the physics of excitons and electron-hole
plasma has been a subject of great interest in the
60's and 70's, the physics of trions [1,2],
\emph{i}.\
\emph{e}., excitons bound to an electron or a hole,
developed recently only: The exciton
being a neutral object, its possible binding to
a carrier is indeed quite weak compared to the
binding of an electron to a hole. However, due to
the development of nanostructure technology, it is
now possible to experimentally study these trions
because, all binding energies being enhanced by the
reduction of dimensionality, trions, not seen in
bulk samples, can appear as line well below the
exciton line in the absorption spectra of doped
semiconductor quantum wells (see for instance
ref.\ [3-9]).

From a theoretical point of view [10-19], the study
of these trions  still faces major difficulties:
(i) Being the eigenstates of two electrons and one
hole -- or two holes and one electron -- in Coulomb
interaction, their energies and wave functions are
not analytically known, the corresponding
three-body Hamiltonian having no analytical
solution; (ii) because these trions are bound
states, there is no way to reach them from a
(finite) perturbative approach; (iii) while
many-body procedures have been developed in the
60's to treat interactions between fermions (or
bosons), we will show that these many-body
procedures are completely inappropriate to
approach the bound states resulting from the
\emph{exact} summation of all Coulomb processes
between more than two fermions, as in the trion
case.

With respect to this last point, a good idea
can be to bind one electron to the hole in
order for the trion to appear as a two-body
system: one electron interacting with one
exciton [18]. However, since the electrons are
indistinguishable, the electron which is bound to
the hole to make the exciton, is \emph{a priori}
arbitrary. Moreover --- and physically linked
to this arbitrariness --- the exciton is not an
exact boson so that standard many-body
procedures, designed to treat interactions
between exact fermions or exact bosons, cannot
be used for this interacting electron-exciton
system.

In spite of these obvious problems, the
description of a trion as an electron
interacting with an exciton, is physically
appealing because it allows to immediately see
that, as the exciton is neutral, its attraction is
very weak, so that the trion binding has to be much
weaker than the exciton binding. It is thus
worth to find a way to cope with the difficulties
this electron-exciton (e-X) description raises.

In order to use it, we first have
to identify a quantity corresponding to the
electron-exciton Coulomb interaction, although,
due to the composite nature of the exciton, there
is no way to extract such an interaction from the
Coulomb potential between individual carriers.
In addition, we must find a way to take
care of the electron indistinguishability when
constructing the exciton. This
indistinguishability somehow adds novel exchange
``scatterings'' to the somewhat normal
electron-exciton Coulomb scatterings, with
specific rules which have to be
determined and which are the novel part of the
diagrams corresponding to electrons interacting
with excitons.

This paper in fact deals with the simplest possible
problem on trions, namely the absorption
of one photon in the presence of \emph{one}
carrier. The photon creating one electron-hole
(e-h) pair, we will first consider it in the
framework of a two electron-one hole system, and
show that the corresponding response function,
written with standard electron and hole diagrams,
is so complicated that there is no hope to
identify and sum up the processes responsible for
the trion bound states.

We will then show that this response function
appears extremely simply if we bind one electron to
the hole and have this exciton scattered by its
interactions with the additional electron. The spin
conservation of the semiconductor-photon
interaction leads to differentiate absorption in
which the photocreated electron and the initial
electron have different spins, from absorption in
which the two spins are identical. We will show
that electron exchange enters the photon
absorption \emph{explicitly} in this last case
only, \emph{i}.\ \emph{e}., when triplet
trions are the only ones created, in spite
of the fact that triplet trions  can also
be created when the two electrons have
different spins. We will establish the rules for
these electron-exciton diagrams, using our
commutation technique for an exciton interacting
with electrons [18]. It allows to calculate the
response function to a photon field at any order
in the electron-exciton interaction, while formally
keeping the composite nature of the exciton,
\emph{i}.\ \emph{e}., the fact that the exciton
can be made with any of the two electrons.

This first paper allows to establish the problem
of one exciton interacting with one electron on a
firm basis, in order to possibly face a much
harder one, namely one exciton interacting with $N$
electrons, as for the photon absorption in the
presence of a Fermi sea [11,14]. Such an absorption
has been considered long ago by Combescot and
Nozi\`{e}res [20]. In this work, the spin degree of
freedom of the electrons has been dropped from
the first line as well as the electron-electron
(e-e) repulsion. While the first simplification
physically corresponds to have one kind of spin
only in the problem --- as for a $\sigma_-$ photon
absorbed by a quantum well having a
$(+1/2)$ polarized Fermi sea so that
$(S_z=1)$ triplet trions are the only ones possibly
formed ---, the second simplification is dramatic:
Once it is made, the trion physics is
irretrievably lost. Indeed, the e-e repulsion
partly compensates the e-h attraction, making
the trion binding energy much weaker than the
exciton one, whatever the electron spins.

Let us stress that this crucial effect
of the e-e interaction cannot be included through
a na\"{\i}ve screening of the e-h attraction, as
suggested by Hawrylak in his 2D
extension [11] of Combescot-Nozi\`{e}res's work.
Indeed, such a screening in the same way reduces
the e-h attraction responsible for the exciton
binding,  so that the trion binding energy this
procedure generates, is weak but \emph{equal} to
the exciton one, not smaller as it should. This
screening procedure in fact generates one electron
energy level, not two, this level being possibly
occupied by
\emph{two} electrons if their spins are
\emph{different}. This procedure thus
misses the whole physics of the trion, with an
electron very weakly bound, compared to the other.

In order to find a trion with a binding energy
much weaker than the exciton one, we must include
the e-e interaction independently from any
e-h process, not just through a screening of this
e-h attraction. The trouble is that, within the
many-body procedures at hand up to now, there is
no hope to add this e-e repulsion to the set of
already complicated processes which, summed up to
all orders, give rise to
the so called ``Fermi edge singularities''. In
this paper, we explicitly show that the Feynman
diagrams, written with electrons and holes, are
already inappropriate to describe the three-body
problem corresponding to just one trion. A new
many-body formalism is thus highly necessary if we
want to properly describe the trion absorption
change with doping observed experimentally. Its
presentation is in fact \emph{the underlying
purpose of this work}, with a particular attention
paid to the interplay between the somewhat normal
e-X direct Coulomb scattering and the
far more subtle electron exchange inside the
e-X pair.

The paper is organized as follows:

In section 2, we briefly recall the usual procedure
to calculate photon absorption and we try to
calculate it in the case of trion formation, using
the standard many-body procedure which leads to
expand it through the well known electron-hole
Feynman diagrams. We show that, even in the
simplest case of a photon creating an electron
with a spin different from the spin of the initial
electron, the summation of these diagrams
with all possible Coulomb processes between two
electrons and one hole, is totally hopeless.

In section 3, we reconsider our description of
a trion as an electron interacting with an exciton.
The derivations
given here of some important results on
this description in terms of e-X pairs,
are more direct than the somewhat pedestrian ones
we gave in previous works [17,18]: Once a result
is mathematically established and physically
understood, it becomes easy to find
``straightforward '' derivations and enlightening
links...

In section 4, we calculate
the trion response function in terms of these
interacting e-X pairs. We show that
it appears as a sum of
e-X ladder diagrams, with possibly
\emph{one} --- but no more than one --- electron
exchange between e and X, if --- and only if ---
the spins of the photocreated electron and the
initial electron are the same. This can be
surprising at first since, at each e-X scattering,
the exciton can
\emph{a priori} be constructed with any of the two
electrons, whatever their spins are.

\section{The trion as two electrons plus one hole}

A quite direct way to reach the trion is to look
at the photon absorption when the semiconductor
has one electron already present in the sample.
Let us first briefly recall how photon
absorption is usually calculated and how it
appears in the case of exciton formation, since
its comparison with trion is quite enlightening.

\subsection{Standard procedure to calculate
photon absorption}

The linear absorption of a photon field is given
by the Fermi golden rule,
\begin{equation}
\frac{2\pi}{\hbar}\,\sum_F|\langle F|W^\dag|I
\rangle|^2\,\delta(\maj{E}_F-\maj{E}_I)=-\frac{2}{\hbar}\,
\mathrm{Im}S\ ,
\end{equation}
where, due to $1/(x+i\eta)=\mathcal{P}(1/x)-i\pi
\delta(x)$, the response function $S$ reads
\begin{equation}
S=\langle
I|W\,\frac{1}{\maj{E}_I-\hat{H}+i\eta}\,W^\dag
|I\rangle\ ,
\end{equation}
as easy to show by inserting the closure relation
for the $\hat{H}$ eigenstates, $|F\rangle$, in
front of $W^\dag$.

$\hat{H}$ is the Hamiltonian of the uncoupled
matter-photon system,
$\hat{H}=H+\omega_p\alpha^\dag\alpha$,
with
$H$ being the matter Hamiltonian and
$\alpha^\dag$ the creation operator of
$(\omega_p,\v Q_p)$ photons. The coupling
$W^\dag$ between the photon field and the matter
physically corresponds to annihilate one photon
while creating one electron-hole pair, $W
^\dag=U^\dag\,\alpha$. For circularly polarized
photons $(\pm)$ absorbed in quantum wells,
momentum conservation allows to write
$U^\dag$ in terms of free electrons and holes as
$$U_{\pm}^\dag(\v Q_p) = \lambda^\ast\, \sum_{\v
p}a_{\v p+\alpha_e
\v Q_p,\mp}^\dag\,b_{-\v p+\alpha_h\v Q_p,
\pm}^\dag \ ,$$
\begin{equation}
\alpha_e=1-\alpha_h=\frac{m_e}{m_e+m_h}\ .
\end{equation}
$a_{\v
p,\mp}^\dag$ creates an electron with momentum
$\v p$ and spin
$\mp 1/2$, while $b_{\v p,\pm}^\dag$ creates a hole
with spin $\pm 3/2$. The reason for splitting $\v
Q_p$ between the electron and the hole as
$\alpha_e\v Q_p$ and $\alpha_h\v Q_p$, will
become apparent below.

For $N_p$ photons
and a matter
initial state $|i\rangle$ with energy $\maj{E}_i$,
the initial state in eq.\ (2.2) reads
$|I\rangle=|N_p\rangle\otimes |i\rangle$, the
initial energy
$\maj{E}_I$ being
$N_p
\omega_p+\maj{E}_i$. Since
$\alpha|N_p\rangle=\sqrt
{N_p}\,|N_p-1\rangle$, we can rewrite this response
function in terms of matter quantities only, as
\begin{equation}
S_{\pm}=N_p\,\langle i|U_{\pm}(\v Q_p)\,\frac{1}{\omega_p
+\maj{E}_i-H+i\eta}\,U_{\pm}^\dag(\v Q_p)|i\rangle\ .
\end{equation}

In eq.\ (2.4), $H$ acts on the photocreated
electron-hole pair plus the initial carriers.
Many-body effects between them follow from the
identity
\begin{equation}
\frac{1}{a-H}=\frac{1}{a-H_0}+\frac{1}{a-H}V\frac
{1}{a-H_0}\ ,
\end{equation}
valid for $H=H_0+V$, which can be iterated as
\begin{equation}
\frac{1}{a-H}=\frac{1}{a-H_0}+\frac{1}{a-H_0}V\frac{1}
{a-H_0}+\cdots\ .
\end{equation}
In the case of semiconductors, the free part $H_0$
reads
$$H_0=\sum_{\v k,s}\epsilon_{\v
k}^{(e)}a_{\v k,s} ^\dag a_{\v
k,s}+\sum_{\v k,m}\epsilon_{\v k}
^{(h)}b_{\v k,m}^\dag b_{\v k,m}\ ,$$
\begin{equation}
\epsilon_{\v k}^{(e,h)}=\frac{\hbar^2\v
k^2}{2m_{e,h}} \ ,
\end{equation}
while $V$ is
the Coulomb potential between carriers,
\begin{eqnarray}
V=\frac{1}{2}\sum_{\v q\neq\v 0}V_{\v q}\left[
\sum_{\v k,\v k',s,s'}a_{\v k+\v q,s}^\dag
a_{\v k'-\v q,s'}^\dag a_{\v k',s'}a_{\v k,s}+
\sum_{\v k,\v k',m,m'}b_{\v k+\v q,m}^\dag
b_{\v k'-\v q,m'}^\dag b_{\v k',m'}b_{\v
k,m}\right.\nonumber \\ \left.
-2\sum_{\v k,\v k',s,m'}a_{\v k+\v q,s}^\dag
b_{\v k'-\v q,m'}^\dag b_{\v k',m'}a_{\v k,s}
\right]\ ,
\end{eqnarray}
with $V_{\v q}=2\pi e^2/\Omega q$ in 2D, $\Omega$
being the sample volume.

\subsection{Photon absorption with exciton
formation, using electron-hole diagrams}

For $\sigma_+$ photons absorbed in an \emph{empty}
quantum well, the response function given in eq.\
(2.4) reads, due to eq.\ (2.3),
\begin{equation}
S_X=N_p|\lambda|^2\sum_{\v p',\v p}
\langle v|b_{-\v p'+\alpha_h\v Q_p,+}\,a_{\v
p'+\alpha_e\v Q_p,-}\,\frac{1}{\omega_p-H+i\eta}\,
a_{\v p+\alpha_e\v Q_p,-}^\dag\,b_{-\v p+\alpha_h
\v Q_p,+}^\dag|v\rangle\ .
\end{equation}
$|v\rangle$ is the electron-hole vacuum state, its
energy being chosen to be zero.
$S_X$ can be calculated using the expansion (2.6).
By noting that
$\epsilon_{\v p+\alpha_e\v Q_p}
^{(e)}+\epsilon_{-\v p+\alpha_h\v
Q_p}^{(h)}=
\epsilon_{\v p}^{(X)}+\mathcal{E}_{\v
Q_p}^{(X)}$, with
\begin{equation}
\epsilon_{\v p}^{(X)}=\frac
{\hbar^2\v
p^2}{2(m_e^{-1}+m_h^{-1})^{-1}}\hspace{2cm}
\mathcal{E}_{\v
Q}^{(X)}=\frac{\hbar^2\v
Q^2}{2(m_e+m_h)}\ ,
\end{equation}
--- which is the reason for the $(\alpha_e\v
Q_p,\alpha_h\v Q_p)$ splitting --- the zero order
term reduces to
\begin{equation}
S_X^{(0)}=N_p|\lambda|^2\sum_{\v
p}\frac{1}{\omega _p-\epsilon_{\v
p}^{(X)}-\mathcal{E}_{\v Q_p}^{(X)}
+i\eta}=N_p|\lambda|^2\sum_{\v
p}G^{(eh)}(\omega_p,\v Q_p;\v p)
\ ,
\end{equation}
which can be used as a definition of
$G^{(eh)}(\omega_p,\v Q_p;\v p)$. The first order
term in Coulomb interaction then appears as
\begin{equation}
S_X^{(1)}=N_p|\lambda|^2\sum_{\v p',\v p}G^{(eh)}
(\omega_p,\v Q_p;\v p')\,V_{\v p'-\v p}\,
G^{(eh)}(\omega_p,\v Q_p;\v p)\ ,
\end{equation}
while the second order term contains three
$G^{(eh)}$ and two Coulomb potentials; and so
on\ldots This shows that the response function in
the case of exciton corresponds to the well known
set of electron-hole ladder diagrams [21] shown in
fig.\ (1), since $G^{(eh)}$ is nothing but
\begin{equation}
G^{(eh)}(\omega_p,\v
Q_p;\v p)=
\int\frac{id\omega}{2\pi}\,g^{(e)}(\omega
+\omega_p,\v p+\alpha_e\v Q_p)\,g^{(h)}(-\omega,
-\v p+\alpha_h\v Q_p)\ ,
\end{equation}
where $g^{(e)}(\omega,\v
k)=[\omega-\epsilon_{\v
k}^{(e)}+i\eta]^{-1}$ and
$g^{(h)}(\omega,\v k)=
[\omega-\epsilon_{\v
k}^{(h)}+i\eta]^{-1}$ are the electron
propagator and hole propagator respectively,
both bands being initially empty.

\subsection{Photon absorption with trion
formation, using electron-hole diagrams}

For the $X^-$ trion to be formed, the
semiconductor initial state must have one electron.
If
$\v k_i$ and $s_i$ are its momentum and spin, this
initial state reads $|i\rangle=a_{\v
k_i,s_i}^\dag|v\rangle$, with
$\maj{E}_i=\epsilon_{\v k_i}^{(e)}$. We then
note that, while the photon polarization is
unimportant when the semiconductor is empty, it is
crucial when it already has electrons: Indeed, in a
quantum well, if the spins of the photocreated
electron and the initial electrons are different,
the hole can only recombine with the photocreated
electron, while it can recombine with any of them
if their spins are identical.
Consequently, the response functions differ if the
electron spins are identical or not. Let us start
with the simplest case.

\subsubsection{Photocreated electron with spin
different from the initial one}

This happens when a $\sigma_+$ photon is absorbed
in a quantum well having a $s_i=+1/2$ electron.
The zero order term of the response function
appears, using eqs.\ (2.3,2.4), as
\begin{eqnarray}
\tilde{S}_{\neq}^{(0)}=N_p|\lambda|^2\sum_{\v p',\v
p}\langle v| a_{\v k_i,+}\,b_{-\v p'+\alpha_h\v
Q_p,+}\, a_{\v p'+\alpha_e\v Q_p,-}\,\left
(\frac{1}{\omega_p+
\epsilon_{\v
k_i}^{(e)}-H_0+i\eta}\right)\nonumber\\
\times\ a_{\v p+\alpha_e\v Q_p,-}^\dag\,
b_{-\v p+\alpha_h\v Q_p,+}^\dag\, a_{\v k_i,+}
^\dag|v\rangle\ .
\end{eqnarray}
We readily find that $\tilde{S}_{\neq}^{(0)}$
reduces to the exciton zero order response function
$S_X^{(0)}$, so that it corresponds to the
diagram of fig.\ (1a). If we now turn to
the first order term, it is given by
\begin{eqnarray}
\tilde{S}_{\neq}^{(1)}=N_p|\lambda|^2\sum_{\v p',\v
p}G^{(eh)} (\omega_p,\v Q_p;\v
p')G^{(eh)}(\omega_p,\v Q_p;
\v p)\hspace{5cm}\nonumber
\\ \times\,\langle v|a_{\v k_i,+}\,b_{-\v
p'+\alpha_h
\v Q_p,+}\,a_{\v p'+\alpha_e\v Q_p,-}\,
V\,a_{\v p+\alpha_e\v Q_p,-}^\dag\,
b_{-\v p+\alpha_h\v Q_p,+}^\dag\,
a_{\v k_i,+}^\dag|v\rangle\ ,
\end{eqnarray}
where $V$ is the Coulomb potential given in eq.\
(2.8). As this Coulomb potential only contains
$\v q\neq\v 0$ excitations, the above
matrix element differs from 0 for $\v
p'=\v p+\v q$ only: This first order term is thus
equal to $S_X^{(1)}$ and corresponds to the ladder
diagram of fig.\ (1b).

The higher order terms are not as simple: When
more than one Coulomb excitation take place, in
addition to processes in which the
photocreated electron is scattered several times
by the hole, other processes involving the initial
electron become possible: In addition to
the exciton ladder diagrams shown in fig.\ (1c)
at second order in
$V$ and in fig.\ (1d) at third order, we also
have the 4 diagrams of fig.\ (2) at second order
in $V$ and the 20 diagrams of fig.\ (3) at third
order. Note that, since the
$(+1/2)$ electron band
has one electron only, these diagrams
have one conduction-hole line only, without any
possible scattering,
\emph{i}.\
\emph{e}., one electron line only going backward,
from left to right.

As the $\mathrm{X}^-$trion corresponds to the
\emph{bound} state of two electrons and one hole
resulting from their Coulomb interaction, it is
necessary to take the Coulomb potential
\emph{exactly} into account
\emph{i}.\ \emph{e}., to include it at all orders,
to possibly generate the bound state poles
in the response function. In view of the
third order processes shown in fig.\ (3), it is
obviously hopeless to write down the diagrams
corresponding to all possible Coulomb interactions
at any order in $V$ and to sum them up to get the
trion.

\subsubsection{Photocreated electron with spin
identical to the initial one}

The situation is worse when the photocreated
electron has the spin of the initial
electron, because the hole can now recombine with
any of the two electrons. While the zero order
term, given in eq.\ (2.14), with $a_{\v
k_i,+}^\dag$ replaced by
$a_{\v k_i,-}^\dag$, stays essentially unchanged,
$\v p$ being just different from $\v
k_i-\alpha_e\v Q_p$, new diagrams with exchange
processes between the two electrons appear at
higher orders. From eq.\ (2.15) with $a_{\v
k_i,+}^\dag$ replaced by $a_{\v k_i,-}^\dag$, it
is easy to see that, beside $\v p'=\v p+\v q$
which gives the first order ladder diagram of
fig.\ (1b), we can also have $\v p=\v p'=\v k_i-\v
q$ which gives the first order exchange diagram
shown in fig.\ (4a). In a similar way, beside the
second order direct diagrams already contained in
$\tilde{S}_{\neq} ^{(2)}$,
we also have the six diagrams of fig.\ (4b) which
result from exchange processes between the
photocreated electron and the initial electron;
and so on\ldots

Consequently, the response
function for photocreated and initial electron
having the same spin, is even more
complicated than the one for
different spins: This leads us to conclude that
the representation of a trion as two electrons plus
one hole, and its associated standard electron-hole
Feynman diagrams, are completely inappropriate.

We are now going to show that the description of a
trion as an electron interacting with an exciton
is far better. For that, let us first
recall the main steps of the
many-body procedure on which it is based and which
allows an exact treatment of the tricky part of
this description, namely the electron
indistinguishability. This procedure ultimately
leads to represent the trion through novel
electron-exciton diagrams, with quite specific
rules for the new ``scattering'' coming from pure
exchange, \emph{i}.\ \emph{e}., exchange without
Coulomb, not easy to guess at first.

\section{The trion as one electron plus one
exciton}

By considering the trion as an electron
interacting with an exciton, we tend to put the
trion and the exciton on equal footing --- with the
hole of the exciton just replaced by an exciton.
This is actually quite misleading because, due to
the electron indistinguishability, the trion is
definitely far more subtle than the exciton. In
order to grasp the deep differences which exist
between them, it appears to us useful to first
recall a few well known results on exciton. In
doing so, we will settle some important notations,
also useful for trion.

\subsection{A few results on exciton}

\subsubsection{First quantization}

The exciton can be seen as a two-body object, made
of one electron $(m_e,\v r_e)$ and one hole
$(m_h,\v r_h)$. If we extract its center of mass,
characterized by
\begin{equation}
M_X=m_e+m_h\hspace{2cm}\v R_X=(m_e\v
r_e+m_h\v r_h)/M_X\ ,
\end{equation}
we are left with its
relative motion, characterized by
\begin{equation}
\mu_X^{-1}=m_e^{-1}+m_h^{-1}\hspace{2cm}
\v r=\v r_e-\v r_h\ .
\end{equation}
The exciton Hamiltonian in first quantization reads
\begin{equation}
H_X=\frac{\v p_e^2}{2m_e}+\frac{\v p_h^2}{2m_h}
-\frac{e^2}{|\v r_e-\v r_h|}=\frac{\v P_X^2}{2M_X}
+h_X\ ,
\end{equation}
where $h_X=h_X^{(0)}-v(\v r)$ is the exciton
relative motion hamiltonian composed of a
free-particle part $h_X^{(0)}=\v p_{\v
r}^2/2\mu_X$ and a Coulomb attraction $v(\v r)=e^2
/r$.

The eigenstates of the relative
motion free part are the plane waves $|\v
p\rangle$ of energies $\epsilon_{\v p}^{(X)}$
given in eq.\ (2.10):
$(h_X^{(0)}-\epsilon_{\v p}^{(X)})|\v p\rangle=0$.
They are such that $\langle\v
p'|\v p\rangle=\delta_{\v p',\v p}$, their closure
relation being $\sum_{\v p}|\v p\rangle\langle\v
p|=I$.

The eigenstates $|\nu\rangle$ of the full relative
motion Hamiltonian, $(h_X-\varepsilon_\nu^{
(X)})|\nu\rangle=0$, are made of bound \emph{and}
diffusive states. They are such that
$\langle\nu'|\nu\rangle=\delta_{\nu',\nu}$, their
closure relation being
$\sum_\nu|\nu\rangle\langle\nu|=I$.

In terms of $|\v p\rangle$, these $|\nu\rangle$
states read $|\nu\rangle=\sum_{\v p}|\v
p\rangle\langle\v p|\nu\rangle$, so that the
projection over $\langle\v p|$ of the
Schr\"{o}dinger equation for $|\nu\rangle$ leads
to
\begin{equation}
(\epsilon_{\v p}^{(X)}-\varepsilon_\nu^{(X)})
\langle\v p|\nu\rangle-\sum_{\v p'}\langle\v p|
v(\v r)|\v p'\rangle\,
\langle\v p'|\nu\rangle=0\ ,
\end{equation}
the coupling being such that
\begin{equation}
\langle\v p|v(\v r)|\v p'\rangle
=V_{\v p'-\v p}\ ,
\end{equation}
where $V_{\v q}$ is the Fourier transform of
the Coulomb potential
$v(\v r)$ appearing in $h_X$.

If we come back to the full exciton Hamiltonian,
its eigenstates are the $|\nu,\v Q\rangle$'s with
wave functions $\langle\v r_e,\v r_h|\nu,\v
Q\rangle=\langle\v r|\nu\rangle
\langle\v R_X|\v Q\rangle$, where $|\v Q\rangle$
is the plane wave of energy $\mathcal{E}_{\v
Q}^{(X)}$ given in eq.\ (2.10): It is indeed easy
to check that
$(H_X-E_{\nu,\v Q}^{(X)})|\nu,\v
Q\rangle=0$, the total exciton energy being
\begin{equation}
E_{\nu,\v Q}^{(X)}=\varepsilon_\nu^{(X)}+
\mathcal{E}_{\v Q}^{(X)}\ .
\end{equation}

\subsubsection{Second quantization}

In second quantization, the semiconductor
Hamiltonian (which has the same form whatever the
number of electrons and holes), is given by eqs.÷
(2.7-8). In terms of these free electron and free
hole creation operators, the creation operator for
an exciton $(\nu,\v Q)$ with electron spin $s$ and
hole momentum $m$ is given by
\begin{equation}
B_{\nu,\v Q,s,m}^\dag=\sum_{\v p}\langle\v
p|\nu\rangle\,a_{\v p+\alpha_e\v Q,s}^\dag\,b
_{-\v p+\alpha_h\v Q,m}^\dag\ .
\end{equation}
Indeed, we do have $(H-E_{\nu,\v Q}^{(X)})
B_{\nu,\v Q,s,m}^\dag|v\rangle=0$, as easy to
check directly from eqs.\ (2.7-8), (3.4) and (3.7).
Note that, with this splitting of the exciton
momentum
$\v Q$ between the electron and the hole, the
remaining momentum $\v p$ is just the momentum of
the exciton relative motion. It will also be
useful to note that, in the same way as excitons
read in terms of electron-hole pairs,
electron-hole pairs can be written in terms of
excitons as
\begin{equation}
a_{\v p+\alpha_e\v Q,s}^\dag\,b_{-\v p+\alpha_h\v
Q,m}^\dag=\sum_\nu\langle\nu|\v p\rangle\,B_{\nu,
\v Q,s,m}^\dag\ ,
\end{equation}
easy to check from eq.\ (3.7).

\subsection{Trion in first quantization}

The trion is a three-body object (two
electrons and one hole) or (two holes and one
electron). To simplify the notations, we will here
consider the two-electron-one-hole case only.

A first difficulty with trions, compared to
excitons, arises from the spin variables. They are
unimportant for excitons if we neglect
``electron-hole exchange'', \emph{i}.\ \emph{e}.,
valence-conduction Coulomb excitations, their
energies being then degenerate with respect to
$(s,m)$. On the opposite, the spin variables are
crucial for trions because they differenciate
their possible states through the parity of the
orbital part of the wave functions with respect to
the electron positions. Indeed, this parity is
directly linked to the trion total
\emph{electronic} spin: As singlet states are odd
while triplet states are even, their associate
orbital wave functions must be even and odd
respectively, due to the symmetry principle for
the whole wave function of fermions.

Another difficulty, less obvious at first, comes
from the appropriate variables to describe the
trion. The center of mass $(M_T,\v R_T)$ is surely
one of these appropriate variables. For two
electrons $(m_e,\v r_e,\v r_{e'})$ and one hole
$(m_h,\v r_h)$, it reads
\begin{equation}
M_T=2m_e+m_h\hspace{2cm} \v R_T=(m_e\v r_e+m_e
\v r_{e'}+m_h\v r_h)/M_T\ .
\end{equation}

While for excitons, there is only one spatial
variable, namely $\v r$, which along with $\v R_X$
forms a good set of variables, \emph{i}.\
\emph{e}., for which
$[r_i,p_j]=i\hbar\delta_{ij}$, there are many ways
[17] to choose the two other spatial variables
which, along with $\v R_T$, form a good set for
trions. Among them, \emph{the convenient ones for
physical understanding} turn out to be $\v r$,
associated to $\mu_X$, defined in eq.\ (3.2), and
$\v u$, associated to $\mu_T$, defined as
\begin{equation}
\mu_T^{-1}=m_e^{-1}+M_X^{-1}\hspace{2cm}
\v u=\v r_{e'}-\v R_X\ .
\end{equation}
$\v u$ is the distance between $e'$ and the center
of mass of $(e,h)$, while $\mu_T$ is the relative
motion mass of this $e'$ electron and the $(e,h)$
exciton. Of course, due to the electron
indistinguishability, variables as good as $(\v
r,\v u)$ are
\begin{equation}
\v r'=\v r_{e'}-\v r_h=\v u+\alpha_e\v
r\hspace{2cm} \v u'=\v r_e-\v
R_X'=(1-\alpha_e^2)\v r-\alpha_e\v u\ ,
\end{equation}
since $\v R_X'=(m_e\v r_{e'}+m_h\v r_h)/M_X$. This
possible change from $(\v r,\v u)$ to $(\v r',\v
u')$, which corresponds to exchange the two
electrons of the trion, is present all over the
trion representation in terms of one electron plus
one exciton. We will show below how to handle it
in a simple way.

In terms of these variables, the trion Hamiltonian
in first quantization reads
\begin{eqnarray}
H_T &=& \frac{\v p_e^2}{2m_e}+\frac{\v
p_{e'}^2}{2m_e}+\frac{\v p_h^2}{2m_h}-\frac{e^2}
{|\v r_e-\v r_h|}-\frac{e^2}{|\v r_{e'}-\v r_h|}
+\frac{e^2}{|\v r_e-\v r_e'|}\nonumber
\\ &=& \frac{\v P_T^2}{2M_T}+h_T\ .
\end{eqnarray}
Like for excitons, the relative motion Hamiltonian
$h_T$ splits into a free part $h_T^{(0)}=h_X+\v
p_{\v u}^2/2\mu_T$, made of one exciton plus one
effective free particle of mass $\mu_T$, and an
interaction $w(\v r,\v u)$ which corresponds to the
Coulomb interaction between $e'$ and $(e,h)$,
\begin{equation}
w(\v r,\v u)=\frac{e^2}{|\v r_{e'}-\v r_e|}-
\frac{e^2}{|\v r_{e'}-\v r_h|}=\frac{e^2}{|\v u
-\alpha_h\v r|}-\frac{e^2}{|\v u+\alpha_e\v r|}\ .
\end{equation}

The eigenstates of the relative motion free part
are just the $|\nu,\v p\rangle$'s since we do
have $h_T^{(0)}|\nu,\v p\rangle=
(\varepsilon_\nu^{(X)}+\epsilon_{\v p}^{(eX)})
|\nu,\v p\rangle$, with
\begin{equation}
\epsilon_{\v p}^{(eX)}=\frac{\hbar^2\v
p^2}{2\mu_T}\ .
\end{equation}
The corresponding wave functions are $\langle\v
r,\v u|\nu,\v p\rangle=\langle\v r|\nu\rangle
\langle\v u|\v p\rangle$. These $|\nu,\v p\rangle$
states are such that $\langle\nu',\v p'|\nu,\v
p\rangle=\delta_{\nu',\nu}\delta_{\v p',\v p}$,
while their closure relation reads $\sum_{\nu,\v p}
|\nu,\v p\rangle\langle\nu,\v p|=I$.

Let us call $|\eta\rangle$ the eigenstates of the
full relative motion Hamiltonian,
$(h_T-\varepsilon_\eta^{(T)})|\eta\rangle=0$.
Using them, the eigenstates of the trion
Hamiltonian $H_T$ are the $|\eta,\v K\rangle$'s,
since we do have $(H_T-E_{\eta,\v K}
^{(T)})|\eta,\v K\rangle=0$, with
\begin{equation}
E_{\eta,\v
K}^{(T)}=\varepsilon_\eta^{(T)}+\mathcal{E}_{\v K}
^{(T)},\hspace{2cm} \mathcal{E}_{\v K}^{(T)}=
\frac{\hbar^2\v K^2}{2M_T}\ ,
\end{equation}
the corresponding wave functions being
\begin{equation}
\langle\v r_e,\v r_{e'},\v r_h|\eta,\v K\rangle=
\langle\v r,\v u|\eta\rangle\langle\v R_T|\v
K\rangle\ .
\end{equation}

As $H_T$, given in eq.\ (3.12), stays invariant
under the
$(e\leftrightarrow e')$ exchange, its eigenstates
are odd or even
with respect to this exchange. Since $\v R_T$ is
unchanged, this means that the
$\langle\v r,\v u|\eta\rangle$'s are odd or even.
Let us call $\eta_0$, the $\eta$ indices which
correspond to even functions with respect to the
$(e\leftrightarrow e')$ exchange and $\eta_1$, the
ones which correspond to odd functions. This
parity condition thus reads, for $S=(0,1)$,
\begin{equation}
\langle\v r,\v u|\eta_S\rangle=(-1)^S\langle
\v r',\v u'|\eta_S\rangle=(-1)^S\langle\v
u+\alpha_e\v r,(1-\alpha_e^2)\v r-\alpha_e\v u|
\eta_S\rangle\ ,
\end{equation}
due to eq.\ (3.11). Within these $(\v r,\v u)$
variables, the parity condition of the
$|\eta_S\rangle$'s is obviously
not very appealing. We can rewrite it in a nicer
form for physical understanding, by using other
variables than $(\v r,\v u)$, namely $(\nu,\v p)$.
For that, we introduce the Fourier transform of
$\langle\v r,\v u|\eta_S\rangle$ ``in the exciton
sense'', defined as
\begin{equation}
\langle\nu,\v p|\eta_S\rangle=\int d\v r\,d\v u\,
\langle\nu|\v r\rangle\langle\v p|\v u\rangle
\langle\v r,\v u|\eta_S\rangle\ .
\end{equation}
We then insert eq.\ (3.17) into eq.\ (3.18), and
replace $\langle\v r',\v u'|\eta_S\rangle$ by its
expression in terms of its Fourier transform,
namely $\sum_{\nu',\v p'}\langle\v r'|\nu'\rangle
\langle\v u'|\v p'\rangle\langle\nu',\v p'|\eta_S
\rangle$. Next, we
express all the spatial variables in terms of
$\v r$ and $\v r'$, using $\v u=\v
r'-\alpha_e\v r$ and $\v u'=\v r-\alpha_e\v
r'$. By noting that $\langle\v p|\v
r'-\alpha_e\v r\rangle
\langle\v r-\alpha_e\v r'|\v p'\rangle$ is nothing
but $\langle\v p+\alpha_e\v p'|\v r'\rangle
\langle\v p'+\alpha_e\v p|\v r\rangle$, we
eventually find that the
$|\eta_S\rangle$'s fulfilling eq.\ (3.17) are such
that
\begin{equation}
\langle\nu,\v p|\eta_S\rangle=(-1)^S\sum_{\nu',\v
p'}L_{\nu\v p;\nu'\v p'}\,\langle\nu',\v p'|\eta_S
\rangle\ ,
\end{equation}
where $L_{\nu\v p;\nu'\v p'}$ appears as
\begin{equation}
L_{\nu\v p;\nu'\v p'}=\langle\nu|\v p'+\alpha_e\v p
\rangle\,\langle\v p+\alpha_e\v p'|\nu'\rangle\ .
\end{equation}
We will show below that this $L_{\nu\v p;\nu'\v
p'}$ coefficient is just the exchange
``scattering'' of the ``commutation technique'' for
excitons interacting with electrons. Its link
with electron exchange inside an e-X pair can
however be made apparent right now, by noting that
\begin{equation}
\int d\v r_e\,d\v r_{e'}\,
d\v r_h\,\phi_{\nu',\v Q'}^\ast(\v r_{e'},\v
r_h)\,f_{\v k'}^\ast(\v r_{e})\,\phi_{\nu,\v
Q}(\v r_{e},\v r_h)\, f_{\v k}(\v r_{e'})=
\delta_{\v K',\v K}\,L_{\nu'\v p';\nu\v p}\ ,
\end{equation}
where $\phi_{\nu,\v Q}(\v r_{e},\v
r_h)=\langle\v r_e,\v r_h|\nu,\v Q\rangle$ is the
exciton wave function, $f_{\v k}(\v r)=\langle\v
r|\v k\rangle$ the free electron wave function
while the $(\v Q,\v k)$ and $(\v K,\v p)$ are
linked by
$$\v K=\v Q+\v k,\hspace{2cm}\v k=\v p+\beta_e\v
K$$
\begin{equation}
\beta_e=1-\beta_X=\frac{m_e}{M_T}\ .
\end{equation}
Note that, in eq.\ (3.21), the ``in'' exciton
$(\nu,\v Q)$ and the ``out'' exciton $(\nu',\v
Q')$ are made with \emph{different} electrons, $\v
r_e$ and $\v r_{e'}$. The corresponding process is
shown in fig.\ (5a). As two exchanges reduce to an
identity, we do have
\begin{equation}
\sum_{\nu'',\v p''}L_{\nu'\v p';\nu''\v p''}\,
L_{\nu''\v p'';\nu\v p}=\delta_{\nu',\nu}\,
\delta_{\v p',\v p}\ ,
\end{equation}
easy to check from eq.\ (3.20).

If we now come back to the $|\eta\rangle$
eigenstates of the trion relative motion
Hamiltonian, the $\eta$ index is actually an
$\eta_0$ if the trion state has a total electron
spin $S=0$ and an $\eta_1$ if its total spin is
$S=1$. This means that $\eta$ actually reads
\begin{equation}
\eta=\eta_0\,\delta_{S,0}+\eta_1\,\delta_{S,1}\ ,
\end{equation}
where $S$ is the trion electronic spin.
Let us recall that, as the trion ground state must
have a symmetrical orbital wave function, its
index belongs to the
$\eta_0$ set, while its total electronic spin is
$S=0$.

Since the $|\eta\rangle$'s are eigenstates of an
Hamiltonian, namely $h_T$, they form an orthogonal
basis, so that
$\langle\eta'|\eta\rangle=\delta_{\eta',\eta}$,
while their closure relation reads
$\sum_\eta|\eta\rangle\langle\eta|=I$, the sum
being taken over the $\eta_0$'s and the
$\eta_1$'s, so that $I=I_0+I_1$, with
$I_S=\sum_{\eta_S}|\eta_S\rangle\langle\eta_S|$.
An interesting relation also exists for the
partial sum $I_S$. It reads
\begin{equation}
\langle\nu',\v p'|I_S|\nu,\v p\rangle=\frac{1}{2}
(\delta_{\nu',\nu}\,\delta_{\v p',\v p}+(-1)^S
L_{\nu'\v p';\nu\v p})\ ,
\end{equation}
and can be shown by noting that, due to eq.\
(3.19), $\langle\nu',\v p'|I_S|\nu,\v p\rangle=
(-1)^S\sum_{\nu'',\v p''}L_{\nu'\v p';\nu''\v p''}
\linebreak\langle\nu'',\v p''|I_S|\nu,\v p\rangle$,
so that, while $\langle\nu',\v p'|I_0+I_1|\nu,\v
p\rangle=\delta_{\nu',\nu}\,\delta_{\v p',\v p}$,
we do have $\langle\nu',\v p'|I_0-I_1|\nu,\v
p\rangle=L_{\nu'\v p';\nu\v p}$. Eq.\ (3.25)
follows from the combination of these two results.

Finally, the closure relation for the free states
$|\nu,\v p\rangle$ leads to write $|\eta\rangle$
as $\sum_{\nu,\v p}|\nu,\v p\rangle\langle\nu,\v p|
\eta\rangle$. So that the projection over
$\langle\nu,\v p|$ of the Schr\"{o}dinger equation
for the $|\eta\rangle$'s gives
\begin{equation}
(\varepsilon_\nu^{(X)}+\epsilon_{\v
p}^{(eX)}-\varepsilon_\eta^{(T)})\langle\nu,\v
p|\eta\rangle+\sum_{\nu',\v
p'}\langle\nu,\v p|w(\v r,\v
u)|\nu',\v p'\rangle\,\langle\nu',\v
p'|\eta\rangle=0\ ,
\end{equation}
the coupling being linked to the Fourier
transform ``in the exciton sense'' of the
Coulomb potential $w(\v r,\v u)$ appearing in
$h_T$, namely
\begin{equation}
\langle\nu,\v p|
w(\v r,\v u)|\nu',\v
p'\rangle=W_{\v p'-\v
p}^{\nu\nu'}\equiv\langle\nu|w_{\v p-\v p'}(\v
r)|\nu'\rangle\ ,
\end{equation}
where $w_{\v q}(\v r)=V_{\v q}(e^{-i\alpha_h\v
q.\v r}-e^{i\alpha_e\v q.\v r})$ is
the usual Fourier transform of $w(\v
r,\v u)$ with respect to the
variable $\v u$. Note that these results
for trions are formally similar to the ones for
excitons, given in eqs.\ (3.4-5), except for the
additional exciton quantum number $\nu$.

It turns out
that this coupling is just the
direct Coulomb scattering $C_{\nu\v p;\nu'\v
p'}^\mathrm{dir}$ of the ``commutation technique''
for excitons interacting with electrons, we
will introduce below:
\begin{equation}
W_{\v p'-\v p}^{\nu\nu'}\equiv C_{\nu\v p;\nu'\v
p'}^\mathrm{dir}\ .
\end{equation}
Its link to direct Coulomb processes is
easy to see right now, by noting that
\begin{eqnarray}
\int d\v r_e\,d\v r_{e'}\,d\v r_h\,\phi_{\nu',\v
Q'} ^\ast(\v r_e,\v r_h)\,f_{\v k'}^\ast(\v
r_{e'})\left(
\frac{e^2}{|\v r_{e'}-\v r_e|}-\frac{e^2}{|\v
r_{e'}-\v r_h|}\right)
\phi_{\nu,\v Q}(\v r_e,\v
r_h)\,f_{\v k}(\v r_{e'})\nonumber
\\=\delta_{\v
K',\v K}\, C_{\nu'\v p';\nu\v p}^\mathrm{dir}\ ,
\end{eqnarray}
where $(\v Q,\v k;\v K,\v p)$ and $(\v Q',\v k';\v
K',\v p')$ are again linked by eq.\ (3.22),
the ``in'' exciton
$(\nu,\v Q)$ and the ``out'' exciton $(\nu',\v
Q')$ being here made with the \emph{same} electron
$(\v r_e)$. The corresponding process is shown in
fig.\ (5b).

\subsection{Trions in second quantization}

\subsubsection{Creation operators of
e-X pairs}

If we look at the expression of the exciton
creation operator in terms of e-h pairs
given in eq.\ (3.7), we see that the exciton
center of mass momentum $\v Q$ is split between
the electron and the hole according to their
masses, namely $\alpha_e\v Q_p$ and $\alpha_h\v
Q_p$. In a similar way, we are led to introduce
e-X pair operators with the center
of mass momentum $\v K$ split between the
electron and the exciton according to their masses,
namely
\begin{equation}
\mathcal{T}_{\nu,\v p,\v K;\sigma,s,m}^\dag=a_{\v
p+
\beta_e\v K,\sigma}^\dag\,B_{\nu,-\v p+\beta_X\v
K,s,m}^\dag\ ,
\end{equation}
with $\beta_e,\beta_X$ given in eq.\ (3.22).

In order to calculate the scalar product of these
e-X states, it is convenient to
introduce the ``commutation technique'' for
excitons interacting with electrons. From the
deviation-from-boson operator $D_{n'n}$ defined as
[22,23]
\begin{equation}
[B_{n'},B_n^\dag]=\delta_{n',n}-D_{n'n}\ ,
\end{equation}
where the $B_n^\dag$'s are the exciton creation
operators defined in eq.\ (3.7) and $n$ stands for
$(\nu,\v Q,s,m)$ while $n'$ stands for $(\nu',\v
Q',s',m')$, we find
\begin{equation}
[D_{n'n},a_{\v
k,\sigma}^\dag]=\delta_{m',m}\,\delta_{s',\sigma}\,
\delta_{\v K',\v K}\,L_{\nu'\v p';\nu\v p}\,
a_{\v k',s}^\dag\ ,
\end{equation}
where $(\v Q,\v k;\v K,\v p)$ and $(\v Q',\v k';\v
K',\v p')$ are linked by eq.\ (3.22),
$L_{\nu'\v p';\nu\v p}$ being the
parameter already appearing in eqs.\ (3.19-21)
(see fig.\ (5a)).

From eqs.\ (3.31-32), we then
readily find
\begin{equation}
\langle v|\mathcal{T}_{\nu',\v p',\v
K';\sigma',s',m'}\mathcal{T}_{\nu,\v p,\v
K;\sigma,s,m}^\dag|v\rangle=\delta_{m',m}\,\delta_{\v
K',\v
K}\,(\delta_{\sigma',\sigma}\,\delta_{s',s}\,
\delta_{\nu',\nu}\,\delta_{\v p',\v
p}-\delta_{\sigma',s}\,\delta_{s',\sigma}\,L_{
\nu'\v p';\nu\v p})\ .
\end{equation}
It will also be useful to note that
\begin{equation}
\mathcal{T}_{\nu,\v p,\v K;\sigma,s,m}^\dag=-
\sum_{\nu',\v p'}L_{\nu'\v p';\nu\v p}\,
\mathcal{T}_{\nu',\v p',\v K;s,\sigma,m}^\dag\ ,
\end{equation}
which results from the two possible ways to
construct a trion out of two electrons and one
hole. (Note that the spins of
the electron and the exciton are exchanged in the
right and left hand sides of eq.\ (3.34)).

In the same way as exciton reads in terms of
e-h pairs, trion reads in
terms of e-X pairs. The simplest way
to get this decomposition is to first find how
the semiconductor Hamiltonian $H$ acts on one of
this pair. For that, we again
use the ``commutation technique'' for excitons
interacting with electrons. From the Coulomb
creation operator $V_n^\dag$ defined as [22,23]
\begin{equation}
[H,B_n^\dag]=E_n^{(X)}\,B_n^\dag+V_n^\dag\ ,
\end{equation}
we find
\begin{equation}
[V_n^\dag,a_{\v k,\sigma}^\dag]=\sum_{\nu',\v p'}
C_{\nu'\v p';\nu\v p}^\mathrm{dir}\,
\mathcal{T}_{\nu',\v p',\v K;\sigma,s,m}^\dag\ ,
\end{equation}
where $(\v Q,\v k;\v K,\v p)$ are
linked by eq.\ (3.22), $C_{\nu'\v p';\nu\v
p}^\mathrm{dir}$ being the quantity already
appearing in eqs.\ (3.28-29). From eqs.\ (3.30),
(3.35) (which implies $V_n^\dag|v\rangle=0$)
and (3.36), we readily get
\begin{equation}
H\,\mathcal{T}_{\nu,\v p,\v K;\sigma,s,m}^\dag
|v\rangle=E_{\nu\v p\v K}\,\mathcal{T}_{\nu,\v
p,\v K;\sigma,s,m}^\dag|v\rangle+\sum_{\nu',\v p'}
C_{\nu'\v p';\nu\v p}^\mathrm{dir}\,\mathcal{T}
_{\nu',\v p',\v K;\sigma,s,m}^\dag|v\rangle\ ,
\end{equation}
where $E_{\nu\v p\v K}$ is the energy of the free
e-X pair $(\nu,\v p,\v K)$,
\begin{equation}
E_{\nu\v p\v K}=\varepsilon_\nu^{(X)}+\epsilon_{\v
p}^{(eX)}+\mathcal{E}_{\v K}^{(T)}\ ,
\end{equation}
with $\epsilon_{\v p}^{(eX)}$ and
$\mathcal{E}_{\v K}^{(T)}$ being the relative
motion energy of the e-X pair and the
center of mass energy of this pair, defined in
eqs.\ (3.14-15). Note that, for $(\v Q,\v k;\v K,\v
p)$ linked by eq.\ (3.22), we do have
\begin{equation}
\epsilon_{\v k}^{(e)}+\mathcal{E}_{\v Q}^{(X)}
=\epsilon_{\v p}^{(eX)}+\mathcal{E}_{\v K}^{(T)}\
.
\end{equation}

\subsubsection{Creation operators for $(S_z=\pm1)$
trions}

The trions with a spin projection $S_z=\pm 1$ have
a total spin $S=1$, so that their relative motion
index belongs to the $\eta_1$ set. Moreover, they
have to be constructed from a $\sigma=\pm 1/2$
electron and a $s=\pm 1/2$ exciton. Let us
introduce the operator [18]
\begin{equation}
\mathrm{T}_{\eta_1,\v K;1,\pm
1,m}^\dag=\frac{1}{\sqrt{2}}
\sum_{\nu,\v p}\langle\nu,\v p|\eta_1\rangle\,
\mathcal{T}_{\nu,\v p,\v K;\pm 1/2,\pm
1/2,m}^\dag\ ,
\end{equation}
which is similar to the exciton creation
operator defined in eq.\ (3.7), except for the
$1/\sqrt{2}$ prefactor which is made for this
operator to create a normalized
two-electron-one-hole state. Indeeed, from
eqs.\ (3.19) and (3.33), we can check that
\begin{equation}
\langle v|\mathrm{T}_{\eta_1,\v K;1,\pm
1,m}\,\mathrm{T}_{\eta_1,\v K;1,\pm
1,m}^\dag|v\rangle=\frac{1}{2}\,2\,\sum_{\nu,\v
p}|\langle\nu,\v p|\eta_1\rangle|^2=1\ ,
\end{equation}
This $\mathrm{T}_{\nu,\v K;1,\pm
1,m}^\dag$ is actually the creation operator
for
$(S=1)$,
$(S_z=\pm 1)$ trions, since, due to eqs.\ (3.26)
and (3.37), we do have
\begin{equation}
H\,\mathrm{T}_{\eta_1,\v K;1,\pm
1,m}^\dag|v\rangle= E_{\eta_1,\v
K}^{(T)}\,\mathrm{T}_{\eta_1,\v K;1,\pm
1,m}^\dag|v\rangle\ ,
\end{equation}
with
\begin{equation}
E_{\eta_S,\v K}^{(T)}=\varepsilon_{\eta_S}^{(T)}
+\mathcal{E}_{\v K}^{(T)}\ .
\end{equation}

\subsubsection{Creation operators for $(S_z=0)$
trions}

$(S_z=0)$ trions have a total spin $S$ either equal
to 0 or to 1, so that their relative motion
indices can be either an $\eta_0$ or an $\eta_1$.
Moreover, they can be built either from
a $(\sigma=1/2)$ electron and a $(s=-1/2)$ exciton,
or the reverse. However, as the two corresponding
e-X operators are linked
by the relation (3.34), we are led to introduce
just the first type of e-X pairs [18],
\begin{equation}
\mathrm{T}_{\eta_S,\v K;S,0,m}^\dag=\sum_{\nu,\v
p}\langle\nu,\v
p|\eta_S\rangle\,\mathcal{T}_{\nu,\v p,\v
K;+1/2,-1/2,m}^\dag\ ,
\end{equation}
since from eqs.\ (3.19) and (3.34), this operator
also reads
\begin{equation}
\mathrm{T}_{\eta_S,\v
K;S,0,m}^\dag=-(-1)^S\,\sum_{\nu,\v
p}\langle\nu,\v
p|\eta_S\rangle\,\mathcal{T}_{\nu,\v p,\v
K;-1/2,+1/2,m}^\dag\ ,
\end{equation}
which makes clear the fact that such a $(S_z=0)$
trion can be built either from a $(\sigma=1/2)$
electron and a
$(s=-1/2)$ exciton or from a $(\sigma=-1/2)$
electron and a $(s=1/2)$ exciton.

Using again eqs.\ (3.26) and (3.37), it is
straightforward to check that this
$\mathrm{T}_{\eta_S,\v K;S,0,m}^\dag$ is indeed a
trion creation operator, since we do have
\begin{equation}
H\,\mathrm{T}_{\eta_S,\v
K;S,0,m}^\dag|v\rangle=E_{\eta_S,\v K}^{(T)}\,
\mathrm{T}_{\eta_S,\v K;S,0,m}^\dag|v\rangle\ ,
\end{equation}
while, from eqs.\ (3.19) and
(3.33), one can check that it indeed creates a
normalized
$(S_z=0)$ trion state.

Let us end by noting that the trion states defined
above form an orthogonal basis, since we do have
\begin{equation}
\langle v|\mathrm{T}_{\eta'_{S'},\v
K';S',S_z',m'}\,\mathrm{T}_{\eta_S,\v
K;S,S_z,m}^\dag|v\rangle=
\delta_{S',S}\,\delta_{S_z',S_z}\,
\delta_{m',m}\,\delta_{\eta'_{S'},\eta_S}\,\delta_{\v
K',\v K}\ ,
\end{equation}
while their closure relation reads
\begin{equation}
1=\sum_{S,S_z,m,\eta_S,\v K}\mathrm{T}_{\eta_S,\v
K;S,S_z,m}^\dag|v\rangle\,\langle
v|\mathrm{T}_{\eta_S,\v K;S,S_z,m}\ .
\end{equation}

Finally, it is easy to check that, in the same way
as trions can be written in terms of e-X pairs
according to eqs.\ (3.40) and (3.44), e-X pairs
can be written in terms of trions, according to
\begin{equation}
\mathcal{T}_{\nu,\v p,\v K;\pm 1/2;\pm 1/2,m}^\dag
=\sqrt{2}\sum_{\eta_1}\langle\eta_1|\nu,\v
p\rangle\,\mathrm{T}_{\eta_1,\v K;1,\pm 1,m}^\dag\
,
\end{equation}
\begin{equation}
\mathcal{T}_{\nu,\v p,\v K;+1/2;-1/2,m}^\dag
=\sum_{S,\eta_S}\langle\eta_S|\nu,\v
p\rangle\mathrm{T}_{\eta_S,\v K;S,0,m}^\dag\ ,
\end{equation}
\begin{equation}
\mathcal{T}_{\nu,\v p,\v K;-1/2;+1/2,m}^\dag
=-\sum_{S,\eta_S}(-1)^S\langle\eta_S|\nu,\v
p\rangle\,\mathrm{T}_{\eta_S,\v K;S,0,m}^\dag\ .
\end{equation}
Although somewhat more complicated due to the
importance of spins for trions, these equations
are the analogues of eq.\ (3.8) relating free
e-h pairs to excitons.

\subsection{Many-body effects between electrons
and excitons}

In usual many-body problems, the Hamiltonian
splits as $H=H_0+V$, so that the many-body effects
result from eq.\ (2.5) and its iteration (2.6). In
the case of many-body effects with excitons, such
a separation of the Hamiltonian is not possible,
due to the composite nature of the exciton.
Attempts have been made to produce a potential
$V_{XX}$ between excitons by bosonizing them.
However, quite recently, we have
shown that these procedures, although rather
sophisticated, fail to give the correct answer to
physical quantities such as the exciton lifetime
and the exciton-exciton scattering rate [25],
whatever the X-X scattering used in $V_{XX}$ is.
We have also shown that they fail to give the
correct nonlinear susceptibilities [26], because
they miss purely Pauli many-body effects,
\emph{i}.\
\emph{e}., scattering processes which exist in the
absence of any Coulomb interaction. This is why we
will not here use the bosonization procedures to
describe the interaction of one electron with one
exciton.

It is actually possible to handle many-body
effects with exact excitons properly, by noting
that eq.\ (3.35) leads to [24]
\begin{equation}
\frac{1}{a-H}\,B_n^\dag=B_n^\dag\,\frac{1}{a-H-E_n
^{(X)}}+\frac{1}{a-H}\,V_n^\dag\,\frac{1}{a-H-E_n
^{(X)}}\ .
\end{equation}
The above equation, which is the equivalent of
eq.\ (2.5) for usual many-body effects, is the key
equation for many-body effects involving excitons.
It cannot be iterated as simply as
eq.\ (2.6). It is however possible to generate such
an iteration by having eq.\ (3.52) acting
on excitons or on electrons and by using either
$[V_n^\dag,B_n^\dag]$ given in eq.\ (3) of ref.\
[22] for many-body effects between
excitons, or
$[V_n^\dag,a_{\v k,\sigma}^\dag]$, given in eq.\
(3.36), for many-body effects
between excitons and electrons.

\subsubsection{$(a-H)^{-1}$
acting on e-X pairs}

Equations (3.36) and (3.52) give $(a-H)^{-1}$
acting on one e-X pair as [18]
\begin{equation}
\frac{1}{a-H}\,\mathcal{T}_{\nu,\v
p,\v
K;\sigma,s,m}^\dag|v\rangle=\left[\mathcal{T}_
{\nu,\v p,\v K;\sigma,s,m}^\dag|v\rangle+\sum_
{\nu',\v p'}\,\frac{1}{a-H}\,\mathcal{T}_
{\nu',\v p',\v K;\sigma,s,m}^\dag|v\rangle
C_{\nu'\v p';\nu\v p}^{\mathrm{dir}}\right]
\frac{1}{a-E_{\nu\v p\v K}}\ ,
\end{equation}
where $E_{\nu\v p\v K}$ is the free e-X pair
energy given in eq.\ (3.38).

If we now iterate eq.\ (3.53), we find
\begin{equation}
\frac{1}{a-H}\,\mathcal{T}_{\nu,\v
p,\v K;\sigma,s,m}
^\dag |v\rangle
=\sum_{\nu',\v p'}
A_{\nu'\v p';\nu\v p}(a,\v K)\,
\mathcal{T}_
{\nu',\v p',\v
K;\sigma,s,m}^\dag |v\rangle\ ,
\end{equation}
where the prefactor $A_{\nu'\v p';\nu\v p}(a,\v K)$
expands on the e-X \emph{direct} Coulomb
scatterings only, through
\begin{eqnarray}
A_{\nu'\v p';\nu\v p}(a,\v
K)=\frac{\delta_{\nu',\nu}\,\delta_{\v p',\v p}}
{a-E_{\nu\v p\v K}}+\frac{C_{\nu'\v p';\nu\v p}
^{\mathrm{dir}}}{(a-E_{\nu'\v p'\v K})
(a-E_{\nu\v p\v K})}\hspace{4cm}\nonumber
\\ +\sum_{\nu_1,\v p_1}
\frac{C_{\nu'\v p';\nu_1\v p_1}^{\mathrm{dir}}
C_{\nu_1\v p_1;\nu\v p}^{\mathrm{dir}}}
{(a-E_{\nu'\v p'\v K})(a-E_{\nu_1\v p_1\v K})
(a-E_{\nu\v p\v K})}+\cdots
\end{eqnarray}
It corresponds to the ladder processes between
electron and exciton shown in fig.\ (6).

Just as the summation of the e-h ladder processes
producing the exciton reads in terms of
exciton energies and wave functions, the summation
of these e-X ladder processes producing the trion
should read in terms of trion energies
and wave functions. Let us now show it.

\subsubsection{Sum of e-X ladder processes}

If we take the
scalar product of
$\langle v|
\mathcal{T}_{\nu'',\v p'',\v K;+1/2,-1/2,m}$ with
eq.\ (3.54) taken for $\sigma=-s=1/2$, we find from
equation (3.33)
\begin{equation}
\langle v|\mathcal{T}_{\nu'',\v p'',\v
K;+1/2,-1/2,m}
\frac{1}{a-H}\,\mathcal{T}_{\nu,\v p,\v
K;+1/2,-1/2,m}^\dag|v\rangle=A_{\nu''\v p'';\nu\v
p}(a,
\v K)\ .
\end{equation}
If we now insert the trion closure relation
(3.48) in the LHS of the above equation, we get,
using eqs.\ (3.42) and (3.46),
\begin{eqnarray}
A_{\nu''\v p'';\nu\v p}(a,\v
K)=
\sum_{S',S_Z',m',\eta_{S'}',\v
K'}\hspace{9cm}\nonumber\\
\frac{\langle v|\mathcal{T}_{\nu'',\v
p'',\v K;+1/2,-1/2,m}\,\mathrm{T}_{\eta_{S'}',\v
K';S',S_z';m'}^\dag |v\rangle\langle
v|\mathrm{T}_{\eta_{S'}',\v K';S',S_z';m'}\,
\mathcal{T}_{\nu,\v p,\v
K;+1/2,-1/2,m}^\dag|v\rangle} {a-E_{\eta_{S'}',\v
K'}^{(T)}}\ .
\end{eqnarray}
From the expansion of e-X pairs in terms of trions
given in eq.\ (3.50), we immediately find, since
trions form an orthogonal basis,
\begin{equation}
A_{\nu''\v p'';\nu\v p}(a,\v K)=\sum_{S=0,1}
A_{\nu''\v p'';\nu\v p}^{(S)}(a,\v K)\ ,
\end{equation}
\begin{equation}
A_{\nu''\v p'';\nu\v p}^{(S)}(a,\v
K)=\sum_{\eta_S}
\frac{\langle\nu'',\v p''|\eta_S\rangle
\langle \eta_S|\nu,\v p\rangle}
{a-E_{\eta_S,\v K}^{(T)}}\ .
\end{equation}
We see that $A_{\nu''\v p'';\nu\v p}(a,\v K)$
contains contributions from both, singlet trions
$(S=0)$ and triplet trions $(S=1)$.

For physical understanding, it can be of interest
to note that, if we set
$a'=a-\mathcal{E}_{\v K}^{(T)}$, the
compact expression of
$A_{\nu''\v p'';\nu\v p}(a,\v K)$ in terms of
trions, given in eqs.\ (3.58-59), reads
\begin{equation}
A_{\nu''\v p'';\nu\v p}(a,\v K)=\hat{A}_{\nu''\v
p'';
\nu\v p}(a')=\langle\nu'',\v
p''|\frac{1}{a'-h_\mathrm{T}}|\nu,\v p\rangle
\ ,
\end{equation}
while its expansion in e-X Coulomb scatterings
given in eq. (3.55) just corresponds to
\begin{equation}
\hat{A}_{\nu''\v p'';\nu\v p}(a')=\langle\nu'',\v
p''|
\frac{1}{a'-h_T^{(0)}}|\nu,\v p\rangle
+\langle\nu'',\v
p''|\frac{1}{a'-h_T^{(0)}} w(\v r,\v u)
\frac{1}{a'-h_T^{(0)}}|
\nu,\v p\rangle+\cdots\ ,
\end{equation}
due to the link between $C_{\nu'\v p';\nu\v
p}^\mathrm{dir}$ and $w(\v r,\v u)$ given in eqs.\
(3.27-28). It is then obvious that eq.\ (3.61) just
follows from eq.\ (3.60), since for
$h_T=h_T^{(0)}+w(\v r,\v u)$, we do have
\begin{equation}
\frac{1}{a'-h_T}=\frac{1}{a'-h_T
^{(0)}}+\frac{1}{a'-h_T^{(0)}}\,w(\v r,\v
u)\,
\frac{1}{a'-h_T^{(0)}}+\cdots\ .
\end{equation}

Let us stress that the summation of e-X ladder
processes has first been established
from a quite formal procedure designed to treat
electrons interacting with excitons. It can
\emph{a priori} be used in
\emph{any} other problem dealing with electrons and
excitons, not just in the case of one
electron plus one exciton, \emph{i}.\
\emph{e}., one trion. It is however clear that we
can also use it in this simple problem too, for
which a formulation in first quantization through
the trion Hamiltonian
$H_T(\v R_T,\v r,\v u)$ is simple enough to be of
practical use. This first quantization formulation
of the one-trion problem actually provides
enlightening foreshortenings to some results on
e-X pairs. In more complicated situations however,
second quantization along with the formal
definitions of the direct Coulomb and
exchange scatterings
between electrons and excitons it allows, are the
only possible way to treat problems dealing with
one hole and more than two electrons.  The two
aspects of the same results are however of
interest for the understanding of the trion
physics.

\section{Photon absorption using electron-exciton
diagrams}

In section 2, we have derived the photon
absorption using e-h diagrams in the
case of exciton formation and in the case of trion
formation. We have shown that these standard
Feynman diagrams are in fact totally inappropriate
for trions. In section 3, we have derived all the
tools necessary to propose a new diagrammatic
procedure for photon absorption in terms of
excitons, while taking into account the fact that
the exciton can also be made with any
of the electrons present in the sample, through the
exchange ``scatterings'' generated by the
``commutation technique''.

This new formulation of photon absorption is in
fact quite natural: Indeed, the
semiconductor-photon interaction, given in eq.\
(2.3) in terms of free electrons and free holes,
can also be written in terms of excitons. From
eq.\ (3.8) and the fact that
$\sqrt{\Omega}\langle\v p|\v r=\v 0\rangle=1$, we
readily get
\begin{equation}
U_{\pm}^\dag(\v Q_p)=\sum_\nu\lambda_\nu^\ast\,
B_{\nu,\v Q_p,\mp,\pm}^\dag\ ,
\end{equation}
\begin{equation}
\lambda_\nu^\ast=\lambda^\ast\sqrt{\Omega}\langle
\nu|\v r=\v 0\rangle\ .
\end{equation}
This just corresponds to the well known
enhancement factor of the coupling to exciton
compared to free pairs, as $|\langle\nu|\v r=\v
0\rangle|^2$ is of the order of the inverse
exciton volume: With respect to this enhanced
coupling already, the exciton representation
appears to us somewhat better.

\subsection{Photon absorption with exciton
formation}

From the response function given in eq.\ (2.4) in
the case of a semiconductor with no carrier,
irradiated with $\sigma_+$ photons, we get
\begin{equation}
S_X=N_p\sum_{\nu',\nu}\lambda_{\nu'}\,
\langle v|B_{\nu',\v Q,-,+}\,
\frac{1}{\omega_p-H+i\eta}\,B_{\nu,\v Q_p,-,+}
^\dag|v\rangle\,\lambda_\nu^\ast\ .
\end{equation}
This readily gives
\begin{equation}
S_X=N_p\sum_\nu\lambda_\nu\,G^{(X)}(\omega_p,\v
Q_p;\nu)\,\lambda_\nu^\ast\ ,
\end{equation}
in which we have set
\begin{equation}
G^{(X)}(\omega,\v Q;\nu)=\frac{1}{\omega-E_{\nu,\v
Q}^{(X)}+i\eta}\ .
\end{equation}
$G^{(X)}(\omega,\v Q;\nu)$ can be seen as an
exciton propagator; so that we are led to
represent $S_X$ by the exciton diagram of fig.\
(7a), which is already far simpler than the set of
electron-hole ladder diagrams of fig.\ (1).

\subsection{Photon absorption with trion formation}

From eqs.\ (2.4) and (4.1), the response
function for an initial state already having one
electron with momentum $\v k_i$ and spin $s_i$ and
an absorbed photon
$\sigma_{\pm}$, reads
\begin{equation}
S_{\pm,s_i}=N_p\sum_{\nu',\nu}\lambda_{\nu'}\,
\langle v|a_{\v
k_i,s_i}\, B_{\nu',\v
Q_p,\mp,\pm}\,\left(\frac{1}{\omega_p
+\epsilon_{\v k_i}^{(e)}-H+i\eta}\right)\,B_{\nu,
\v Q_p,\mp,\pm}^\dag\,a_{\v k_i,s_i}^\dag|v\rangle
\,\lambda_\nu^\ast\ .
\end{equation}
If we introduce the appropriate momenta $(\v p_i,\v
K_i)$ of the e-X pair made of the initial electron
and the photocreated virtual exciton, defined by
\begin{equation}
\v K_i=\v k_i+\v Q_p,\hspace{3cm}
\v k_i=\v p_i+\beta_e\v K_i,
\end{equation}
we can rewrite $S_{\pm,s_i}$ as
\begin{equation}
S_{\pm,s_i}=N_p\sum_{\nu',\nu}\lambda_{\nu'}\,
\langle v|\mathcal{T}_{
\nu',\v p_i,\v K_i;s_i,\mp,\pm}\,\left(\frac{1}
{\omega_p+\epsilon_{\v k_i}^{(e)}-H+i\eta}\right)
\,\mathcal{T}_{\nu,\v p_i,\v K_i;s_i,\mp,\pm}
^\dag|v\rangle\,\lambda_\nu^\ast\ .
\end{equation}
So that, by using eq.\ (3.54),which gives
$(a-H)^{-1}$ acting on one e-X pair, we find
\begin{equation}
S_{\pm,s_i}=N_p\sum_{\nu',\nu}
\sum_{\nu_1,\v
p_1}\lambda_{\nu'}\,\langle v|\mathcal{T}
_{\nu',\v p_i,\v K_i;s_i,\mp,\pm}\,\mathcal{T}
_{\nu_1,\v p_1,\v
K_i;s_i,\mp,\pm}^\dag|v\rangle\,A_{\nu_1\v p_1;
\nu\v p_i}(a_i,\v K_i)\,\lambda_\nu^\ast\ ,
\end{equation}
with $a_i=\omega_p+\epsilon_{\v k_i}^{(e)}+i\eta$.
In order to go further, we note that the scalar
product of e-X states, given by eq.\ (3.33),
depends if the photocreated electron and the
initial electron have the same spin or not.

\subsection{Photocreated electron with
spin different from the initial one}

When the spins are different, the
scalar product of e-X states appearing in eq.\
(4.9) differs from zero for
$\nu_1=\nu'$ and $\v p_1=\v p_i$ only, so that we
simply have
\begin{equation}
S_{\neq}=N_p\sum_{\nu',\nu}\lambda_{\nu'}\,
 A_{\nu'\v p_i;\nu\v p_i}(a_i,\v
K_i)\,\lambda_\nu^\ast
\ .
\end{equation}
Due to eq.\ (3.58), the response function
contains contributions from singlet trions $(S=0)$
and triplet trions $(S=1)$ as expected, since with
two different electron spins, we can construct the
two types of trions. By using the expansion of
$A_{\nu'\v p';\nu\v p} (a,\v K)$ in direct Coulomb
scatterings given in eq.\ (3.55), we can expand
$S_{\neq}$ as
\begin{equation}
S_{\neq}=\sum_{n=0}^{+\infty}
S_{\neq}
^{(n)}\ ,
\end{equation}
where $S_{\neq}^{(n)}$ has
$n$ electron-exciton scatterings $C_{\nu_1\v
p_1;\nu_2\v p_2}^{\mathrm{dir}}$.

\subsubsection{Zero order term in e-X
interactions}

The term without any e-X scattering is given
by
\begin{equation}
S_{\neq}^{(0)}=N_p\sum_\nu\frac{|\lambda_\nu|^2}
{a_i-E_{\nu\v
p_i\v K_i}}=N_p\sum_\nu\lambda_\nu\,
G^{(X)}(\omega_p,\v Q_p;\nu)\,\lambda_\nu^\ast\ ,
\end{equation}
and corresponds to the diagram of fig.\ (7a).
Before going further, let us note that we can
rewrite this zero order term in a compact form,
without any explicit reference to exciton
states, as
\begin{equation}
S_{\neq}^{(0)}=N_p|\lambda|^2\Omega\,\left\langle
\v r=\v
0\left |\frac{1}{\omega_p-(h_X+\mathcal{E}_{\v
Q_p}^{(X)}) +i\eta}\right |\v r=\v 0\right\rangle\
,
\end{equation}
where $h_X$ is the exciton relative motion
Hamiltonian,
$(h_X-\epsilon_\nu^{(X)})|\nu\rangle =0$.

\subsubsection{First order term}

The first order term in e-X scattering is zero
since direct Coulomb processes impose non-zero
momentum transfers, so that $C_{\nu'\v p';\nu\v p}
^{\mathrm{dir}}=0$ for $\v p'=\v p$ (see eqs.\
(3.27-28)). This has to be contrasted with e-h
diagrams for which a first order term exists. Let
us however stress that this e-h diagram first order
term is just a part of the ladder processes giving
rise to the exciton propagator, so that it is in
fact already included in the zero order exciton
diagram of fig.\ (7a).

\subsubsection{Second order term}

Using eqs.\ (3.28) and (3.55), the second order
term in e-X scatterings reads
\begin{eqnarray}
S_{\neq}^{(2)}=N_p\sum_{\nu',\nu}\lambda_
{\nu'}\,G^{(X)}(\omega_p,\v
Q_p;\nu')\left[\sum_{\nu_1,
\v q_1}
\frac{W_{-\v
q_1}^{\nu'\nu_1}\,W_{\v
q_1}^{\nu_1\nu}}{\Delta_{\nu_1,\v q_1}}
\right]\,G^{(X)}(\omega_p,\v
Q_p;\nu)\,\lambda_\nu^\ast\ ,
\end{eqnarray}
in which we have set
\begin{equation}
\Delta_{\nu,\v q}=\omega_p-\left(E_{\nu,\v Q_p+\v
q}^{(X)}+\epsilon_{\v k_i-\v q}^{(e)}-\epsilon_
{\v k_i}^{(e)}\right)+i\eta\ .
\end{equation}
It is easy to check that the bracket of the above
equation can also be written as
\begin{equation}
\sum_{\nu_1,\v
q_1}\int\frac{id\omega_1}{2\pi}\,W_{-\v
q_1}^{\nu'\nu_1}\,B(\omega_1,\v
q_1)\,G_X(\omega_p+\omega_1,\v Q_p+\v
q_1;\nu_1)\,W_{\v q_1}^{\nu_1\nu}\ ,
\end{equation}
where $B(\omega_1,\v q_1)$ is the standard bubble
for the excitation of one e-h pair in the ``Fermi
sea'', here reduced to the $(\v k_i,s_i)$
electron, namely
\begin{equation}
B(\omega_1,\v q_1)=(-1)\sum_{\v k}\int\frac{id
\omega}{2\pi}\,g_i^{(e)}(\omega,\v
k)\,g_i^{(e)} (\omega-\omega_1,\v k-\v q_1)\ .
\end{equation}
where $g_i^{(e)}(\omega,\v
k)=(\omega-\epsilon_{\v k}^{(e)}+i\eta S_{\v k,\v
k_i})^{-1}$, with $S_{\v k,\v
k_i}=+1$ if $\v k\neq\v k_i$ and $S_{\v k,\v
k_i}=-1$ if $\v k=\v k_i$.
Consequently, this second order term just
corresponds to the diagram of fig.\ (7b), with the
e-X direct scattering vertex given by eq.\ (3.27).

It is also possible to rewrite this second order
term in a compact form, without any reference to
exciton states, as
\begin{eqnarray}
S_{\neq}^{(2)}=N_p|\lambda|^2\Omega\left\langle\v
r=\v 0\left |\frac{1}{\omega_p-(h_X+\mathcal{E}_{\v
Q_p}^{(X)})+i\eta}\right.\right.\nonumber
\\ \times
\left[\sum_{\v q_1}w_{\v q_1}(\v r)
\frac{1}{\omega_p-(h_X+\mathcal{E}_{\v Q_p+\v
q_1}^{(X)}+\epsilon_{\v k_i-\v q_1}^{(e)}
-\epsilon_{\v k_i}^{(e)})+i\eta}\,w_{-\v q_1}(\v r)
\right]\nonumber
\\ \times
\left.\left.\frac{1}{\omega_p-(h_X+\mathcal{E}_{\v
Q_p}^{(X)}) +i\eta}\,\right |\v r=\v
0\right\rangle\ .
\end{eqnarray}

This unique diagram (7b) has to be contrasted to
the six diagrams corresponding to the second
order term in $V_{\v q}$ of the standard e-h
many-body procedure (the
second order ladder diagram of fig.\ (1c) plus the
5 diagrams of fig.\ (2)). Let us again note
that the e-X and e-h many-body procedures are not
strictly equivalent: Diagram (7b) contains
terms in
$V_{\v q}^{(n)}$ with $n>2$, not included in
fig.\ (2). They are somehow ``hidden'' in
the exciton propagator $G^{(X)}$ which in fact
includes all ladder processes between the
photocreated electron and the hole.

\subsubsection{Third order term}

In the same way, the third order term in e-X
interactions reads
\begin{eqnarray}
S_{\neq}^{(3)}=N_p\sum_{\nu',\nu}\lambda_{\nu'}
\,G^{(X)}
(\omega_p,\v
Q_p;\nu')\hspace{3cm}\nonumber
\\ \times \left[\sum_{\nu_1,\nu_2;\v
q_1,\v q_2} \frac{W_{-\v
q_2}^{\nu'\nu_2}\,W_{\v q_2-\v
q_1}^{\nu_2\nu_1}\,W_{\v
q_1}^{\nu_1\nu}}{\Delta_{\nu_2,\v q_2}\,
\Delta_{\nu_1,\v q_1}}\right]\,
G_X(\omega_p,\v
Q_p;\nu)\,\lambda_\nu^\ast\ .
\end{eqnarray}
Using the exciton-photon vertex, the
e-X direct scattering and the exciton propagator
defined above, as well as the $(\omega,\v q)$
conservations standard for diagrams, it is easy to
show that this third order term just corresponds
to the diagram of fig.\ (7c).

This unique diagram has again to be contrasted
with the twenty one diagrams of the e-h Feynman
diagram procedure at third order in Coulomb
interaction (the third
order ladder diagram of fig.\ (1d) plus the 20
diagrams of fig.\ (3)): This again shows that our
e-X diagrams are far more convenient than the usual
e-h Feynman diagrams, for this simple problem on
trions already. In more complicated problems, they
should be even more convenient. This is why, by
using them, we can have some hope to calculate
quantities which may appear as impossible to obtain
from the usual e-h many-body procedure.

Before going further, we can note that this third
order term also takes the compact form \newpage
\begin{eqnarray}
S_{\neq}^{(3)}=N_p|\lambda|^2\Omega\,\left\langle\v
r=
\v 0\left|\frac{1}{\omega_p-(h_X+\mathcal{E}_{\v
Q_p}^{(X)})+i\eta}\right.\right.\nonumber\\
\times
\left[\sum_{\v q_1,\v q_2}
w_{\v q_2}(\v
r)\,\frac{1}{\omega_p-(h_X+\mathcal{E}_{\v Q_p+\v
q_2}^{(X)}+\epsilon_{\v k_i-\v
q_2}^{(e)}-\epsilon_{\v k_i}^{(e)})+i\eta}\right.
\nonumber
\\ \times w_{\v q_1-\v q_2}(\v r)
\left.\frac{1}{\omega_p-(h_X+\mathcal{E}_{\v Q_p+\v
q_1} ^{(X)}+\epsilon_{\v k_i-\v
q_1}^{(e)}-\epsilon_{\v k_i}^{(e)})+i\eta}\,w_{-\v
q_1}(\v r)\right]\nonumber
\\
\times
\left.\left.\frac{1}{\omega_p-(h_X+\mathcal{E}_{\v
Q_p}^{(X)})+i\eta}\right |\v r=\v 0\right\rangle.
\end{eqnarray}

\subsubsection{Higher order terms}

While we did not even dare to draw the e-h Feynman
diagrams at fourth order in Coulomb
interaction, the simplicity of the e-X diagram
procedure may lead us to think that the fourth
order terms in e-X interactions should be given by
the ladder diagram of fig.\ (7d). And similarly
at higher orders. The correct result is somewhat
more subtle as we now show.

The standard diagrammatic procedure with electron
propagators, that we partly use here, for example
in the bubble $B(\omega_1,\v q_1)$ of eq.\ (4.17),
is actually designed to describe e-h
\emph{excitations} of the Fermi sea, \emph{i}.\
\emph{e}., processes in which $\v k\neq\v k-\v
q_n$. While at first and second order in e-X
Coulomb processes, this is automatically fulfilled
due to the $\v q\neq\v 0$ constraint on
$W_{\nu'\nu}(\v q)$ scatterings, this is no more
imposed at higher orders. Indeed, while in the
fourth order term of
$A_{\nu'\v p_i;\nu\v p_i}(a_i,\v K_i)$, the first
and last scatterings still impose $\v q_1\neq\v
0$ and $\v q_3\neq\v 0$, the intermediate
scatterings simply impose
$(\v q_2-\v q_1)\neq\v 0$ and $(\v q_3-\v
q_2)\neq\v 0$ so that we can possibly have $\v
q_2=\v 0$. However, the precise calculation of the
ladder diagram of fig.\ (7d) confirms that this
diagram only has excited Fermi sea e-h pairs $(\v
k,\v k-\v q_n)$ with $\v q_n\neq\v 0$: The $\v
q_2=0$ ones are missing. Since the expansion of
$A_{\nu'\v p_i;\nu\v p_i}(a_i,\v K_i)$ does
contain all possible scatterings of the
photocreated exciton, \emph{i}.\ \emph{e}., the
$\v q_n=\v 0$ scatterings too, \emph{we must
add} the diagram of fig.\ (7e) to the one of fig.\
(7d), in order to have all the fourth order terms
of
$S_{\neq}$. We can then be tempted
to separate this additional diagram (and the
similar ones at higher orders) from the set of
ladder diagrams with 0, 1, 2 scatterings between
the photocreated exciton and the electron, in
which the exciton is always in a state $\v Q\neq\v
Q_p$. As explained in more details below, this
idea, which can appear physically appealing, turns
out to be a bad one.

To conclude, we can say that, when the
photocreated electron and the initial electron
have different spins, the response function
$S_{\neq}$ corresponds to all possible diagrams
between one electron and one exciton, with the
following characteristics:

(i) These diagrams have one exciton
line only, going from right to left, since there
is one deep hole only: the photocreated one.

(ii) The photocreated (virtual) exciton suffers
various scatterings, without any electron exchange
with the initial electron, because, for two
electrons having different spins, the deep hole can
only recombine with the photocreated electron.

(iii) These diagrams have a unique
\emph{conduction-hole} line, going from left to
right: Since the initial state has one electron
$(\v k_i,s_i)$ only, the corresponding
initial ``Fermi sea'' can have one hole only,
with a well defined
momentum, namely $\v k_i$, so that this hole cannot
scatter. With such an essentially empty ``Fermi
sea'', the conduction electron $(s_i)$ has, on the
opposite, plenty of sites to scatter.

\subsubsection{Summation of e-X ladder
diagrams}

In the case of a photocreated electron with spin
different from the initial one, the trion
absorption response function $S_{\neq}$ thus
corresponds to the set of e-X ladder diagrams
shown in fig.\ (7). They contain direct
Coulomb processes only, \emph{the exciton being
always made with the same electron}.

These diagrams can be summed up in
terms of the trion relative motion energies and
wave functions. Using eqs.\ (3.58-59), we find
\begin{equation}
A_{\nu'\v p_i;\nu\v p_i}(a_i,\v
K_i)=\sum_S\sum_{\eta_S}\frac{\langle\nu',\v
p_i|\eta_S\rangle\langle\eta_S|\nu,\v p_i\rangle}
{a_i-E_{\eta_S,\v K_i}^{(T)}}\ ,
\end{equation}
where $\v p_i$ is the relative motion momentum
of the e-X pair made of the initial electron and
the photocreated exciton defined in eq.\ (4.7).
According to eqs.\ (4.2,4.10), this leads to
\begin{equation}
S_{\neq}=N_p|\lambda|^2\Omega\sum_S\sum_{\eta_S}
\frac
{|\langle\v r=\v 0,\v p_i|\eta_S\rangle|^2}
{\omega_p-(E_{\eta_S,\v
K_i} ^{(T)}-\epsilon_{\v k_i}^{(e)})+i\eta}\ ,
\end{equation}
in agreement with our previous work on trion
absorption [19].

By using the trion relative motion Hamiltonian
$h_T$, which is such that
\linebreak $(h_T-\varepsilon_{\eta_S}
^{(T)})|\eta_S\rangle=0$, eq.\ (4.22)
also reads
\begin{equation}
S_{\neq}=N_p|\lambda|^2\Omega\,\left\langle\v r= \v
0,\v p_i\left
|\frac{1}{\omega_p-(h_T+\mathcal{E}_{\v
K_i}^{(T)}-\epsilon_{\v k_i}^{(e)})+i\eta}\right
|\v r=\v 0,\v p_i
\right\rangle\ ,
\end{equation}

From the above result, it is easy to
recover the compact expressions of the
various terms of $S_{\neq}$ given in
eqs.\ (4.13,4.18,4.20). Indeed, since
$h_T=h_T^{(0)}+w(\v r,\v u)$, we get, from eq.\
(4.23) and the expansion of
$1/(a'-h_T)$ given in eq.\ (3.62),
\begin{equation}
S_{\neq}=N_p|\lambda|^2\Omega\left\langle\v r=\v
0,\v p_i\left |\frac{1}{a_i'-h_T^{((0)}}+\frac{1}
{a_i'-h_T^{(0)}}\,w(\v r,\v u)\,\frac{1}{a_i'-h_T
^{(0)}}+\cdots\right |\v r=\v 0,\v
p_i\right\rangle\ ,
\end{equation}
where $a_i'=\omega_p+\epsilon_{\v k_i}^{(e)}-
\mathcal{E}_{\v
K_i}^{(T)}+i\eta$. We then insert the closure
relation for free e-X states $|\nu,\v p\rangle$ in
front of each
$1/(a_i'-h_T^{(0)})$ factor.
The $S_{\neq}$ zero order term is proportional to
\begin{equation}
\left\langle\v r=\v 0,\v p_i\left |\frac{1}
{a_i'-h_T^{(0)}}\right |
\v r=\v 0,\v
p_i\right\rangle =\left\langle\v r=\v
0\left |\frac{1}{a_i'-h_X-\epsilon _{\v
p_i}^{(eX)}}\right |\v r=\v 0\right\rangle\ ,
\end{equation}
which is nothing but the expectation value of
$1/(\omega_p-h_X-\mathcal{E}_{\v
Q_p}^{(X)}+i\eta)$ in the $|\v r=\v 0\rangle$
state, in agreement with eq.\ (4.13).

The first
order term of $S_{\neq}$ is proportional to
\begin{equation}
\sum_{\nu',\v p',\nu,\v p}\left\langle\v r=\v 0,\v
p_i\left |
\frac{1}{a_i'-h_T^{(0)}}\right |\nu',\v
p'\right\rangle
\langle\nu',\v p'|w(\v r,\v u)|\nu,\v
p\rangle\left\langle\nu,\v
p\left |\frac{1}{a_i'-h_T^{(0)}}\right |\v r=\v
0,\v p_i\right\rangle\ ,
\end{equation}
which imposes $\v p'=\v p_i=\v p$. It is thus
equal to zero, since Coulomb scatterings are
restricted to non-zero momentum transfers only.

In a similar way, the second order term is
proportional to
\begin{eqnarray}
\sum_{\nu',\v p',\nu,\v p,\nu_1,\v
p_1,\nu_2,\v p_2}\left\langle
\v r=\v 0,\v
p_i\left |\frac{1}{a_i'-h_T^{(0)}}\right |\nu',\v
p'\right\rangle\langle\nu',\v p'|w(\v r,\v
u)|\nu_2,\v p_2\rangle\hspace{2cm}\nonumber
\\ \times\left\langle\nu_2,\v
p_2\left |\frac{1}{a_i'-h_T^{(0)}}\right |\nu_1,\v
p_1\right\rangle\langle\nu_1,\v p_1|w(\v r,\v
u)|\nu,\v p\rangle\left\langle\nu,\v
p\left |\frac{1}{a_i'-h_T^{(0)}}\right |
\v r=\v 0,\v p_i\right\rangle\ ,
\end{eqnarray}
 which imposes $\v p'=\v p_i=\v p$ and $\v p_2=\v
p_1$. By writing
$\v p_1=\v p_i-\v q_1$, it is straightforward to
check that eq.\ (4.27) indeed leads to eq.\ (4.18).

And so on, for the higher order terms.

\subsubsection{An inappropriate separation}

Let us end this part by explaining why it is
inappropriate to treat separately the diagrams in
which the scattered excitons have a center of mass
momentum $\v Q_p+\v q_n$ equal to its initial
value $\v Q_p$.

With such a separation, we would be led to write
the response function $S_{\neq}$ as in fig.\ (8a),
where the cross represents all topologically
connected e-X ladder diagrams shown in fig.\ (8b),
\emph{i}.\
\emph{e}., diagrams in which all intermediate
excitons have their center of mass momentum $\v
Q_p+\v q_n$ different from $\v Q_p$.
Fig.\ (8a) would lead to write $S_{\neq}$ as
\begin{equation}
S_{\neq}=N_p\sum_{\nu',\nu}\lambda_{\nu'}
[\delta_{\nu',\nu}\,G_X
(\omega_p,\v
Q_p;\nu)+G_X(\omega_p,\v
Q_p;\nu')\,\Gamma_{\nu'\nu}\, G_X(\omega_p,\v
Q_p;\nu)+\cdots]\lambda_\nu^\ast\ .
\end{equation}
From the results obtained in the above
section, it is easy to show that the
exciton scattering $\Gamma_{\nu'\nu}$ associated
to the cross has the form
$\langle\nu'|T(\omega_p,\v Q_p,\v
k_i)|\nu\rangle$; so that all these diagrams can
be summed up as
\begin{eqnarray}
\left\langle\nu'\left |\frac{1}{\omega_p-h_X-
\mathcal{E}_{\v
Q_p}^{(X)} +i\eta}+\frac{1}{\omega_p-h_X-
\mathcal{E}_{\v
Q_p}^{(X)} +i\eta}\,T(\omega_p,\v Q_p,\v
k_i)\right.\right.\hspace{3,5cm}\nonumber \\
\times\ \left.\left.\frac{1}{\omega_p-h_X-
\mathcal{E}_{\v
Q_p}^{(X)} +i\eta}+\cdots\right |\nu\right\rangle
=\left\langle\nu'\left
|\frac{1}{\omega_p-\tilde{h}_X-\mathcal{E}_{\v
Q_p}^{(X)}+i\eta}\right |\nu\right\rangle\ ,
\end{eqnarray}
where $\tilde{h}_X$, equal to $h_X+T(\omega_p,\v
Q_p,\v k_i)$, appears as the Hamiltonian of the
photocreated exciton dressed by its possible
scatterings with the initial electron.

While physically appealing at first, this concept
turns out to be technically useless: Indeed,
the sum of e-X ladder processes can be summed up
easily in terms of singlet and triplet trions when
--- and only when ---
\emph{all} $\v q_n$'s are included. On the
opposite, the sum
of ladder processes restricted to
$\v q_n\neq \v 0$ as the ones appearing in
$T(\omega_p,\v Q_p,\v k_i)$ is not known.
Consequently, although physically nice at first,
eq.\ (4.29) which would lead to write the response
function as
\begin{equation}
S_{\neq}=N_p|\lambda|^2\Omega\left\langle\v r=\v
0\left |
\frac{1}{\omega_p-\tilde{h}_X-\mathcal{E}_{\v
Q_p}^{(X)} +i\eta}\right |\v r=\v 0\right\rangle\ ,
\end{equation}
is of no use. If we now take a mathematical point
of view, we hardly see how the
\emph{one}-body operator $\tilde{h}_X$ can be
transformed into a
\emph{two}-body operator, for the $|\nu,\v
p\rangle$ (or $|\v r,\v p\rangle$) states to
appear in the response function as they should, in
view of eq.\ (4.23). By comparing eqs.\ (4.23) and
(4.30), we note that the matrix elements between
$|\v r=\v 0\rangle$ states of
$(\omega_p-\tilde{h}_X-\mathcal{E}_{\v
Q_p}^{(X)}+i\eta)^{-1}$ and $\langle\v
p_i|[\omega_p-(h_T+\mathcal{E}_{\v
K_i}^{(T)}-\epsilon_{\v k_i}^{(e)})+i\eta]^{-1}|\v
p_i\rangle$  must be equal. This however does not
tell anything about the operators themselves.

\subsubsection{Conclusion}

The response function, in the case of a
photocreated electron with spin different from the
initial one, is very similar to the response
function of an exciton: It just contains e-X
ladder processes. It is however of importance to
stress that, while, in the exciton case, there is
no way to cut these ladder diagrams into
pieces, in the trion case, the exciton --- which
plays the role of the hole in the exciton ladder
processes --- can possibly return into a
$(\nu_n,\v Q_p)$ state after a set of scatterings.
This makes the corresponding diagram topologically
separable. These two types of diagrams have
however to be included in order to find all terms
appearing in the response function (see figs.\
(7d) and (7e)). Another important point is the
fact that no  additional diagram resulting from
possible electron exchanges exists, although, at
any stage, the exciton can
\emph{a priori} be made either with the
photocreated electron or with the initial electron.
This is basically due to the fact that, in a
quantum well, when the electron spins are
different, the hole can only recombine with the
photocreated electron; so that the number of
exchanges which can take place before
recombination, has to be even. Since two exchanges
reduce to an identity according to eq.\ (3.23),
two exchanges are basically the same as
no exchange at all.

A last comment: For most people, ``trion''  in fact
means ``ground state trion''. Such trion
corresponds to singlet state, \emph{i}.\
\emph{e}., state in which the two electrons have
different spins, so that it corresponds to the e-X
ladder diagrams described in this paragraph. The
real challenging difficulty with trions seen as
interacting e-X pairs, in fact arises when the two
electrons have the same spin,
\emph{i}.\
\emph{e}., when the hole can recombine with any of
the two electrons, so that electron exchanges have
to explicitly enter the problem.
Let us end this work by considering these triplet
trions, although they are not the interesting ones
in most experiments.

\subsection{Photocreated electron  with spin
identical to the initial one}

The consequences of the possible electron
exchanges in forming the exciton can appear
quite difficult to handle at first. This is
probably why  trions have not been treated as a
set of interacting e-X pairs up to now. Let us
show how our commutation technique allows to take
care of this electron exchange in a simple and
transparent way.

From a rapid look at eq.\ (4.22), we see
that the response function $S_{\neq}$
for different electron spins
reads in terms of trion states $|\eta_S\rangle$
with both
$S=0$ and $S=1$. This is reasonable since, with
$|+-\rangle$ electrons, we can form singlet and
triplet trions. On the opposite, if the two
electrons have the same spin, $|++\rangle$ for
example, we can only form $(S=1)$
trions, so that only
$(S=1)$ trions should enter the response function.
Since $A_{\nu'\v p';\nu\v p}(a,\v K)$
appearing when $1/(a-H)$ acts on e-X pairs, read
in terms of both $(S=0)$ and $(S=1)$ trions, the
exchange processes, which take place
when the two electron spins are identical, have to
withdraw the $(S=0)$
trion contributions from the final
result. Let us see how they do it.

If we go back to eq.\ (4.9), we find that the
scalar product of e-X states which enters
$S_{\pm,s_i}$, has
two terms instead of one when the electron
spins are the same (see eq.\ (3.33)). Beside the
$\nu_1=\nu'$ and
$\v p_1=\v p_i$ term, we also have an
exchange term, so that eq.\ (4.9) now leads to
\begin{equation}
S_{=}=N_p\sum_{\nu',\nu}\lambda_{\nu'}\,[A_{\nu'\v
p_i;\nu\v p_i}(a_i,\v K_i)-\sum_{\nu_1,\v
p_1}L_{\nu'\v p_i;\nu_1\v p_1}\, A_{\nu_1\v
p_1;\nu\v p_i}(a_i,\v K_i)]\,\lambda_\nu
^\ast\ .
\end{equation}

\subsubsection{Response function in terms of
trions}

From eqs.\ (3.19) and (3.59), we get
\begin{equation}
\sum_{\nu_1,\v p_1}L_{\nu'\v p_i;\nu_1\v p_1}\,
A_{\nu_1\v p_1;\nu\v p_i}^{(S)}=(-1)^S\,
A_{\nu'\v p_i;\nu\v p_i}^{(S)}\ ,
\end{equation}
so that the response function for identical
electron spins reads
\begin{equation}
S_{=}=2N_p|\lambda|^2\Omega\sum_{\eta_1}\frac{|
\langle\eta_1|\v r=\v 0,\v
p_i\rangle|^2} {\omega_p-(E_{\eta_1,\v K_i}
^{(T)}-\epsilon_{\v k_i}^{(e)})+i\eta}\ .
\end{equation}
It thus contains contributions from triplet trions
only, as expected. Let us note that, due to this
restriction, $S_{=}$ cannot be written in a
compact form in terms of the trion relative motion
Hamiltonian $h_T$, as for $S_{\neq}$ in eq.\
(4.23).

\subsubsection{Expansion in e-X diagrams}

We now come back to eq.\ (4.31) and use the
expression of
$A_{\nu'\v p';\nu\v p}(a,\v K)$ in terms of e-X
scaterings given in eq.\ (3.55). Beside the set of
direct ladder diagrams leading to $S_{\neq}$ and
which come from the first term of eq.\ (4.31),
there are additional exchange diagrams coming from
the second term. It is of interest to note that
these exchanges appear through the
$L_{\nu'\v p_i;\nu_1\v p_1}$ factor which takes
place
\emph{``at the end''} of a set of direct e-X
scatterings, just before the exciton
recombination, as $A_{\nu_1\v p_1;
\nu\v p_i}(a_i,\v K_i)$ contains direct Coulomb
scatterings only.   From eq.\ (4.31) we thus
see that, in the case of a photocreated electron
having the same spin as the initial electron, the
response function in fact reads
\begin{equation}
S_{=}=S_{\neq}-\hat{S}\ ,
\end{equation}
where the additional $\hat{S}$ term, coming from
possible exchange processes, is given by
\begin{equation}
\hat{S}=N_p\sum_{\nu,\nu_1,\v p_1}\hat{\lambda}_{\v
p_i;\nu_1,\v p_1}\,A_{\nu_1\v p_1;\nu\v
p_i}(a_i,\v K_i)\,\lambda_\nu^\ast\ .
\end{equation}
$\hat{\lambda}_{\v p;\nu_1,\v p_1}$ appears as
the exciton-photon vertex
$\lambda$ of the bare
semiconductor-photon coupling, renormalized by the
possible exchanges between the electron of the
photocreated exciton and an electron $\v
k_i$ already present in the sample (see fig.\
(9)). It is precisely given by
\begin{equation}
\hat{\lambda}_{\v p;\nu_1,\v
p_1}=\sum_\nu \lambda_\nu\,L_{\nu\v p;\nu_1\v
p_1}
=\lambda\,\langle\v
p+\alpha_e\v p_1|\nu_1\rangle\ .
\end{equation}

The zero order term of $\hat{S}$ in e-X
Coulomb processes comes from the zero order term of
$A_{\nu_1\v p_1;\nu\v p_i}(a_i,\v K_i)$ in
$C^\mathrm{dir}_{\nu'\v p';\nu\v p}$. Using eq.\
(3.55), it reads
\begin{equation}
\hat{S}^{(0)}=N_p\sum_\nu \frac{\hat{\lambda}_{\v
p_i;\nu,\v p_i}\,\lambda_\nu^\ast}{a_i-E_{\nu\v
p_i\v K_i}}=N_p\sum_\nu \hat{\lambda}_{\v
p_i;\nu,\v p_i}\,G_X(\omega_p,\v
Q_p;\nu)\,\lambda_\nu^\ast\ .
\end{equation}
$\hat{S}^{(0)}$ corresponds to the diagram of
fig.\ (10a), since we have
\begin{equation}
\hat{\lambda}_{\v
p_i;\nu,\v p_i}=-\sum_{\v
k}\int\frac{id\omega}{2\pi}\,
\hat{\lambda}_{\beta_X\v k-\beta_e\v
Q_p;\nu,\beta_X\v k-\beta_e\v
Q_p}\,g_i^{(e)}(\omega,\v
k)\ .
\end{equation}
$\hat{S}^{(0)}$ can also be written in a compact
form as
\begin{equation}
\hat{S}^{(0)}=N_p|\lambda|^2\sqrt{\Omega}\,
\left\langle(1+\alpha_e)\v p_i\left |\,
\frac{1}{\omega_p-(h_X+\mathcal{E}_{\v Q_p}^{(X)})
+i\eta}\,\right |\v r=\v 0\right\rangle\ .
\end{equation}

If we now turn to the first order term, it reads
\begin{equation}
\hat{S}^{(1)}=\sum_{\nu}\left[\sum_{\nu_1,\v
q_1}\frac {\hat{\lambda}_{\v p_i;\nu_1,\v p_i-\v
q_1}\, W_{\nu_1\nu}(\v
q_1)}{\Delta_{\nu_1,\v
q_1}}\right]\,G^{(X)}(\omega_p,\v
Q_p;\nu)\,\lambda_\nu^\ast \ .
\end{equation}
It is of importance to note that, when the
photocreated electron has a spin different from
the initial one, the first order term of the
response function $S_{\neq}$ is zero: The
exciton-photon vertex then imposes $\v p=\v p_i$
while $C_{\nu'\v p';\nu\v p}^{\mathrm{dir}}=0$ for
$\v p'=\v p_i=\v p$. On the opposite, first order
e-X scatterings can exist when the two spins are
identical, because the possible electron exchanges
appearing in the dressed exciton-photon vertex do
not impose $\v p'=\v p_i$ anymore. The first order
term of $\hat{S}^{(1)}$ in fact corresponds to
the diagram of fig.\ (10b), since it is easy
to show that, for $\v q_1\neq\v 0$,
\begin{eqnarray}
\frac{\hat{\lambda}_{\v p_i;\nu_1,\v p_i-\v
q_1}}{\Delta_{\nu_1,\v
q_1}}=-\sum_{\v k}\int\frac
{id\omega}{2\pi}\int\frac{id\omega_1}{2\pi}\,
\hat{\lambda}_{\beta_X\v k-\beta_e\v
Q_p;\nu_1,\beta_X\v k-\beta_e\v Q_p-\v
q_1}\hspace{2cm}\nonumber
\\ \times g_i^{(e)}(\omega,\v
k)\,g_i^{(e)}(\omega-\omega_1,\v k-\v
q_1)\,
G_X(\omega_p+\omega_1,\v Q_p+\v
q_1;\nu_1)\ .
\end{eqnarray}
This first order term can also be written in a
compact form as
\begin{eqnarray}
\hat{S}^{(1)}=N_p|\lambda|^2\sqrt{\Omega}\sum_{\v
q_1}\left\langle (1+\alpha_e)\v p_i-\alpha_e\v
q_1\left |\,\frac {1}{\omega_p-(h_X+\mathcal{E}_{\v
Q_p+\v q_1}^{(X)}+\epsilon _{\v k_i-\v
q_1}^{(e)}-\epsilon_{\v k_i}^{(e)})+
i\eta}\right.\right.\nonumber
\\ \times\ \left.\left.w_{-\v q_1}(\v
r)\,\frac{1}{\omega_p- (h_X+\mathcal{E}_{\v
Q_p}^{(X)})+i\eta}\,\right|\v r=\v 0\right\rangle\
.
\end{eqnarray}
Using the same procedure, we can show that the
second order terms of
$\hat{S}$ in e-X Coulomb processes correspond to the diagram of fig.\ (10c)
\emph{plus} the one of fig.\ (10d). Indeed, here
again, the e-X scatterings appearing in $A_{\nu'\v
p_i;\nu\v p_i}(a_i,\v K_i)$ impose $\v q_n-\v
q_{n-1}\neq\v 0$ (with $\v q_0\equiv 0$): This
leads to $\v q_1\neq\v 0\neq\v q_2-\v q_1$, so that
we can have both,
$\v q_2\neq 0$ and
$\v q_2=0$. Since the standard rules for
calculating diagrams only lead to processes in
which the carriers of the Fermi sea are excited,
the diagram of fig.\ (10c) takes into account $\v
q_2\neq 0$ excitations only. The
$\v q_2=0$ ones which would be missing, have to be
included separately through the diagram of fig.\
(10d).

If we now turn to the third order terms of
$\hat{S}$, they correspond to the diagrams of
figs.\ (10e,10f,10g). And so on...

This thus shows that the additional exchange
diagrams which take place when the photocreated
electron and the initial electron have the same
spin, and which are of crucial importance to
withdraw the singlet contributions appearing in
$S_{\neq}$, correspond to a set of ``open'' ladder
diagrams which have the following characteristics:

(i) They all have one conduction-hole line only,
going from left to right, without any hole
scattering since the initial ``Fermi sea'' still
contains one electron only.

(ii) They all have an
exciton possibly scattered by direct e-X processes
without any exchange with the other electron,
although the two electrons have the same spin.

(iii) The
fact that the exciton can be made with the initial
electron or the photocreated electron appears in
this case, once and for all, in the renormalized
exciton-photon interaction $\hat{\lambda}_{\v p_i;
\nu,\v p}$, because the photocreated deep hole can
now recombine with any of the two electrons. These
diagrams thus start with a bare exciton-photon
vertex $\lambda_\nu^\ast$ and end with a dressed
one
$\hat{\lambda}_{\v p_i;\nu,\v p}$.

The reader can be puzzled by the apparent
dissymetry of this new set of diagrams, with the
exchange at the end, and not at the beginning. It
is actually possible to show that the result does
not depend on the position of this exchange.

(iv) Since the scattered exciton can be in
a $(\nu_n,\v Q_p)$ state, while the standard rules
for diagrams lead to processes in which the
initial ``Fermi sea'' --- here the initial
electron --- is excited, we must add to the
connected ``open'' ladder diagrams, diagrams with
an ``open'' part separated from a ladder part, as
in fig.\ (10d,10f,10g).

\subsection{Volume dependence}

From the expressions of the response functions
$S_{\neq}$ and $S_=$ in terms of trions given in
eqs.\ (4.22) and (4.33), we see that the trion
oscillator strength reads simply in terms of the
Fourier transform of the trion relative motion
wave function $\langle\v r,\v u|\eta_S\rangle$,
written with the ``good'' trion variables $\v r$
and $\v u$, defined in eqs.\ (3.2) and (3.10). We
also see that, when the photon polarization is
such that only $(S=1)$ trions can be formed, the
oscillator strength of these $(S=1)$ trions have an
additional factor of 2, compared to the case in
which both $(S=0)$ and $(S=1)$ trions can be
created.

From the expression $|\lambda|^2\Omega|\langle\v
r=\v 0,\v p_i|\eta\rangle|^2$ of the trion
oscillator strength obtained in eq.\ (4.22), it is
easy [19] to compare its size for bound state
$(eeh)$, partially dissociated state $(e+eh)$ and
totally dissociated state $(e+e+h)$.

In the case
of a bound trion, $\langle\v r,\v u|\eta\rangle$
has an extension over $r$ of the order of $a_X$ and
an extension over $u$ of the order of $a_T$, with
$a_T$ much larger than $a_X$ since the trion
binding energy is much weaker than the exciton
binding energy. In 2D, dimensional arguments then
lead to $\langle\v r=\v 0,\v u=\v
0|\eta\rangle\sim (a_Xa_T)^{-1}$. From the spatial
extension $a_T$ of $\langle\v r=\v 0,\v
u|\eta\rangle$, we then find, again from
dimensional arguments, that, for $p\ll a_T^{-1}$,
$\langle\v r=\v 0,\v p|\eta\rangle\sim \langle\v
r=\v 0,\v u=\v 0|\eta\rangle\langle\v p|\v
u\simeq\v 0\rangle a_T^2\sim a_T/a_XL$, where $L$
is the sample size $(L^2=\Omega)$. This leads to a
bound trion oscillator strength $\sim|\lambda|^2
(a_T/a_X)^2$, independent from the sample volume.

If we now consider a partially dissociated trion,
$\langle\v r,\v u|\eta\rangle$ now has an extension
$a_X$ over $r$ but $L$ over $u$, so that
$\langle\v r=\v 0,\v u=\v 0|\eta\rangle\sim(a_X
L)^{-1}$. For $p$ small, we then have $\langle\v
r=\v 0,\v p|\eta\rangle\sim 1/a_X$, so that the
partially disssociated trions have an
oscillator strength $\sim |\lambda|^2L^2/a_X^2$,
which is the one of a free exciton.

Finally, for a totally dissociated trion,
$\langle\v r,\v u|\eta\rangle$ has an extension
$L$ over $r$ and $L$ over $u$, so that $\langle\v
r=\v 0,\v u=\v 0|\eta\rangle\sim L^{-2}$. For
small $p$, this gives $\langle\v r=\v 0,\v
p|\eta\rangle\sim 1/L$, which leads to a sample
volume independent oscillator strength
$|\lambda|^2$ equal to the one of the exciton
diffusive states, \emph{i}.\ \emph{e}., free
carriers. This oscillator strength is however
smaller than the bound trion one, since the trion
extension $a_T$ is somewhat larger than the
exciton extension $a_X$.

We thus conclude that the partially dissociated
trion has essentially the same oscillator strength
as the exciton, while the bound trion oscillator
strength is $a_T^2/L^2$ smaller. It is thus
vanishingly small in the large sample limit,
\emph{i}.\
\emph{e}., the limit of solid state physics. This
is after all not surprising: If we take a large
sample and if we add \emph{just one} electron, we
cannot expect any sizeable change in the photon
absorption spectrum! From a technical point of
view, if we consider the volume dependence of the
diagrams representing the response function, we
find that the dominant one corresponds to the bare
exciton diagram, given in fig.\ (7a), so that in
the large $L$ limit, the trion diagrams reduce to
this unique diagram.

These quite simple arguments lead us to
conclude that, when a line is seen, well below the
exciton line, in the photon absorption spectrum of
a doped semiconductor, it cannot be the line of a
(clean) trion because its oscillator strength would
be a sample volume (or a coherence volume) smaller
than the exciton one. It is most probably a trion
having many-body effects with the $(N-1)$ other
electrons of the semiconductor, or better a
photocreated exciton dressed by the $N$ electrons
already present in the sample. This quite
difficult many-body problem will be adressed in a
further work, using the tools we have established
in sections 3 and 4 of the present paper, which
allow to treat excitons interacting with
electrons through electron-exciton diagrams.

\section{Conclusion}

In order to solve the simplest problem on trion,
namely the photon absorption in the presence of
\emph{one} electron, we have first shown that the
standard Feynman diagrams, which read in terms
of free electrons and free holes, are totally
inappropriate: As the trion is a bound state,
Coulomb processes have to be included at all orders.
While these Coulomb processes can be handled
exactly between the two carriers of an exciton, it
is totally hopeless to draw and sum them all, in
the case of three carriers.

We propose to reduce this three-body problem
to a two-body problem by considering the trion as
an electron interacting with an exciton. Although
physically appealing, this approach however faces
the problem of the electron indistinguishability,
\emph{i}.\ \emph{e}., the fact that the exciton can
\emph{a priori} be made with any of the two
electrons. Our commutation technique, designed to
deal with this problem in an exact way, allows to
overcome this difficulty.

We find that when the photocreated electron has a
spin different from the initial electron spin ---
which is for example what happens if we
want to photocreate a ground state trion ---, we
can forget about these possible electron exchanges:
The response function to the  photon field just
corresponds to a set of e-X ladder  diagrams
between e-X pairs $(\nu,\v p,\v K)$, with
$\v K$ being the center of mass momentum of the pair
--- constant in these ladder processes ---,
$\v p$ the relative motion momentum of the e-X
pair and
$\nu$ characterizing the exciton relative
motion level. The e-X
interaction vertex
$C_{\nu'\v p';\nu\v p}^{
\mathrm{dir}}$ of these novel diagrams
corresponds to \emph{direct} Coulomb processes
between the electron and the exciton, the
``in'' exciton $\nu$ and the ``out'' exciton
$\nu'$ being made with the \emph{same}
electron. $C_{\nu'\v p';\nu\v p}^{
\mathrm{dir}}$ reads in  terms of the Fourier
transform
$w_{\v q}(\v r)$ of the Coulomb potential $w(\v
r,\v u)$ between the electron $e'$ and the exciton
made of $(e,h)$.

The possible electron exchanges are only
important when the spins of the two electrons are
identical, \emph{i}.\ \emph{e}., when only triplet
trions $(S_z=\pm 1)$ are photocreated. It is then
possible to include these exchanges through a
dressed exciton-photon interaction. They thus play
a role, once and for all, when the hole
recombines, so that we are again left with direct
e-X ladder processes only, except that they are now
``open''.

The physical reason for a so trivial consequence of
the possible electron exchange lies in the fact
that two  exchanges reduce to an identity; so that,
either we end with no exchange at all --- as when
the photocreated hole can recombine with the
photocreated electron  only, which is what happens
when the two electron spins are different ---, or
we end with zero and one exchange --- as when the
hole can recombine with either the photocreated
electron or the initial electron,  which is what
happens when the two electrons have  the same spin.

In section 3 of this paper, we have also collected
all important results on trions derived in our
previous works --- plus some unpublished ones. This
``background on trion'', which leads to this
novel  many-body procedure in terms of electrons
and excitons, will be of great help to study the
interaction of trion with carriers in doped
semiconductors: Indeed, the existing
literature on  this very difficult many-body
problem, which relies on standard electron-hole
procedure --- the only one known up to now ---,
is quite unsatisfactory, as we outlined in ref.\
[19].

\vspace{2cm}

\hbox to \hsize {\hfill REFERENCES
\hfill}

\vspace{0.5cm}

\noindent
(1) E. Hylleraas, \emph{Phys.\ Rev}.\
\textbf{75}, 491 (1947).

\noindent
(2) M. Lampert, \emph{Phys.\ Rev.\ Lett}.\
\textbf{1}, 450 (1958).

\noindent
(3) K. Kheng, R.T. Cox, Y. Merle d'Aubign\'{e},
F. Bassani, K. Saminadayar, S. Tatarenko,
\emph{Phys.\ Rev.\ Lett}.\ \textbf{71}, 1752
(1993).

\noindent
(4) G. Finkelstein, H. Shtrikman, I. Bar Joseph,
\emph{Phys.\ Rev.\ Lett}.\ \textbf{74}, 976
(1995).

\noindent
(5) S.A. Brown, J.F. Young, J.A. Brum, P.
Hawrylak, Z. Wailewski, \emph{Phys.\ Rev}.\ B
\textbf{54}, R11082 (1996).

\noindent
(6) G. Astakhov, V.P. Kochereshko, D.R. Yakovley,
W. Ossau, J. Nurnberger, W. Faschinger, G.
Landwehr, \emph{Phys.\ Rev.}\ B \textbf{62}, 10345
(2000).

\noindent
(7) P. Redlinski, J. Kossut, \emph{Solid State
Com}.\ \textbf{118}, 295 (2001).

\noindent
(8) J. Puls, G.V. Mikhailov, F. Henneberger, D.R.
Yakovley, A. Waag, W. Faschinger, \emph{Phys.\
Rev.\ Lett}.\ \textbf{89}, 287402 (2002).

\noindent
(9) D.V. Kulakovskii, Y.E. Lozovik,
\emph{J.E.T.P.\ Lett.}\ \textbf{76}, 516 (2002).

\noindent
(10) B. Stebe, G. Munschy, \emph{Solid State
Com}.\ \textbf{17}, 1051 (1975).

\noindent
(11) P. Hawrylak, \emph{Phys.\ Rev.}\ B
\textbf{44}, 3821 (1991).

\noindent
(12) B. St\'{e}b\'{e}, G. Munschy, L. Stauffer,
F. Dujardin, J. Murat, \emph{Phys.\ Rev.}\ B
\textbf{56}, 12454 (1997).

\noindent
(13) B. St\'{e}b\'{e}, E. Feddi, A. Ainane, F.
Dujardin, \emph{Phys.\ Rev.}\ B \textbf{58}, 9926
(1998).

\noindent
(14) A. Esser, E. Runge, R. Zimmerman, W.
Langbein, \emph{Phys.\ Rev.}\ B \textbf{62}, 8232
(2000).

\noindent
(15) W. Florek, \emph{J.\ Math.\ Phys.}\
\textbf{42}, 5177 (2001).

\noindent
(16) L.C. Dacal, R. Ferreira, G. Bastard, J.A.
Brum, \emph{Phys.\ Rev.}\ B \textbf{65}, 115325
(2002).

\noindent
(17) M. Combescot, \emph{Eur.\ Phys.\ J.\ B}
\textbf{33}, 311 (2003).

\noindent
(18) M. Combescot, O. Betbeder-Matibet,
\emph{Solid State Com.}\ \textbf{126}, 687 (2003).

\noindent
(19) M. Combescot, J. Tribollet, \emph{Solid State
Com.}\ \textbf{128}, 273 (2003).

\noindent
(20) M. Combescot, P. Nozi\`{e}res, \emph{J.\ de
Phys.\ (Paris)}, \textbf{32}, 913 (1971).

\noindent
(21) We draw diagrams with arrows from right to
left, while one usually draws them from left to
right. Since diagrams basically represent matrix
elements, our way makes these
diagrams to topologically appear
just as the algebraic expressions
they represent. This is why we
find this drawing more convenient
to read than the one found in usual
textbooks on many-body effects.

\noindent
(22) M. Combescot, O.
Betbeder-Matibet, \emph{Europhys.\
Lett.}\ \textbf{58}, 87 (2002).

\noindent
(23) O. Betbeder-Matibet, M.
Combescot, \emph{Eur.\ Phys.\ J.\
B}, \textbf{27}, 505 (2002).

\noindent
(24) M. Combescot, O.
Betbeder-Matibet, \emph{Europhys.\
Lett.}\ \textbf{59}, 579 (2002).

\noindent (25) M. Combescot, O. Betbeder-Matibet,
Cond-mat/0402087, submitted to Phys. Rev. Lett..

\noindent
(26) M. Combescot, O. Betbeder-Matibet, K. Cho, H.
Ajiki, \emph{Cond-mat}/0311387, submitted to Phys.\ Rev.\ Lett.

\newpage

\begin{figure}
\centerline{ \scalebox{0.22}{\includegraphics{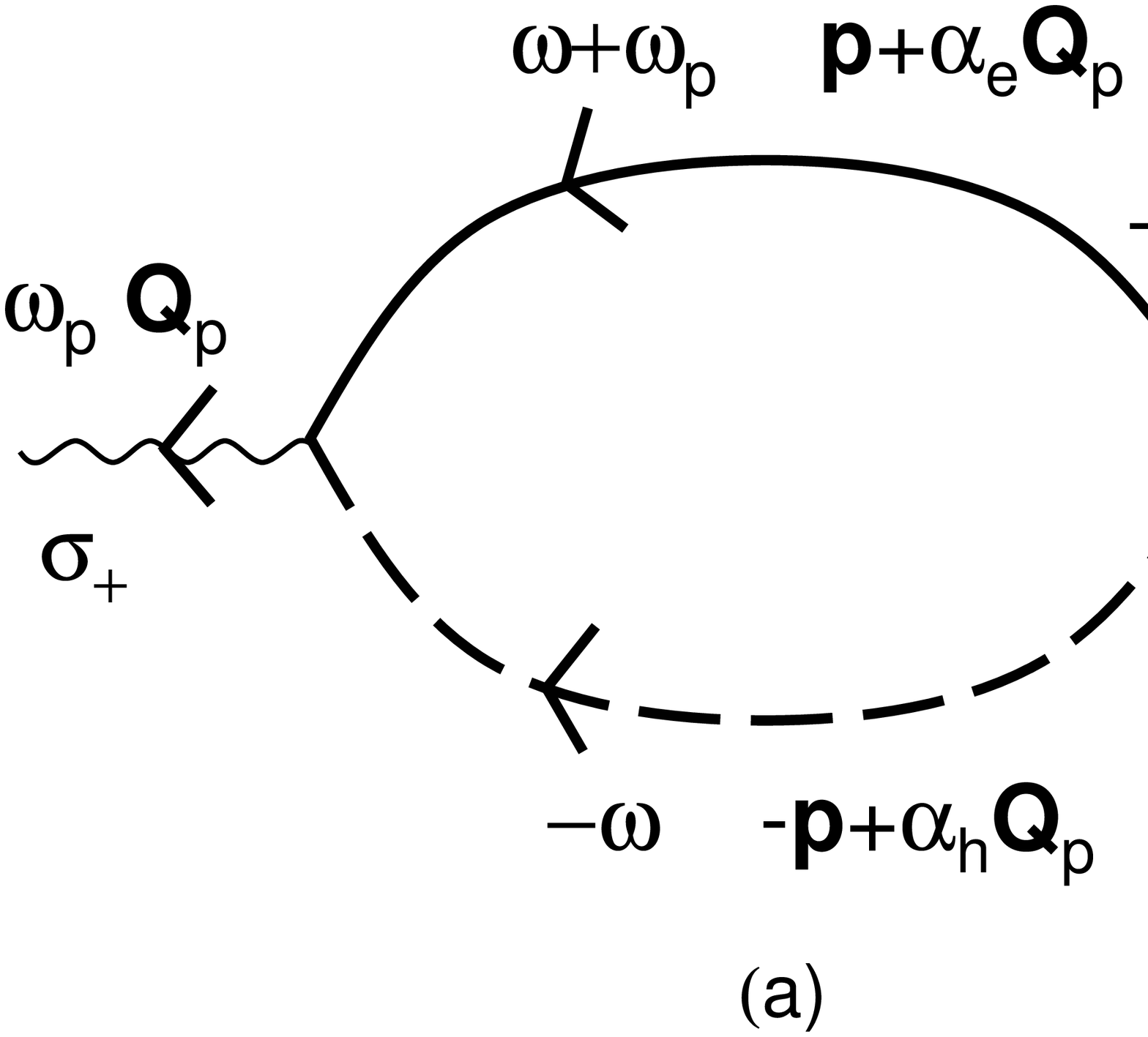}}
\hspace*{0.5cm} \scalebox{0.22}{\includegraphics{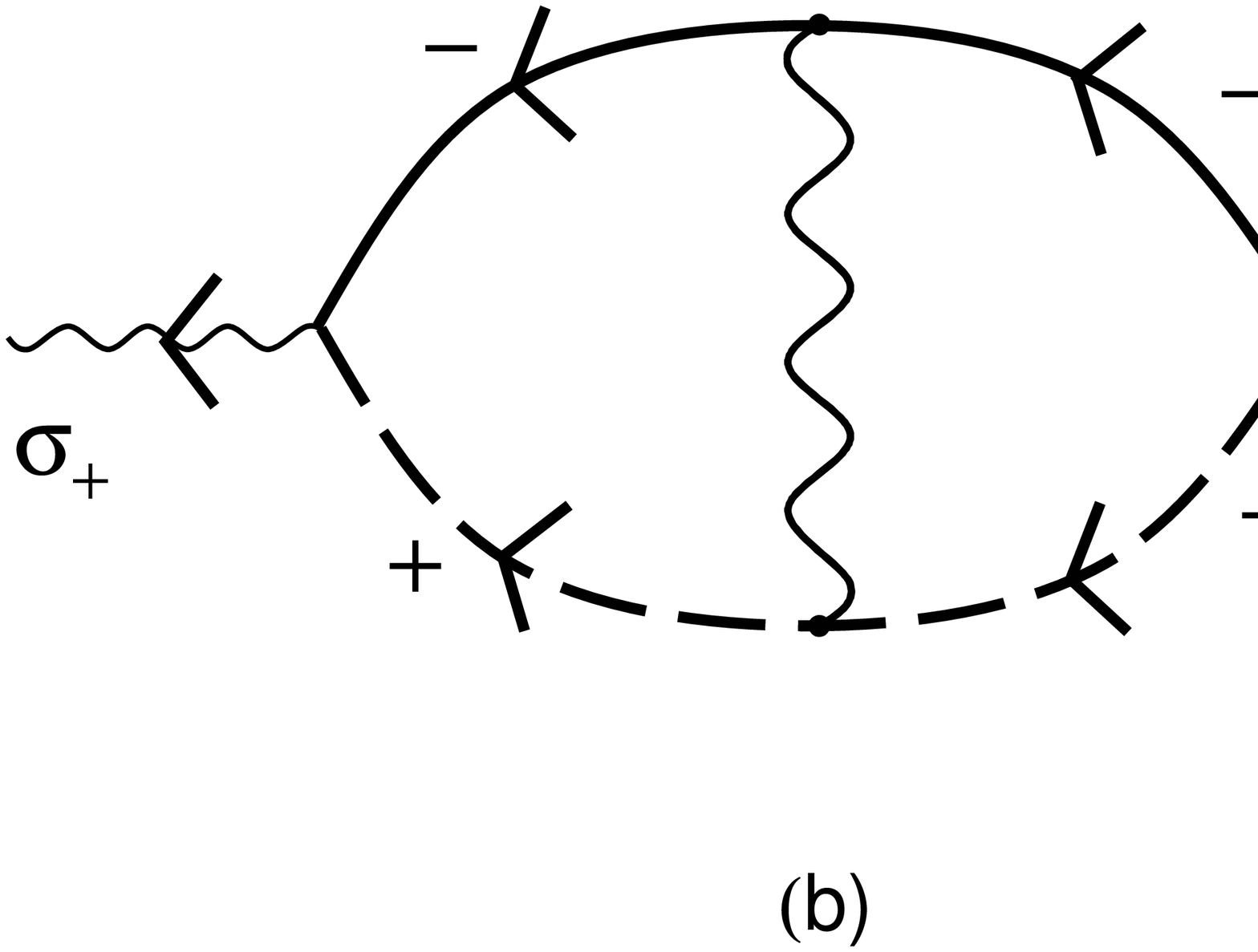}}}
\vspace*{0.5cm} \centerline{
\scalebox{0.22}{\includegraphics{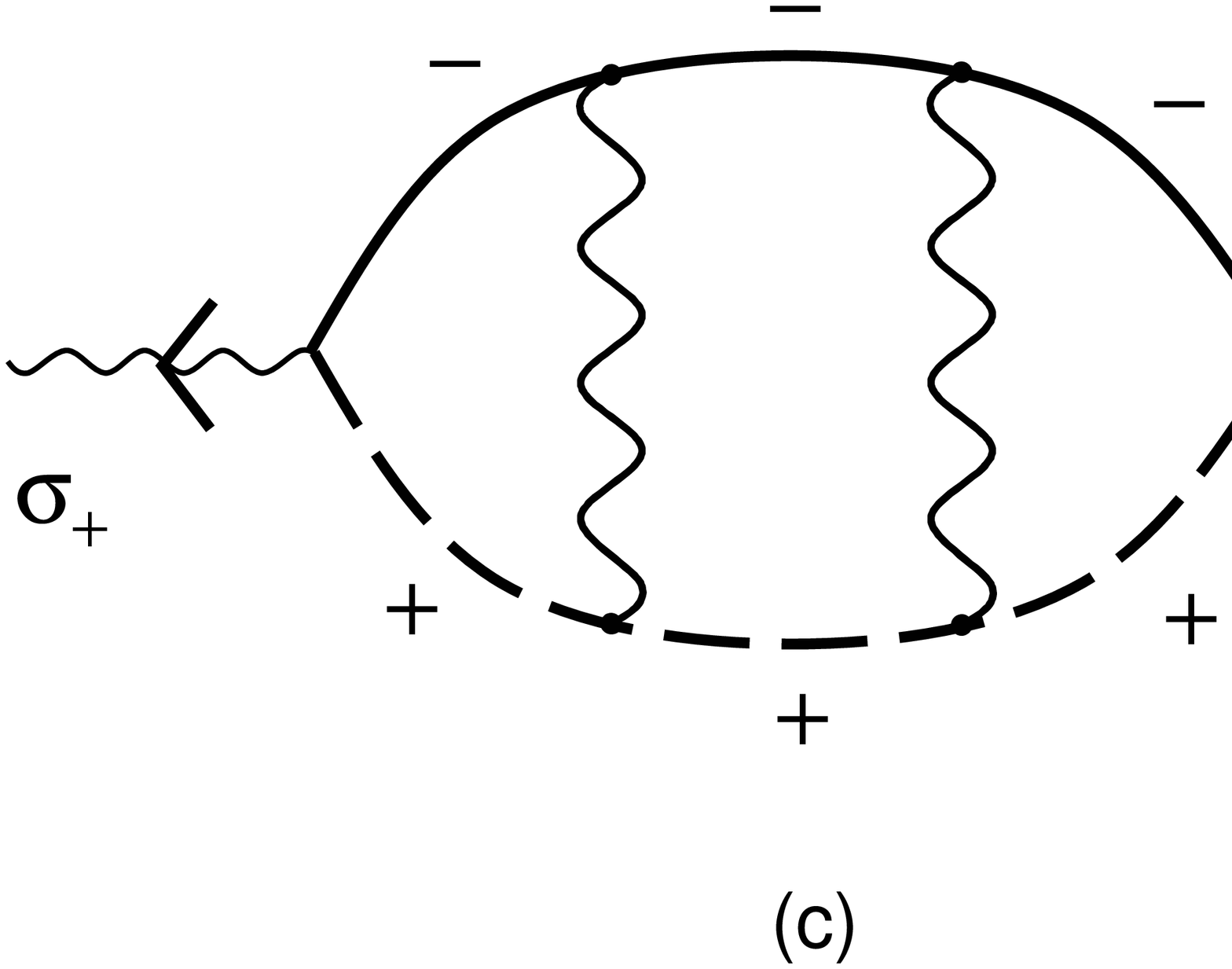}} \hspace*{0.5cm}
\scalebox{0.22}{\includegraphics{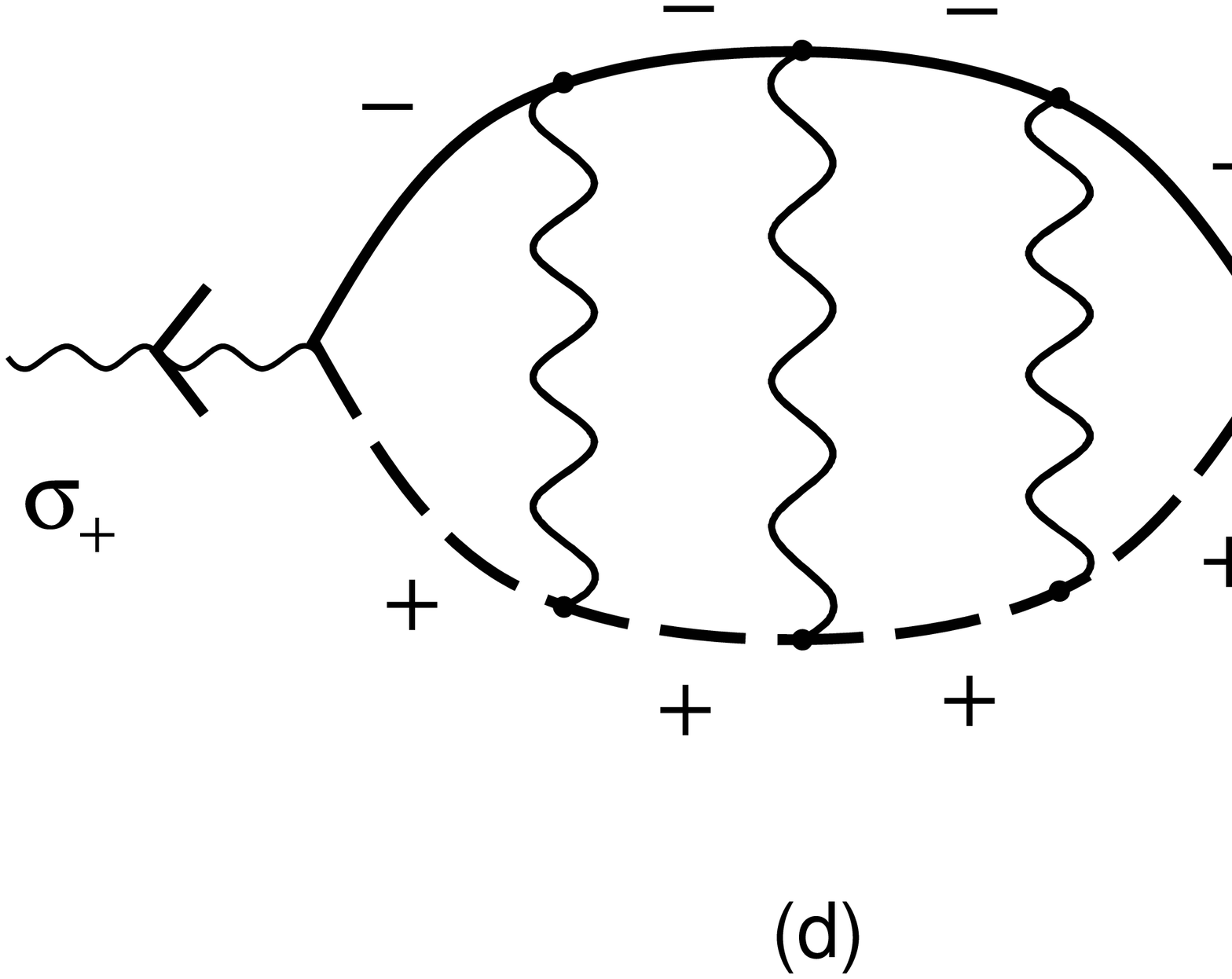}}} \caption{Photon
absorption with exciton formation: well-known set of electron-hole
ladder diagrams with 0, 1, 2, 3\ldots electron-hole Coulomb
interactions, (a), (b), (c), (d) respectively. Solid line:
electron. Dashed line: hole. Wavy line: Coulomb interaction
between electron and hole. In a quantum well, a $\sigma_+$ photon
creates a hole with momentum (+3/2), noted +, and an electron with
spin ($-1/2$), noted $-$.}
\end{figure}

\newpage

\begin{figure}
\centerline{ \scalebox{0.22}{\includegraphics{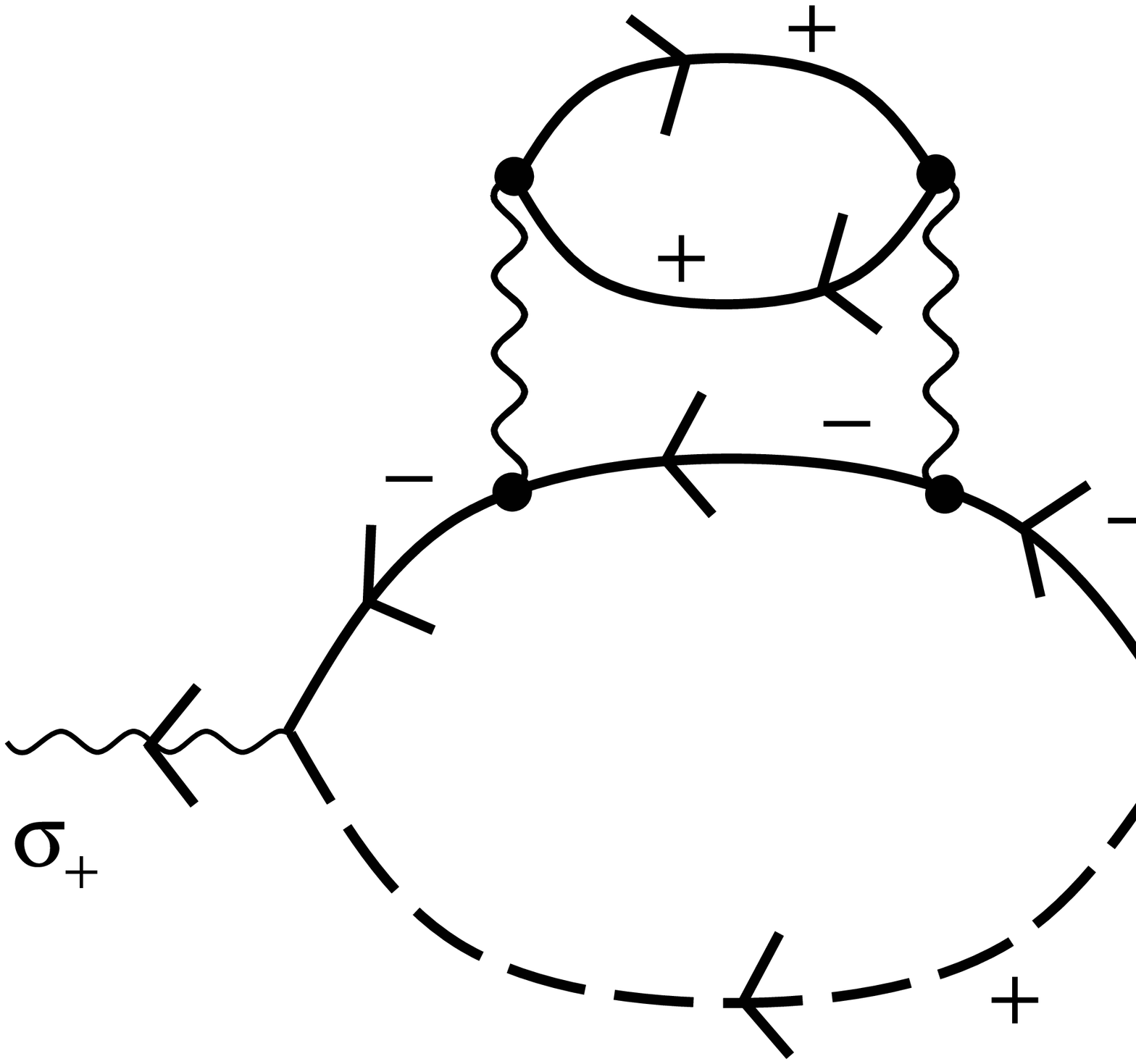}}
\hspace*{0.5cm} \scalebox{0.22}{\includegraphics{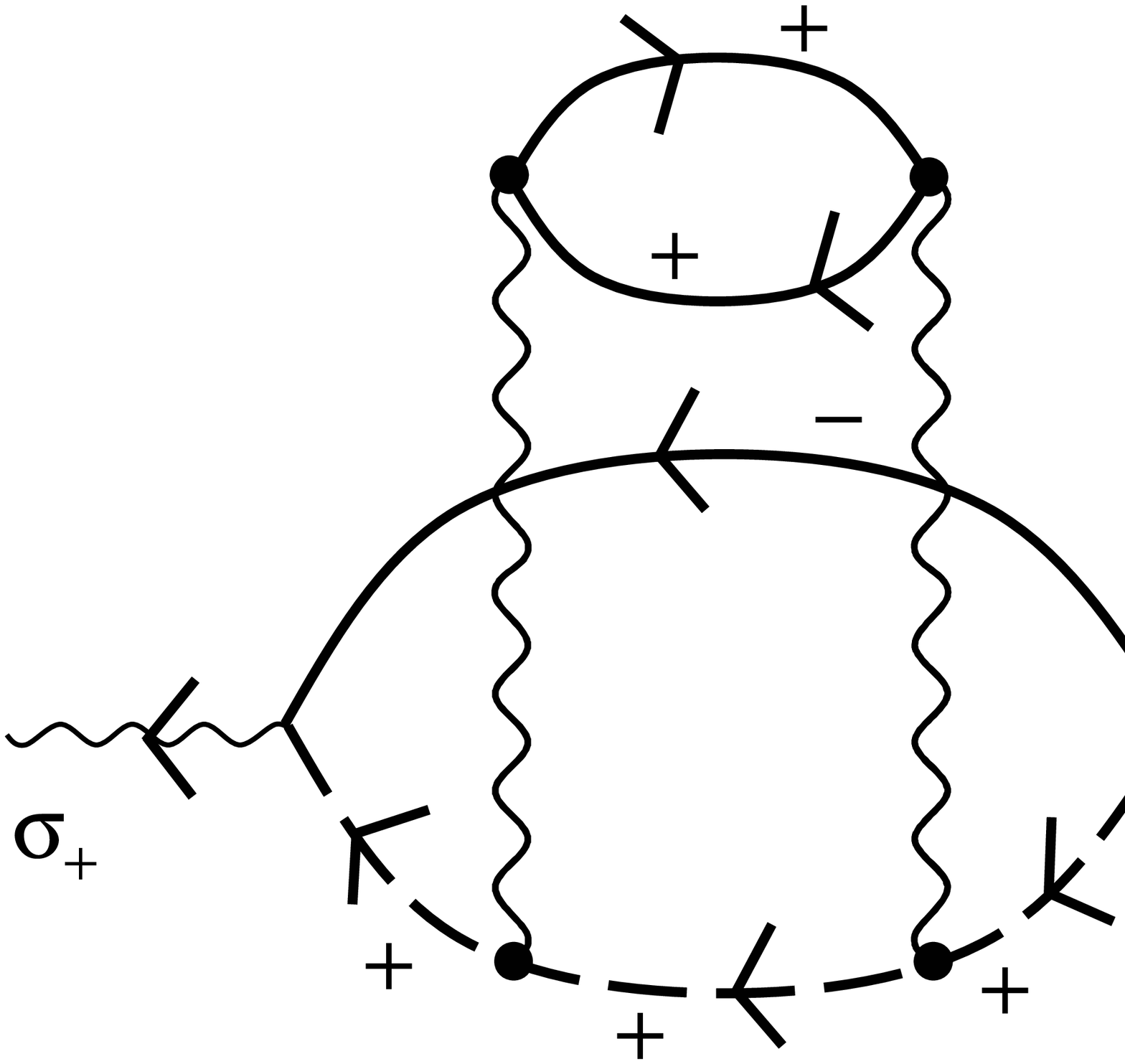}}}
\vspace*{0.5cm} \centerline{
\scalebox{0.22}{\includegraphics{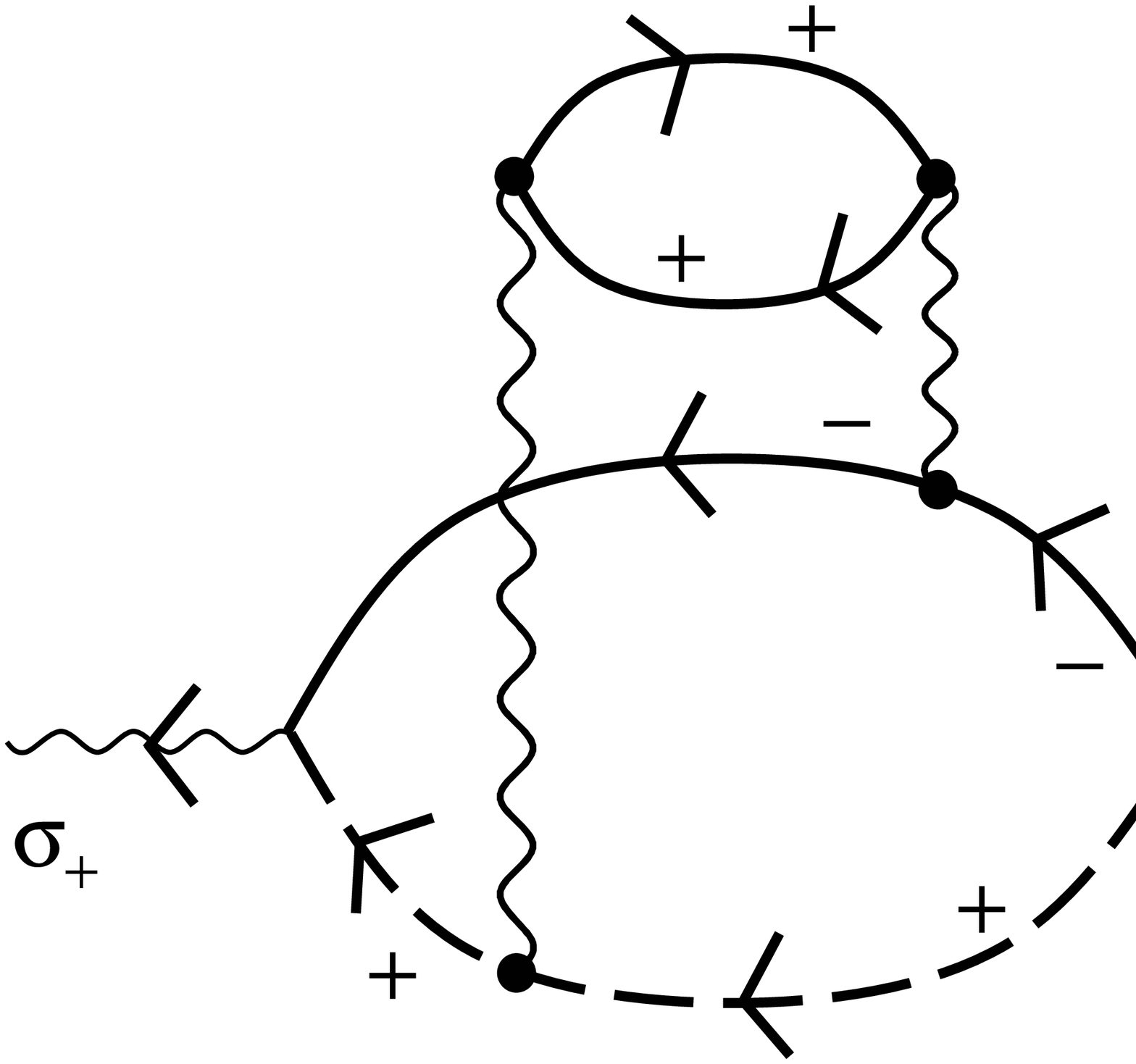}} \hspace*{0.5cm}
\scalebox{0.22}{\includegraphics{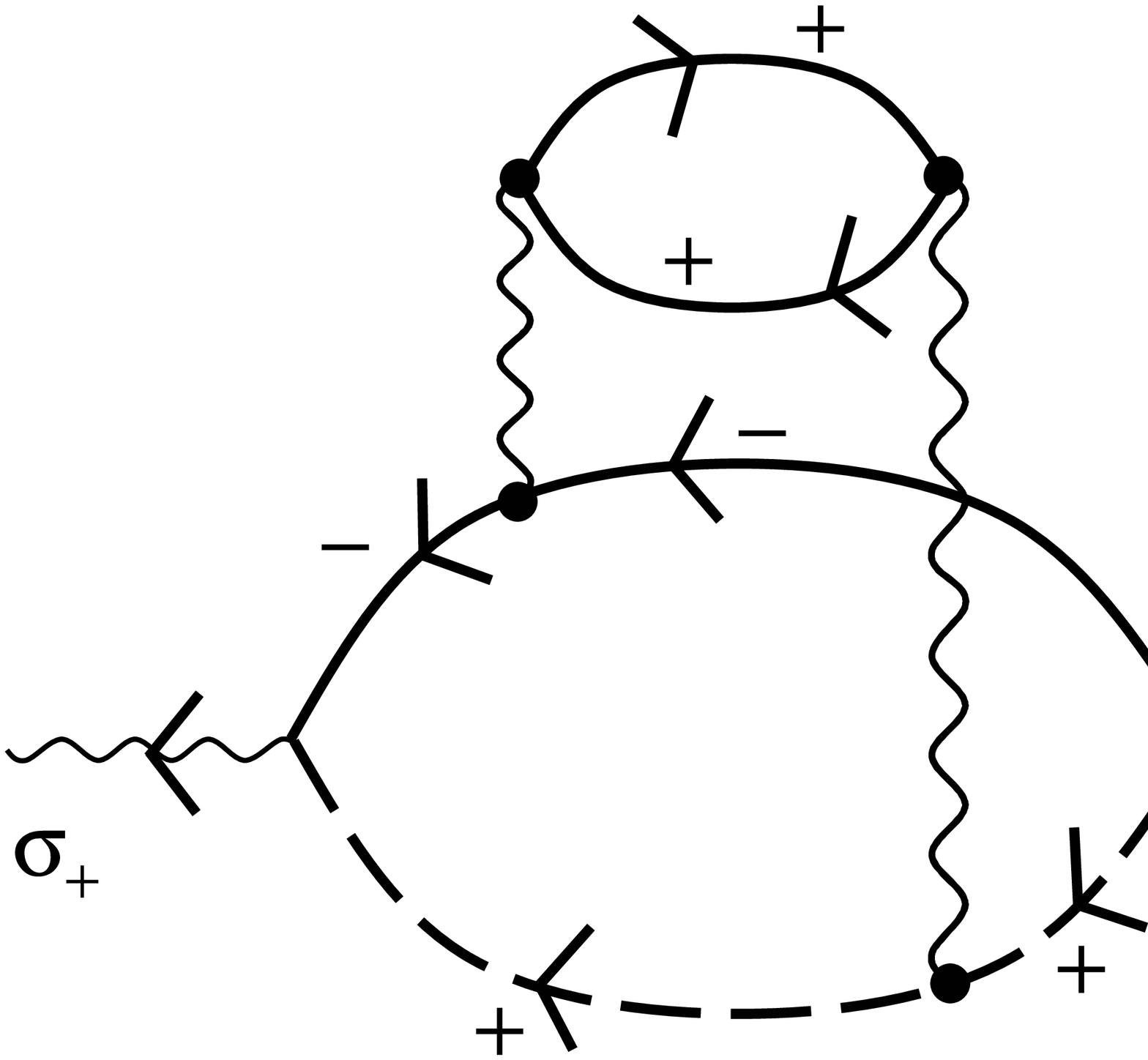}}} \caption{Absorption
of a $\sigma_+$ photon with trion formation, in the case of a
photocreated electron and an initial electron having different
spins: second order in Coulomb interaction. The diagrams of this
figure have to be added to the one of fig.\ (1c).}
\end{figure}

\newpage

\begin{figure}
\centerline{\scalebox{0.15}{\includegraphics{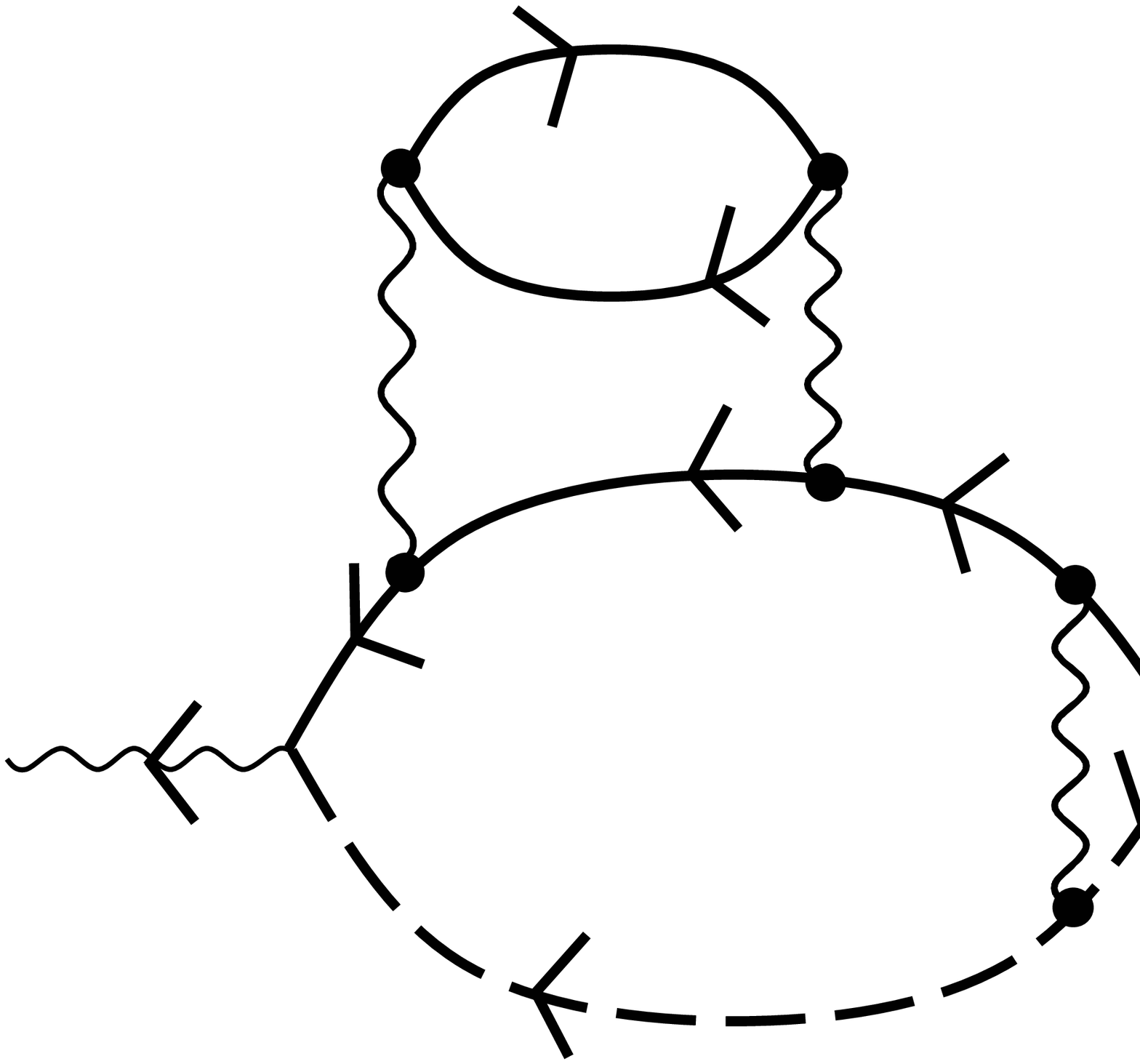}}
\hspace*{0.5cm} \scalebox{0.15}{\includegraphics{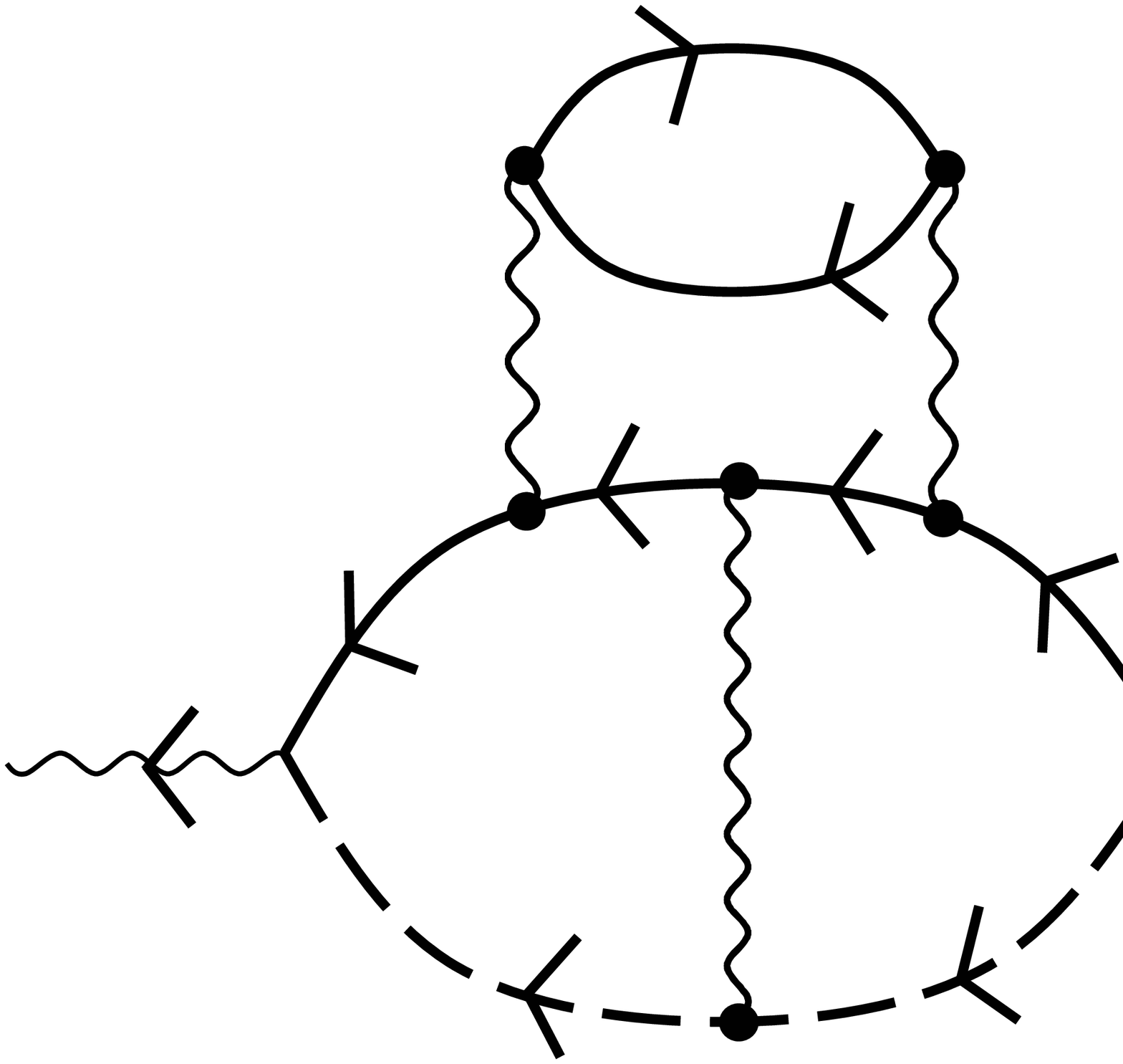}}
\hspace*{0.5cm} \scalebox{0.15}{\includegraphics{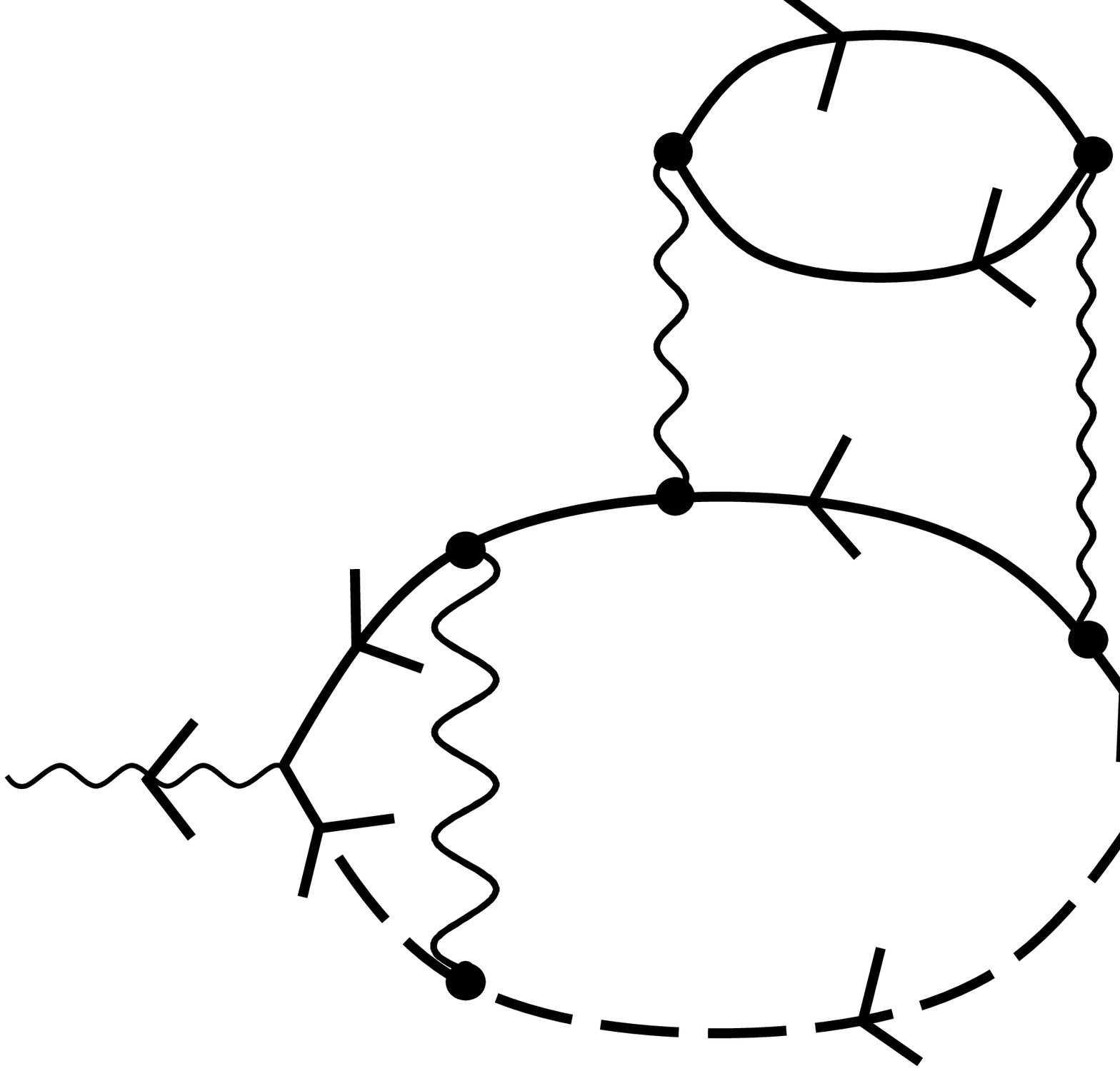}}
\hspace*{0.5cm} \scalebox{0.15}{\includegraphics{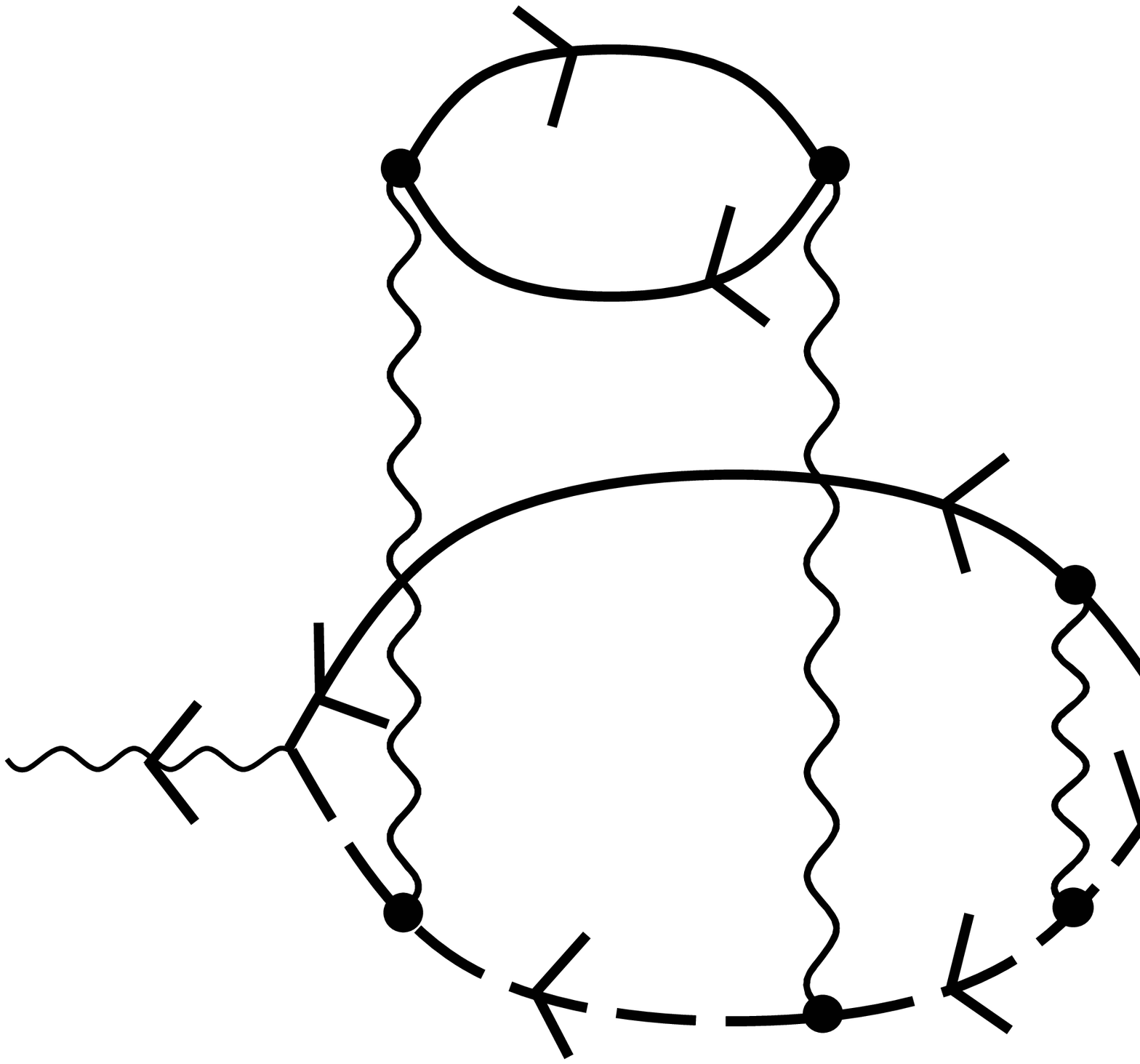}}}
\vspace*{0.5cm}
\centerline{\scalebox{0.15}{\includegraphics{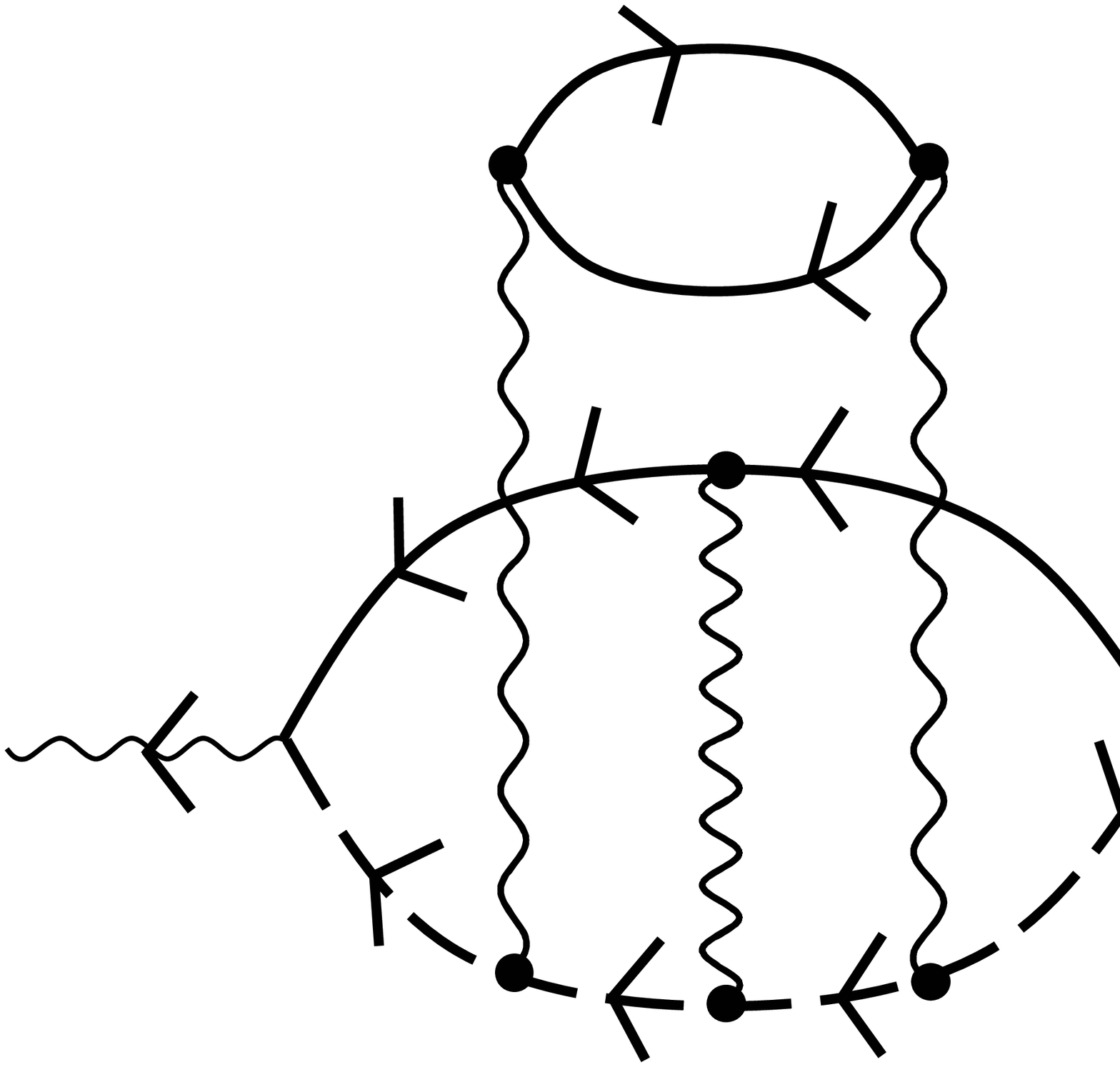}}
\hspace*{0.5cm} \scalebox{0.15}{\includegraphics{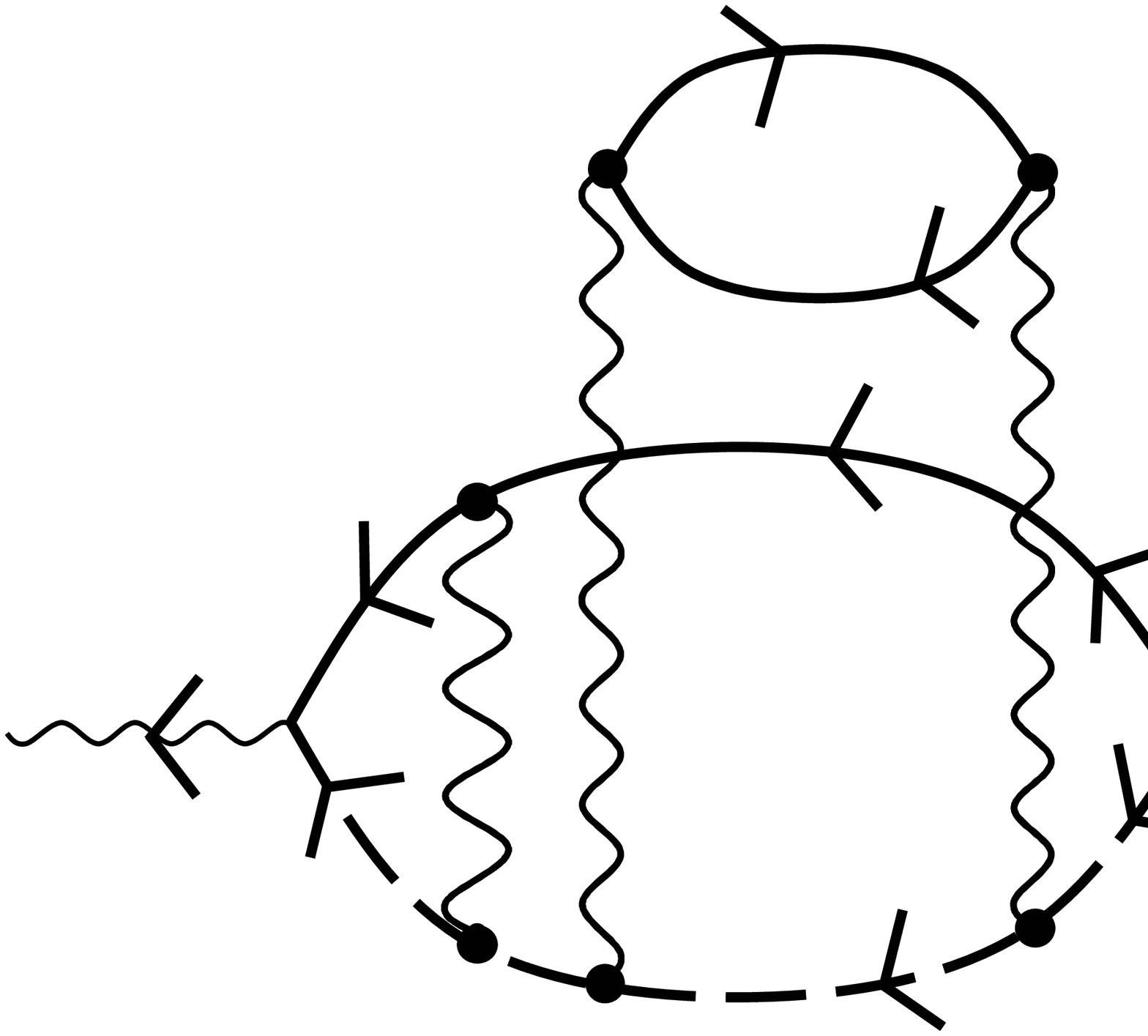}}
\hspace*{0.5cm} \scalebox{0.15}{\includegraphics{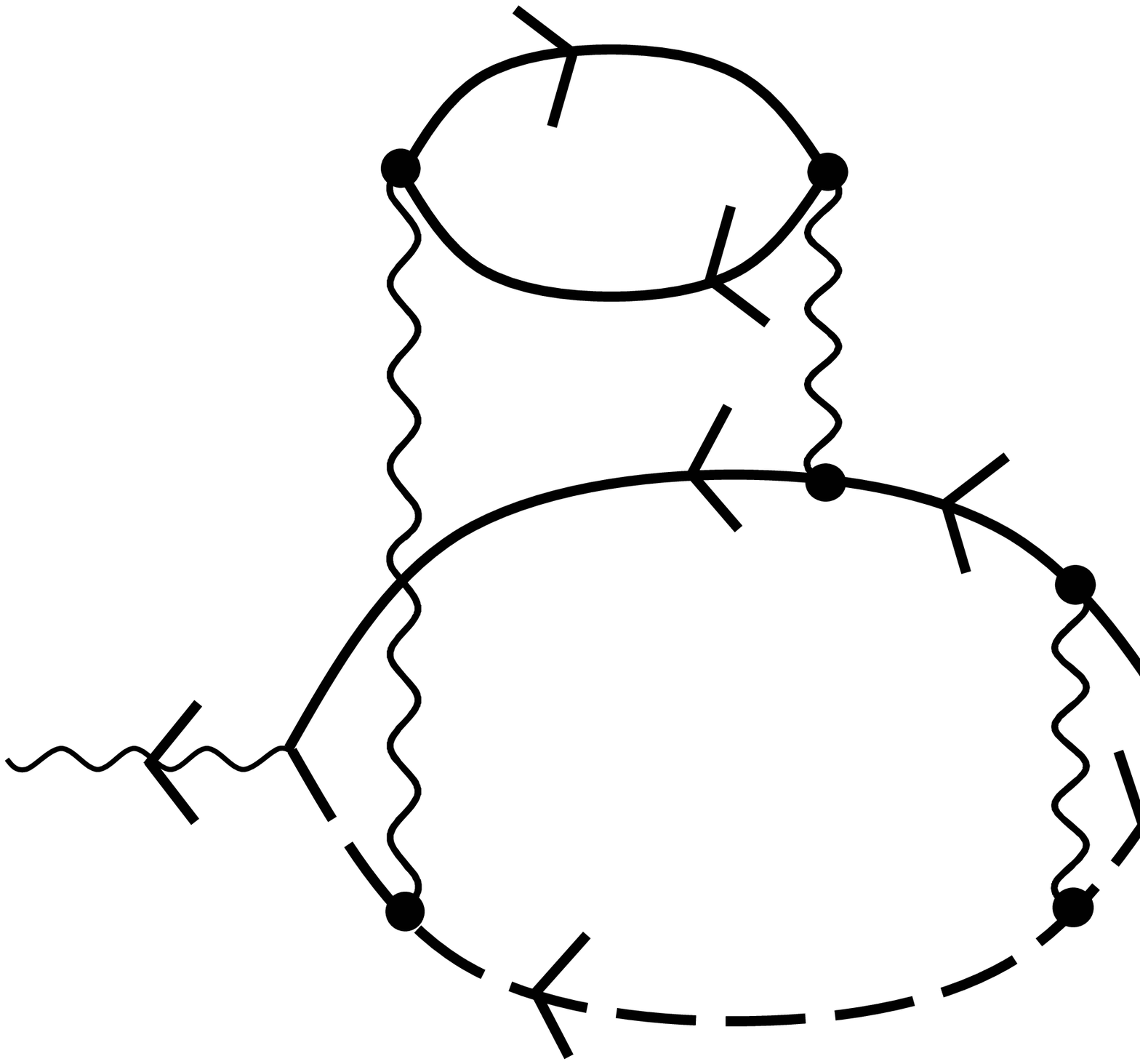}}
\hspace*{0.5cm}\scalebox{0.15}{\includegraphics{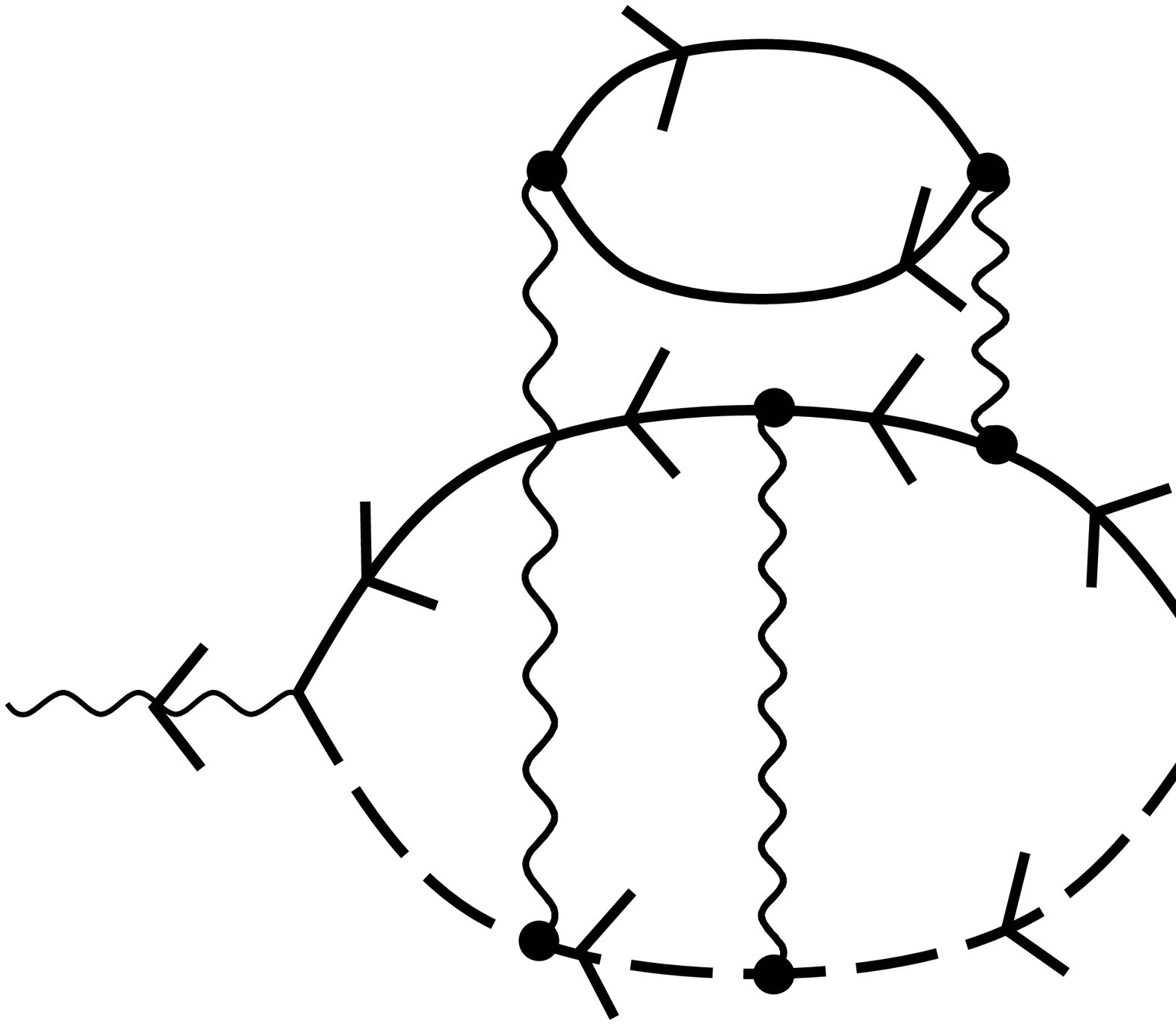}}}
\vspace*{0.5cm}
\centerline{\scalebox{0.15}{\includegraphics{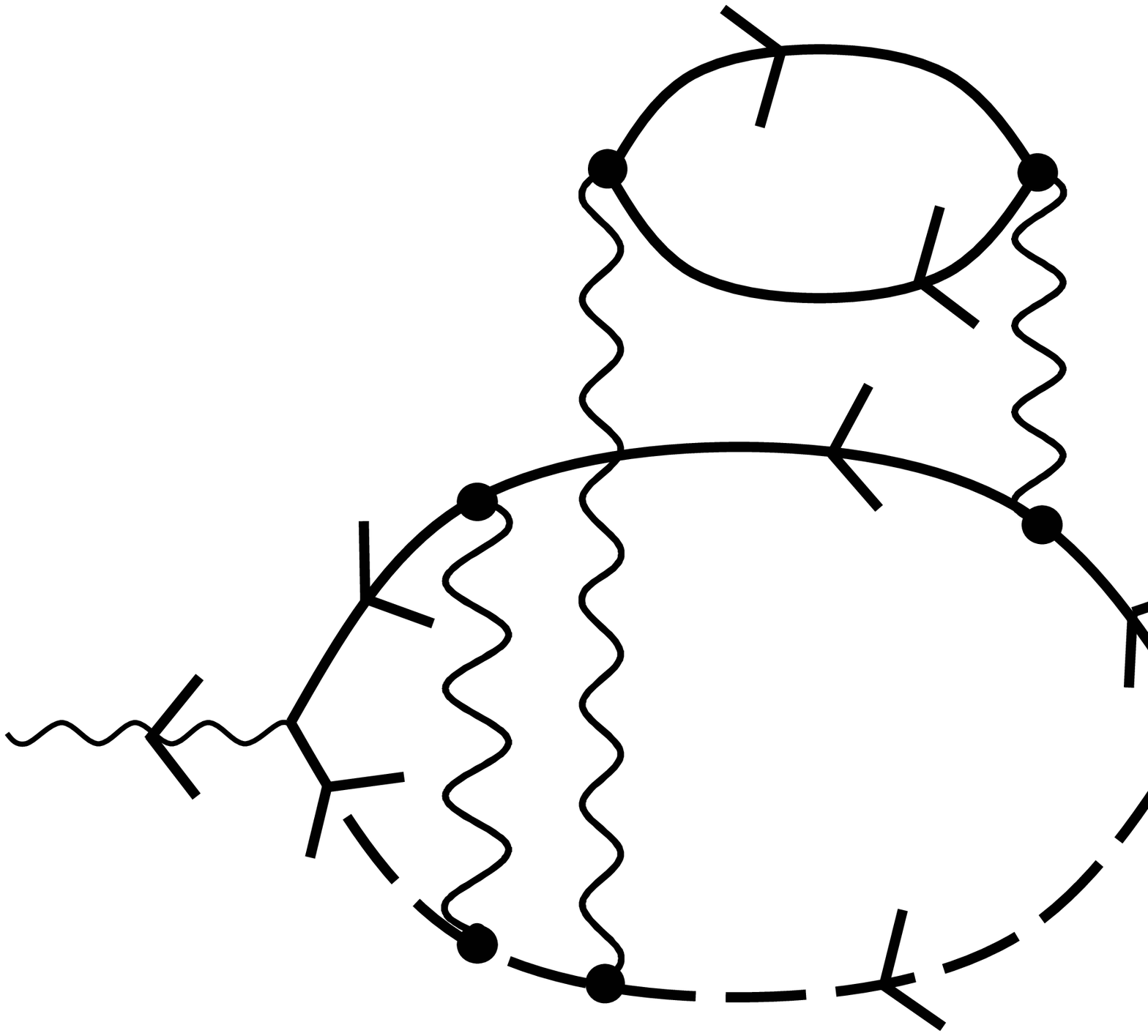}}
\hspace*{0.5cm} \scalebox{0.15}{\includegraphics{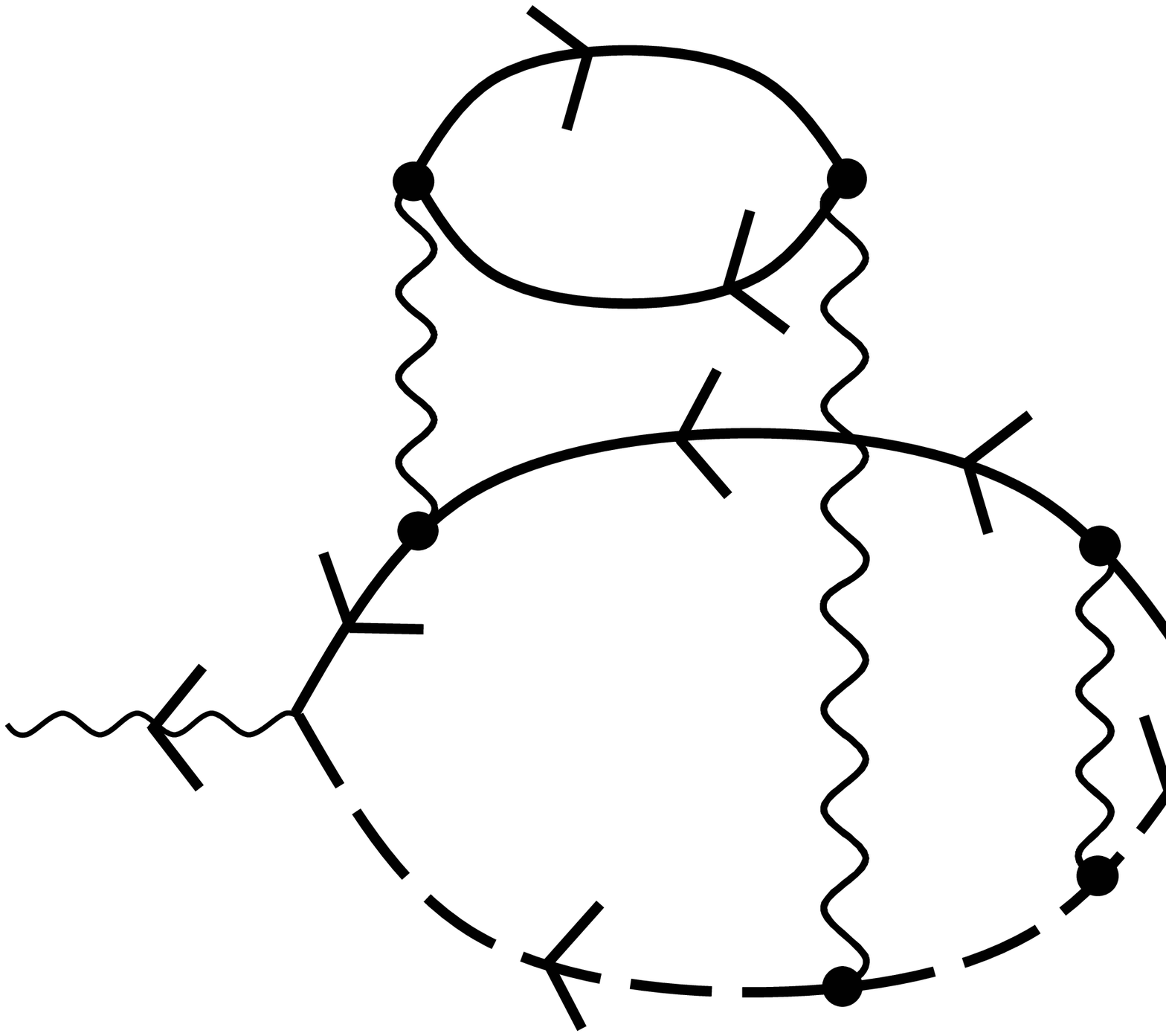}}
\hspace*{0.5cm} \scalebox{0.15}{\includegraphics{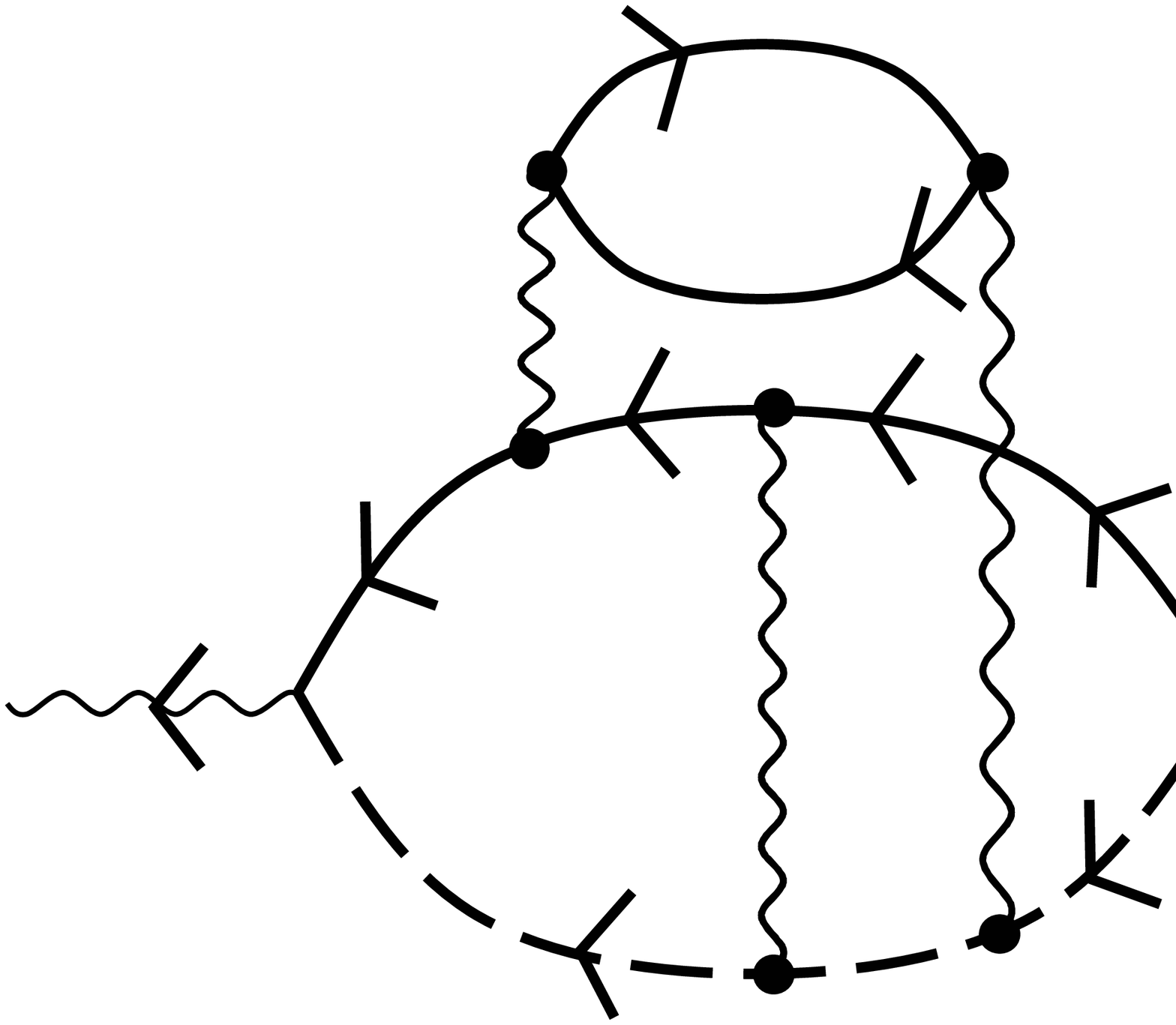}}
\hspace*{0.5cm} \scalebox{0.15}{\includegraphics{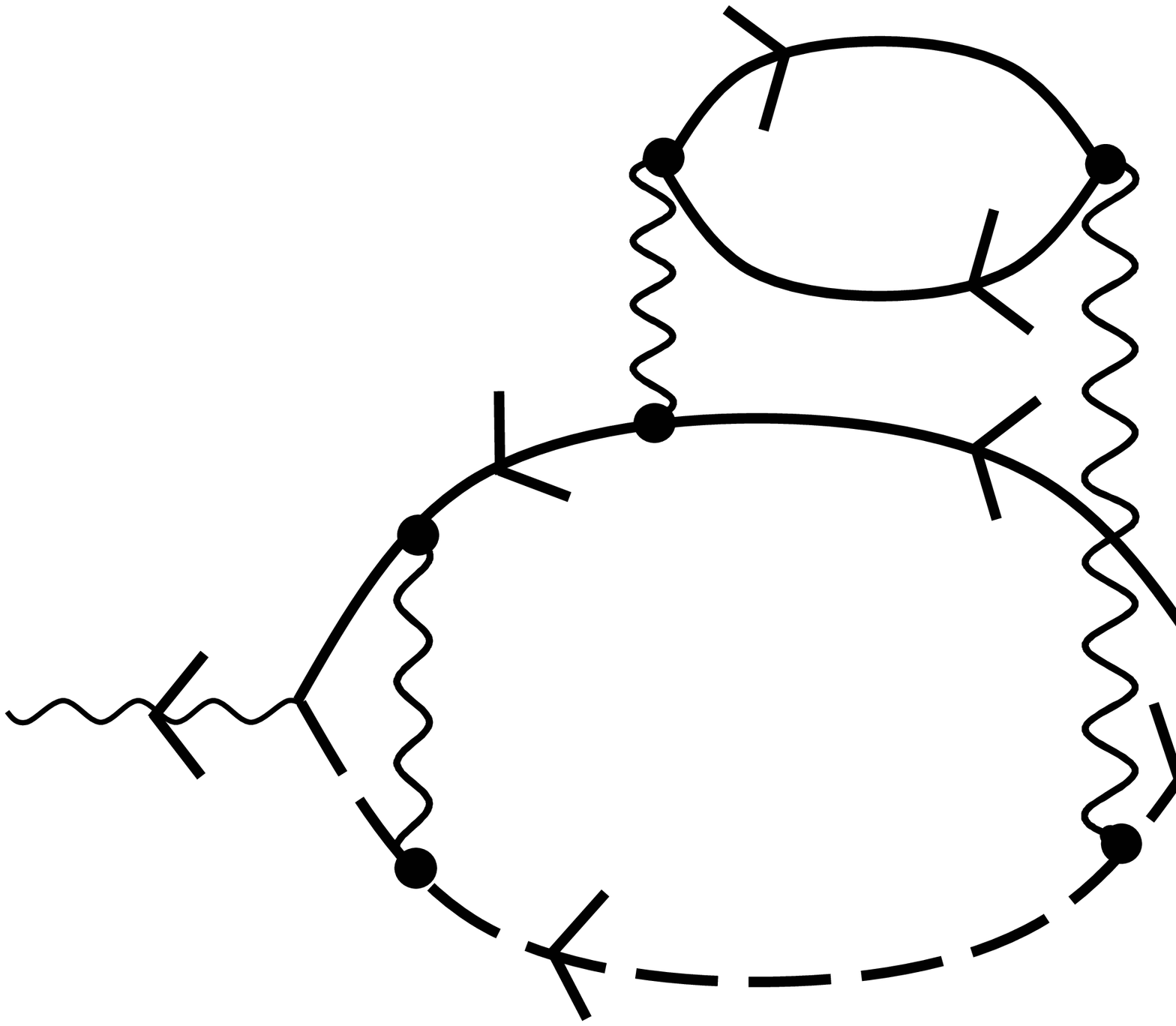}}}
\vspace*{0.5cm}
\centerline{\scalebox{0.15}{\includegraphics{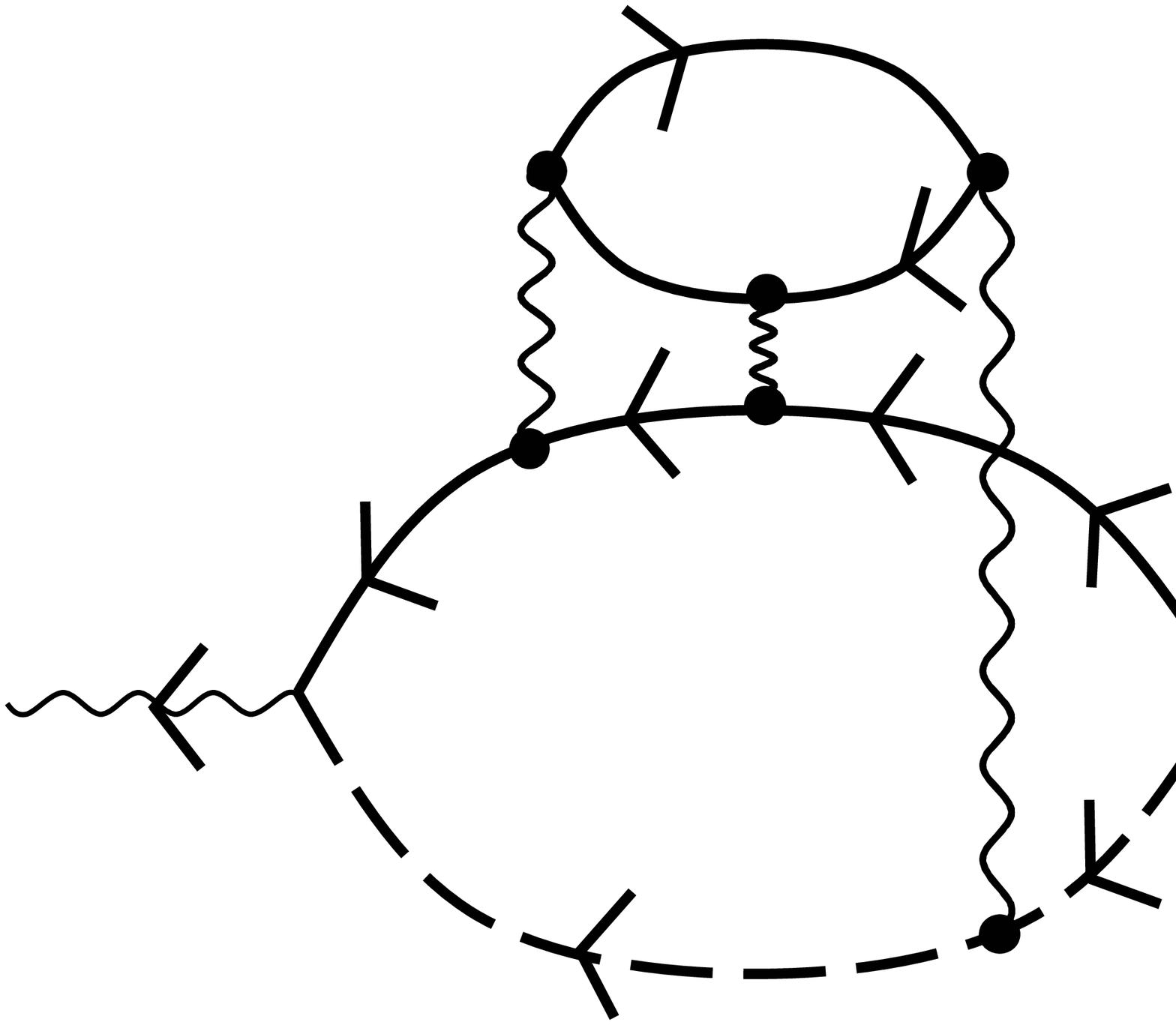}}
\hspace*{0.5cm} \scalebox{0.15}{\includegraphics{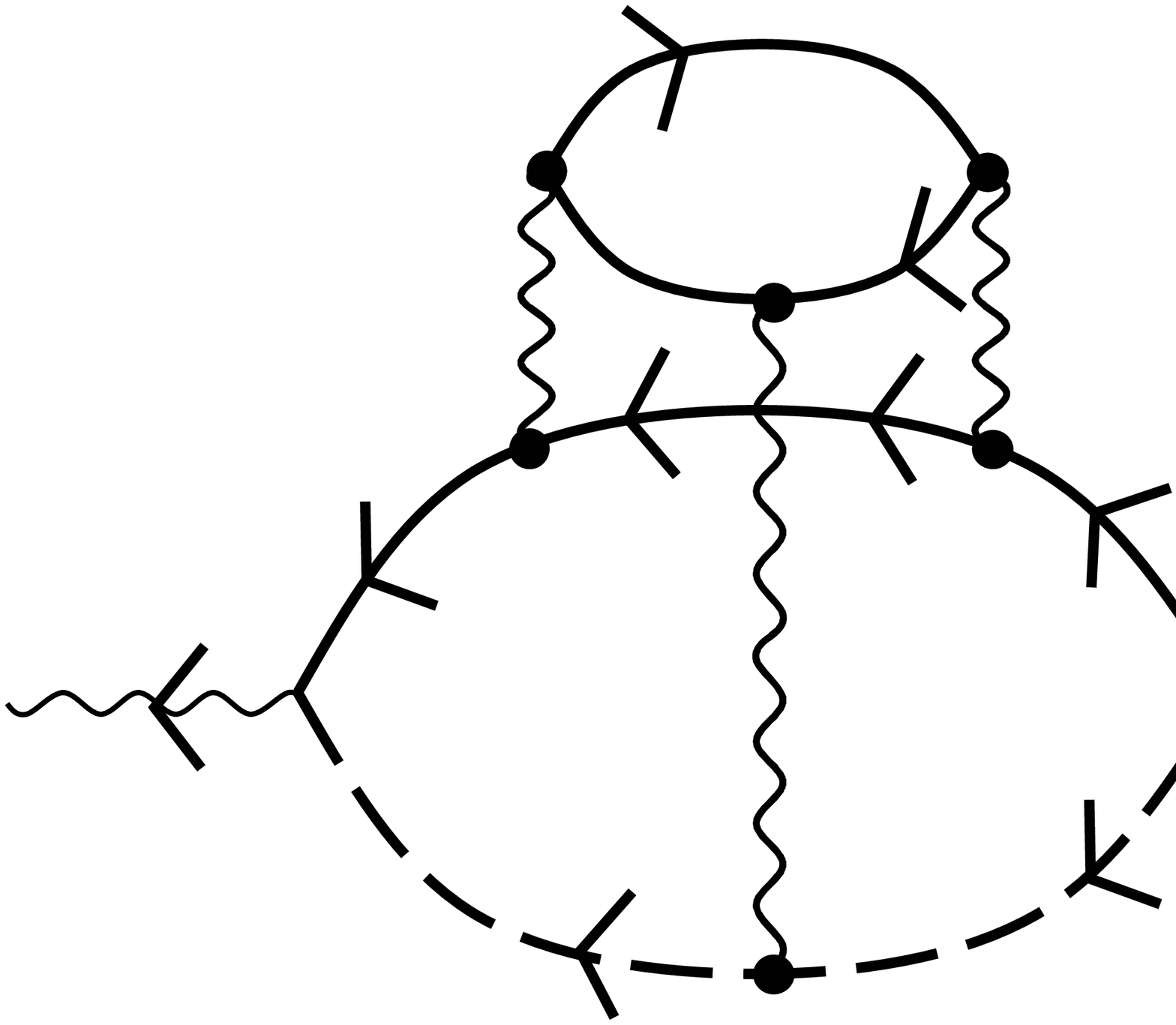}}
\hspace*{0.5cm} \scalebox{0.15}{\includegraphics{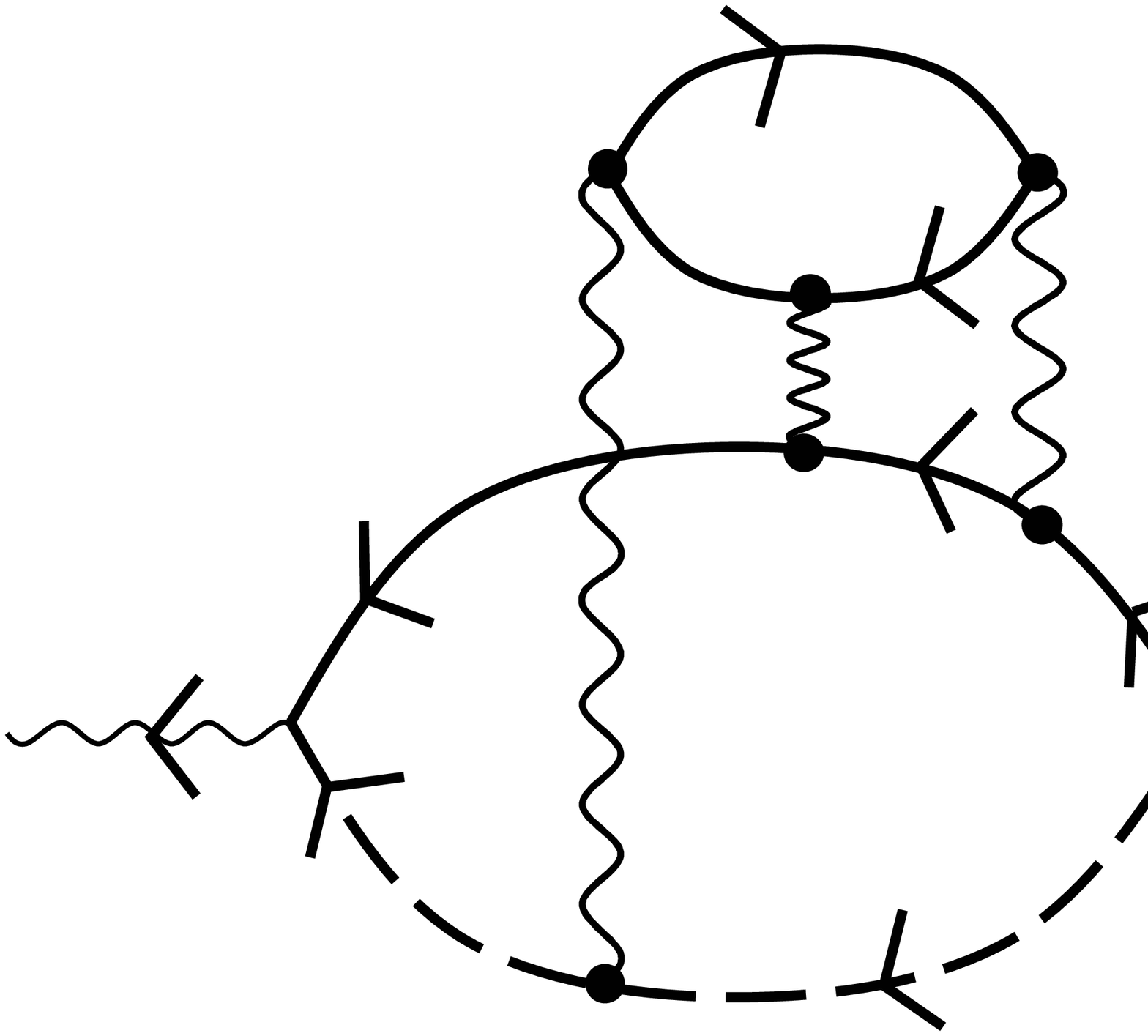}}
\hspace*{0.5cm} \scalebox{0.15}{\includegraphics{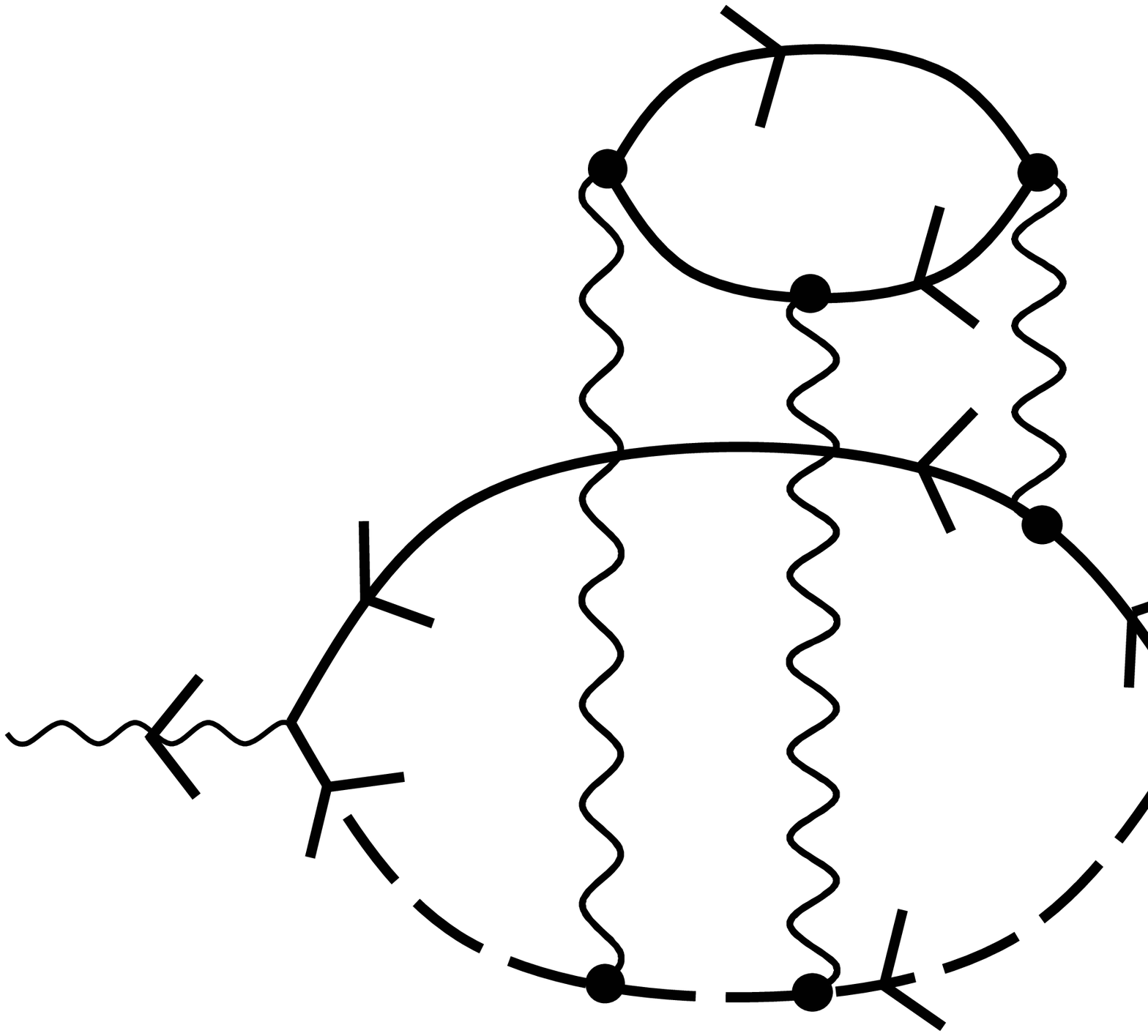}}}
\vspace*{0.5cm}
\centerline{\scalebox{0.15}{\includegraphics{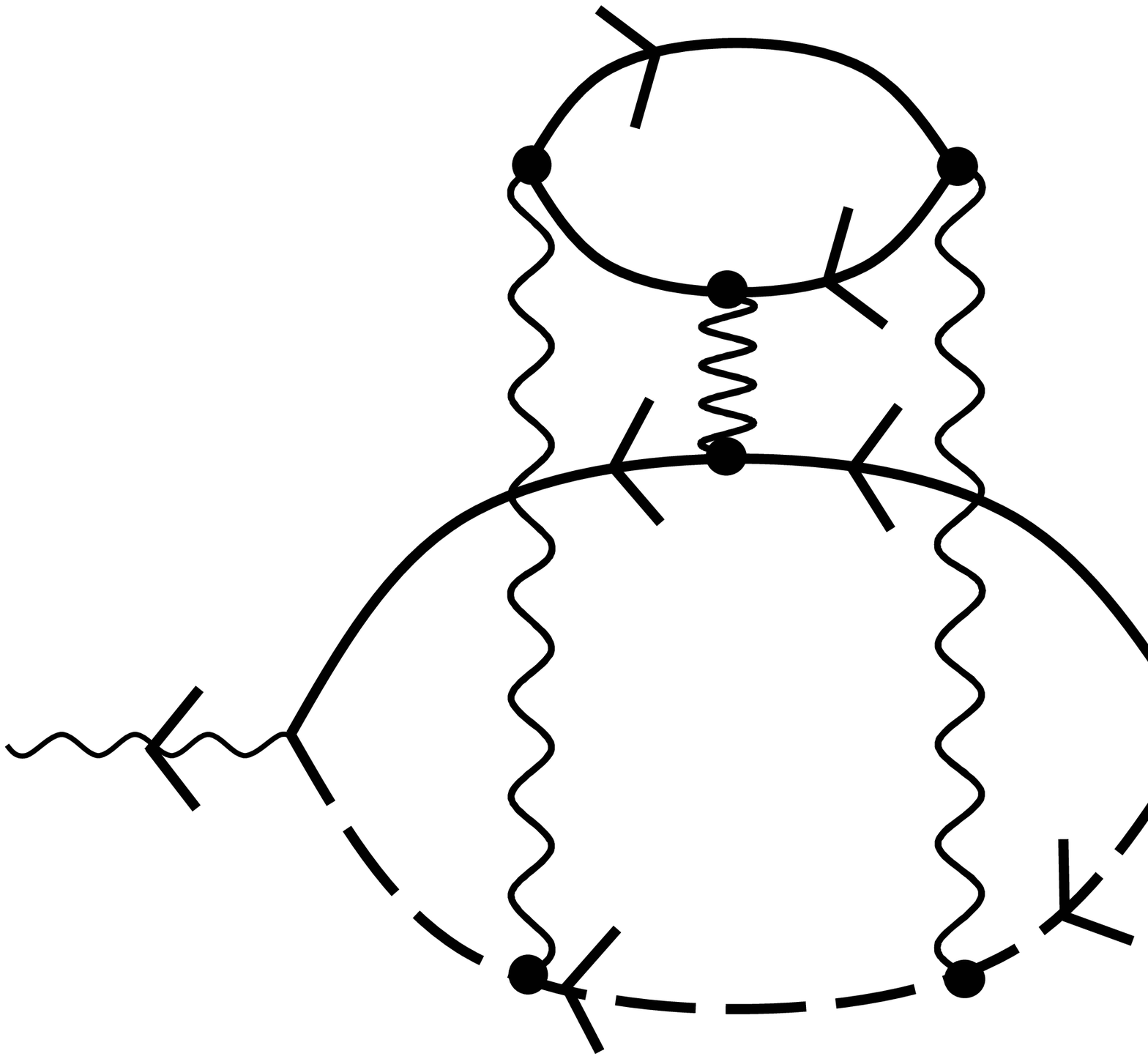}}
\hspace*{0.5cm} \scalebox{0.15}{\includegraphics{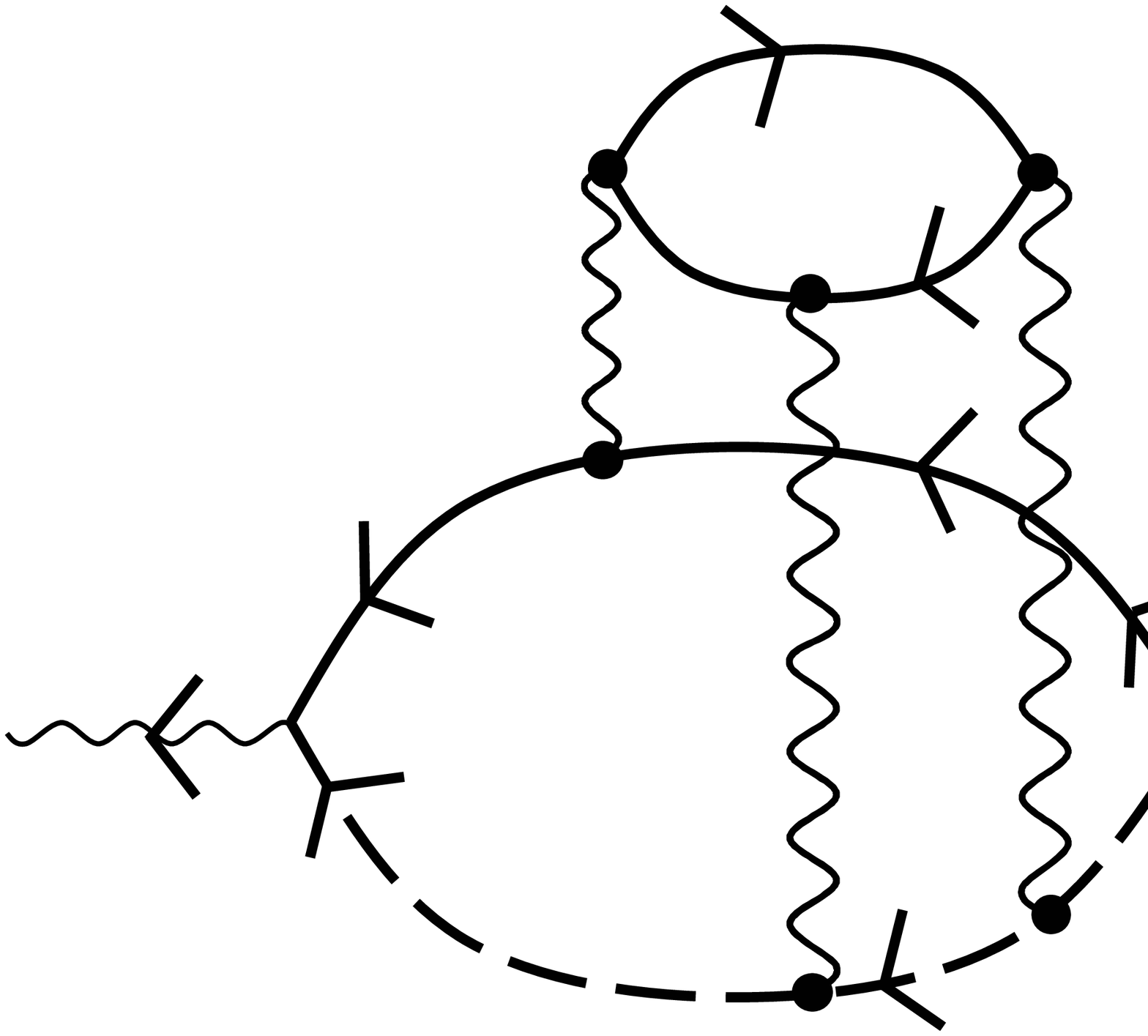}}
\hspace*{0.5cm} \scalebox{0.15}{\includegraphics{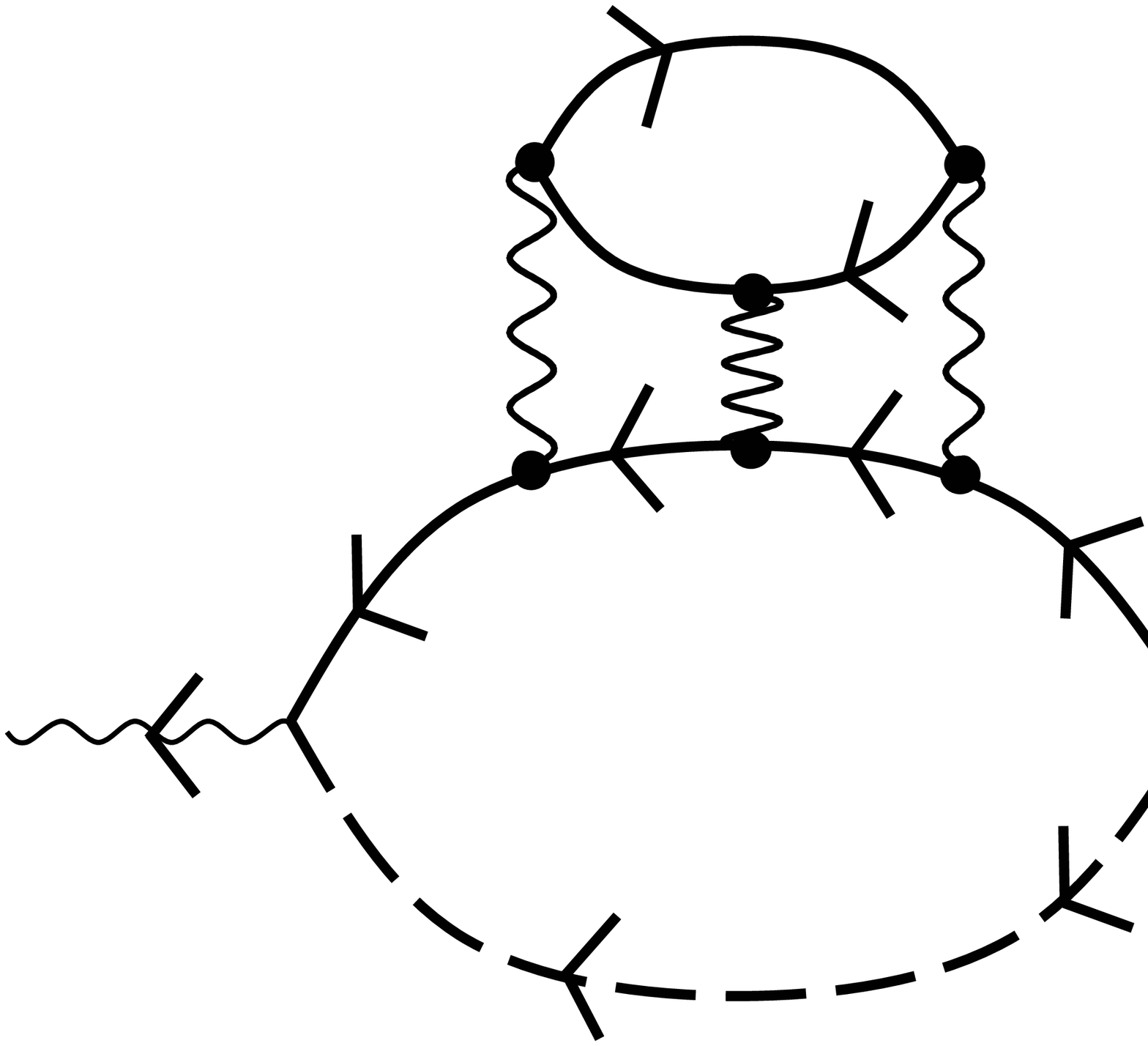}}
\hspace*{0.5cm}
\scalebox{0.15}{\includegraphics{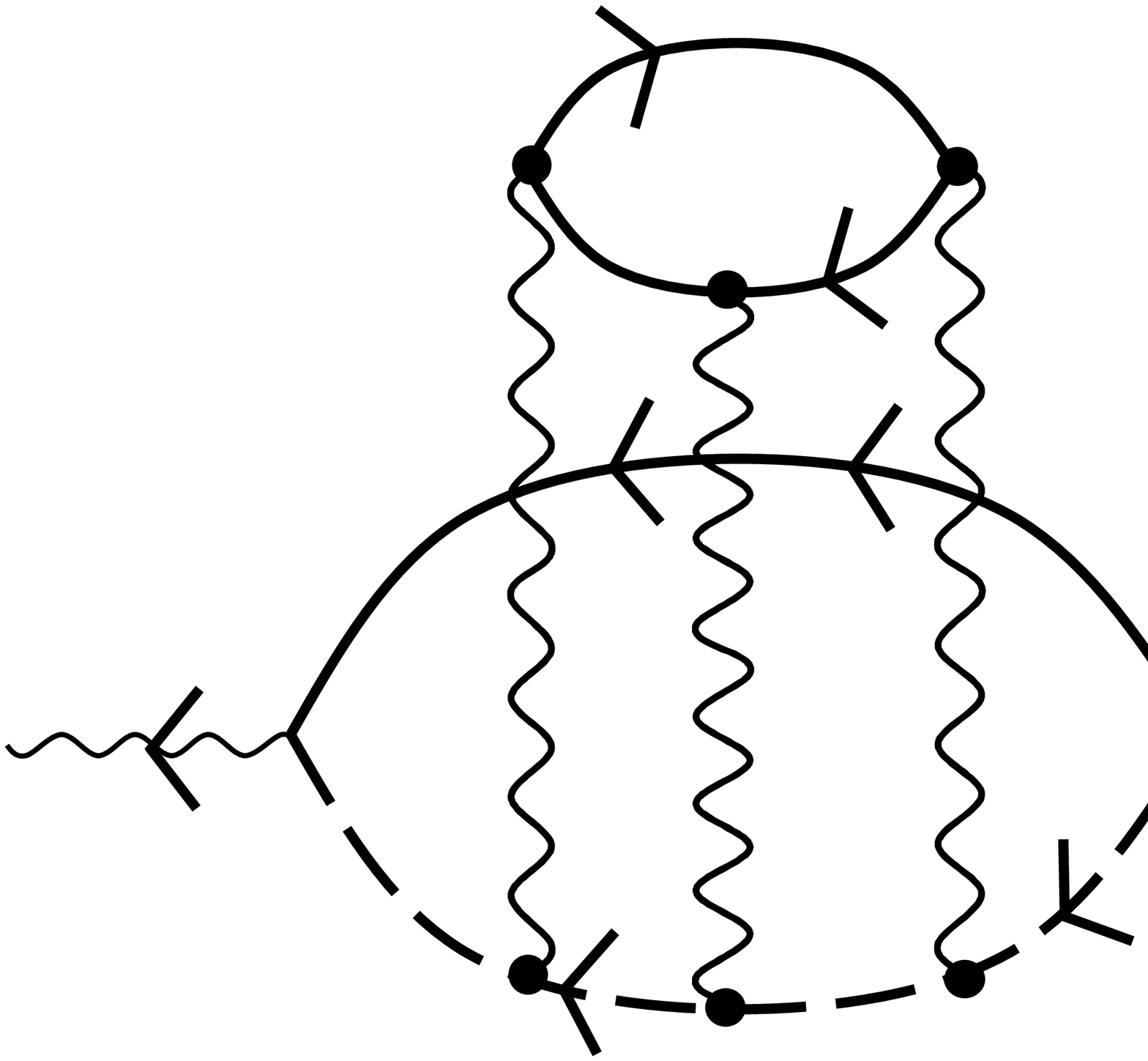}}}\caption{Same as
fig.\ (2), with three Coulomb interactions. The diagrams of this
figure have to be added to the one of fig.\ (1d).}
\end{figure}

\newpage

\begin{figure}
\begin{center}
\centerline{ \scalebox{0.2}{\includegraphics{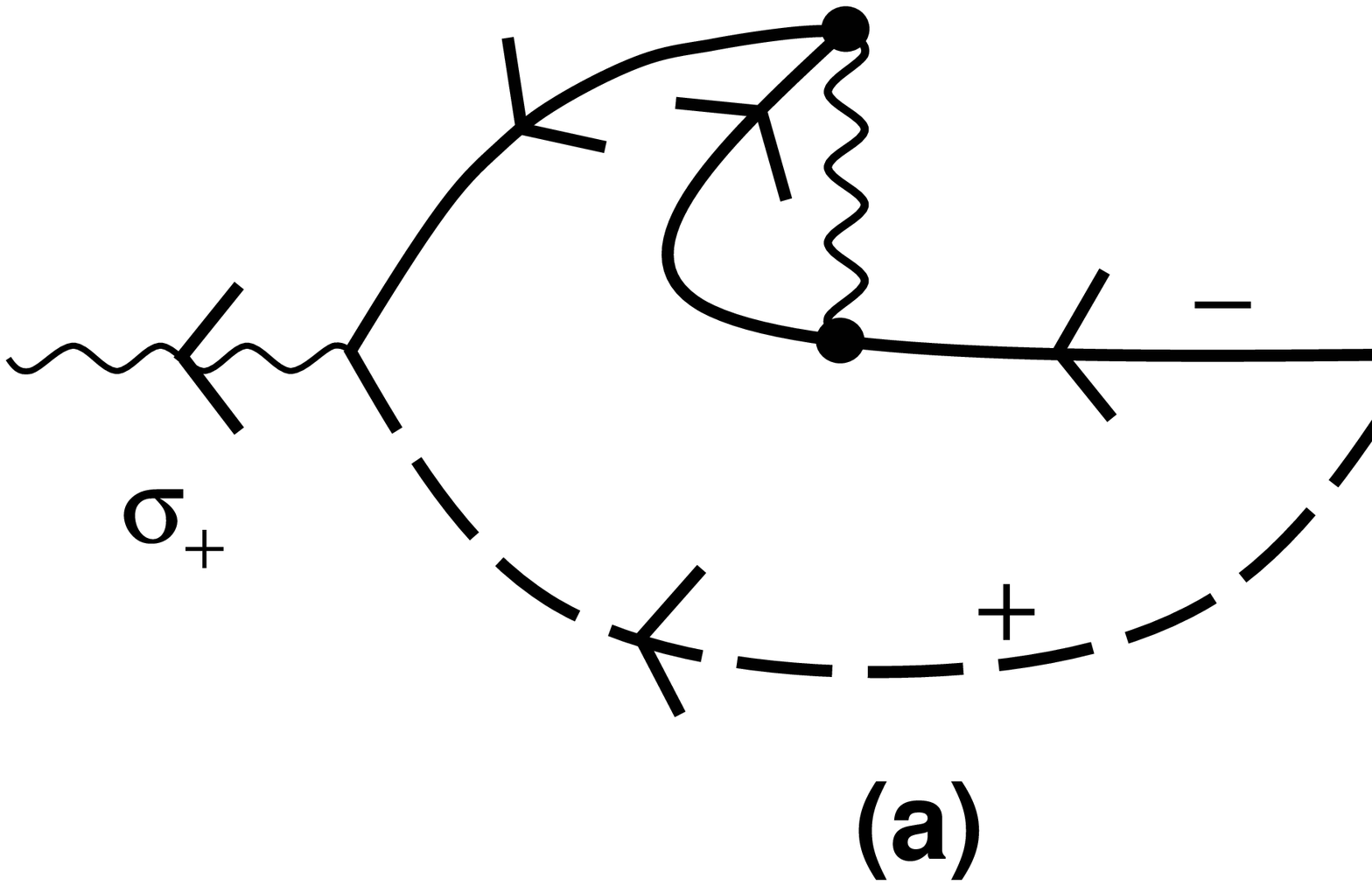}}}
\end{center}
\centerline{ \scalebox{0.2}{\includegraphics{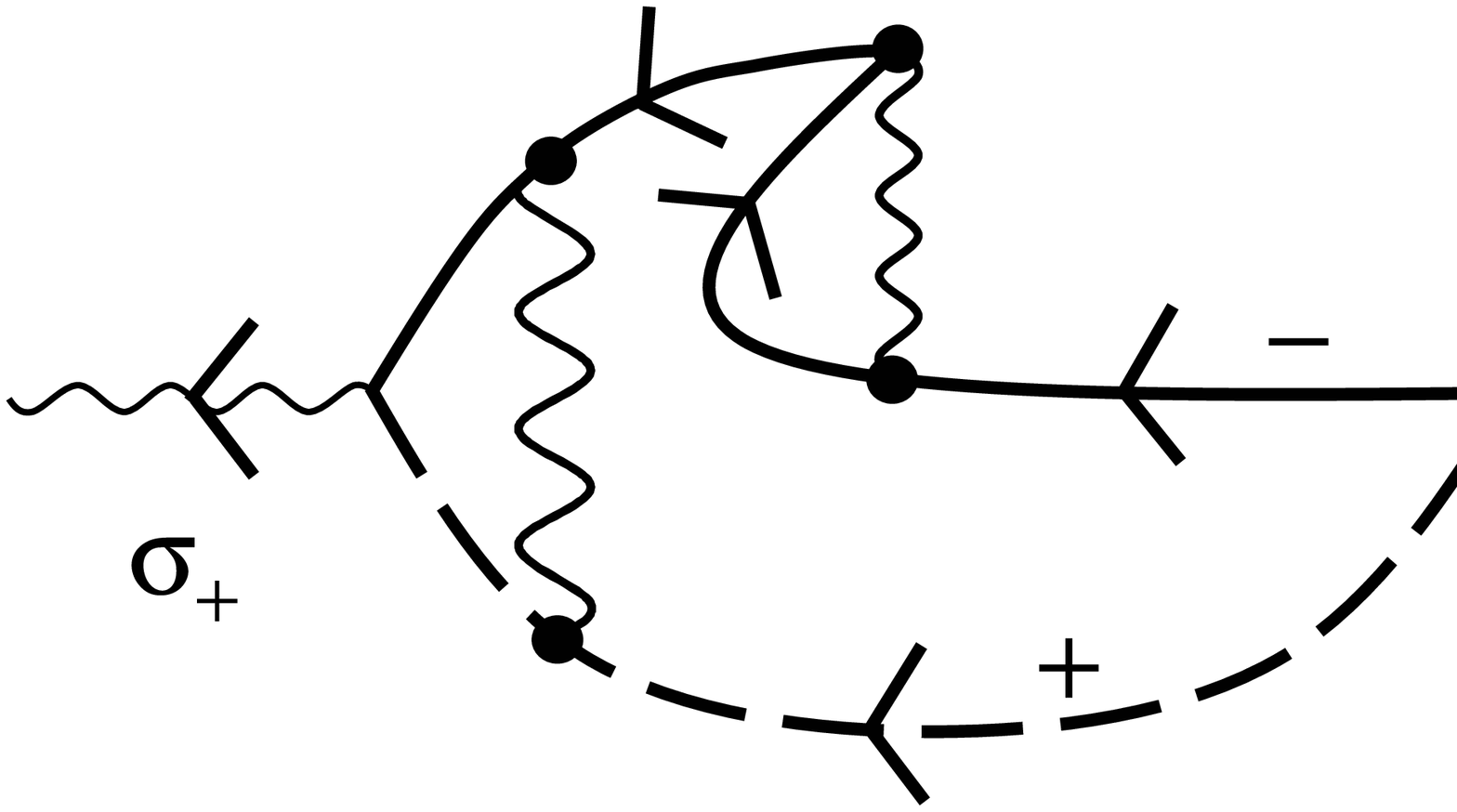}}
\hspace*{0.5cm} \scalebox{0.2}{\includegraphics{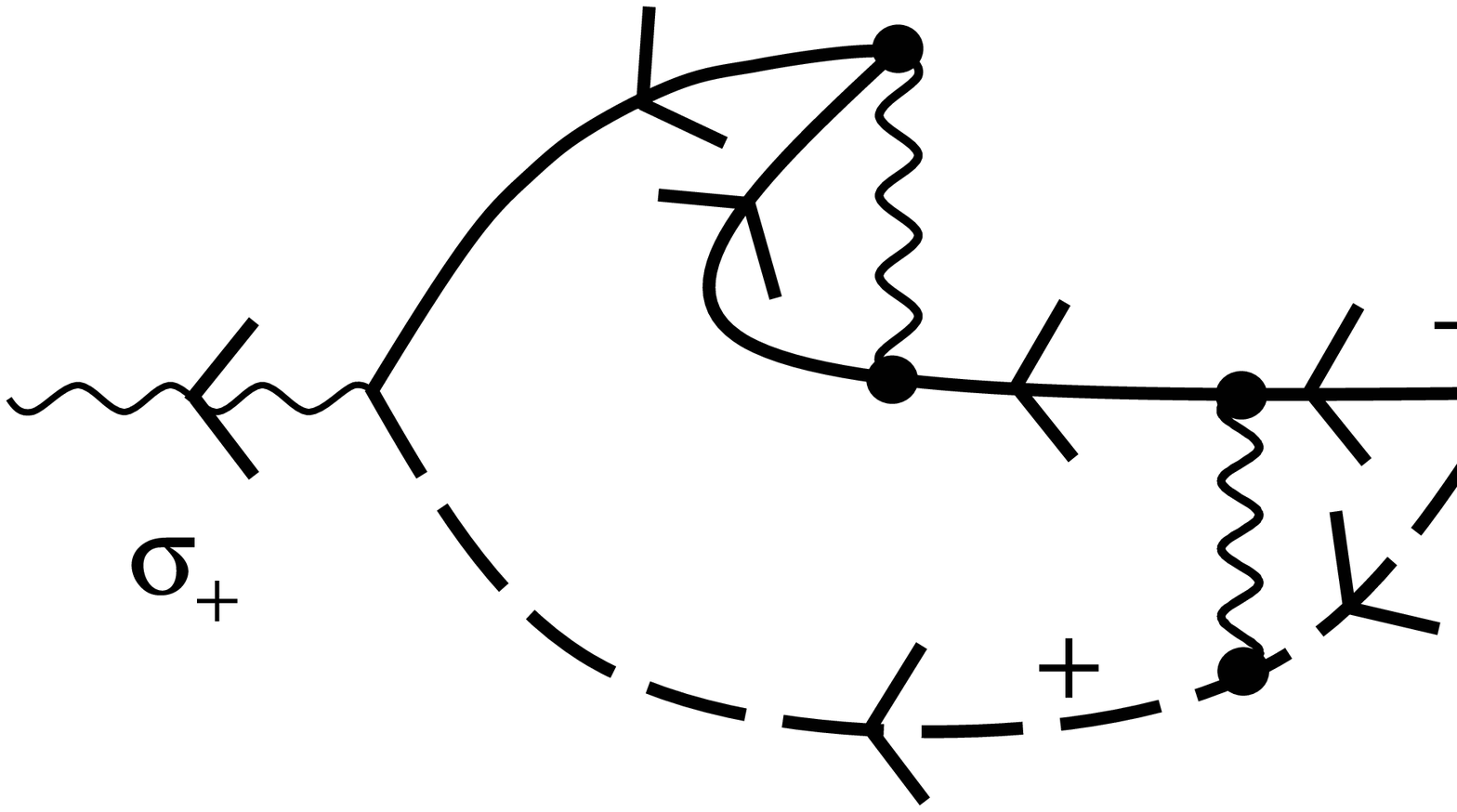}}
\hspace*{0.5cm} \scalebox{0.2}{\includegraphics{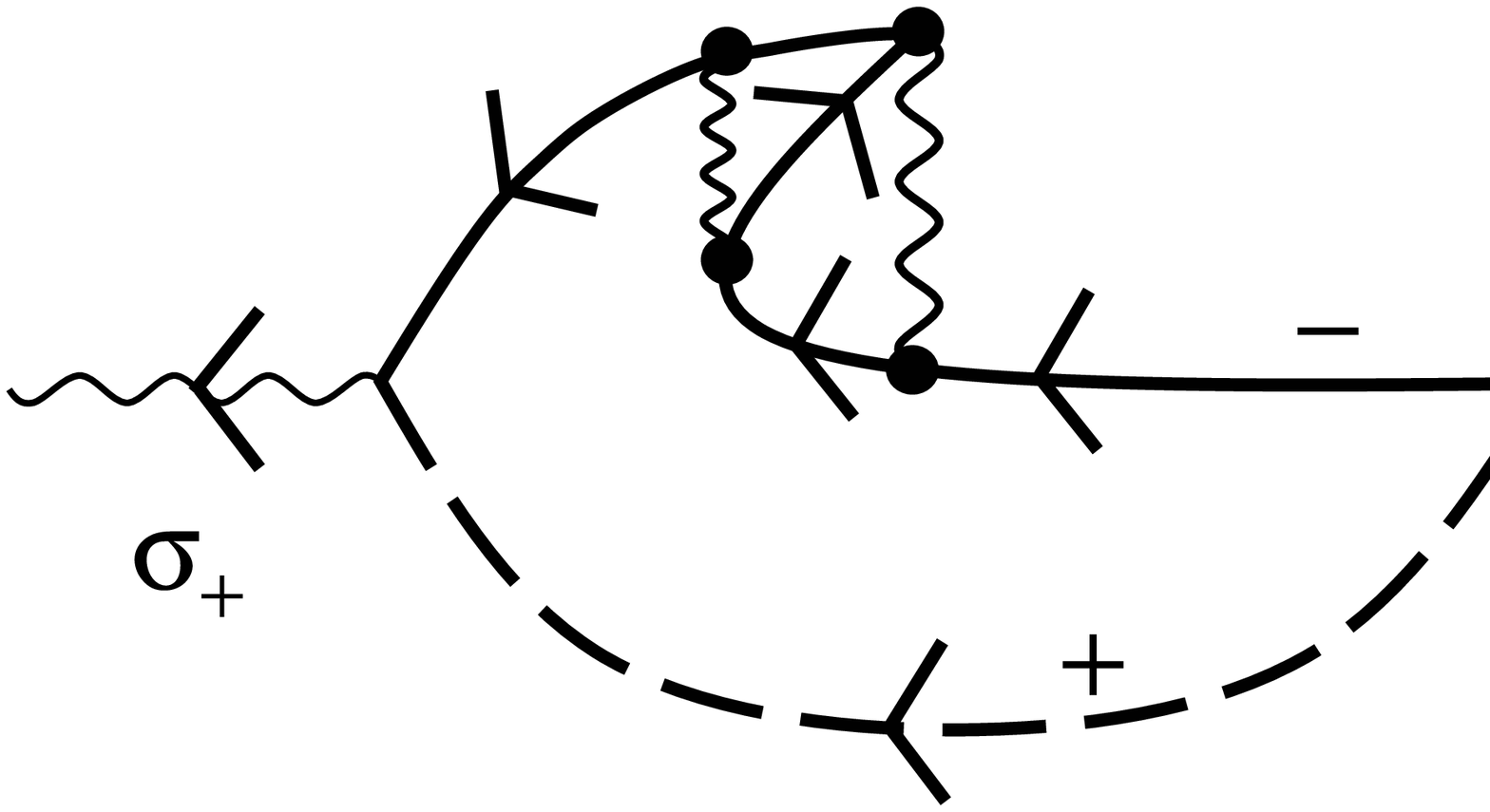}}}
\vspace*{0.5cm} \centerline{
\scalebox{0.2}{\includegraphics{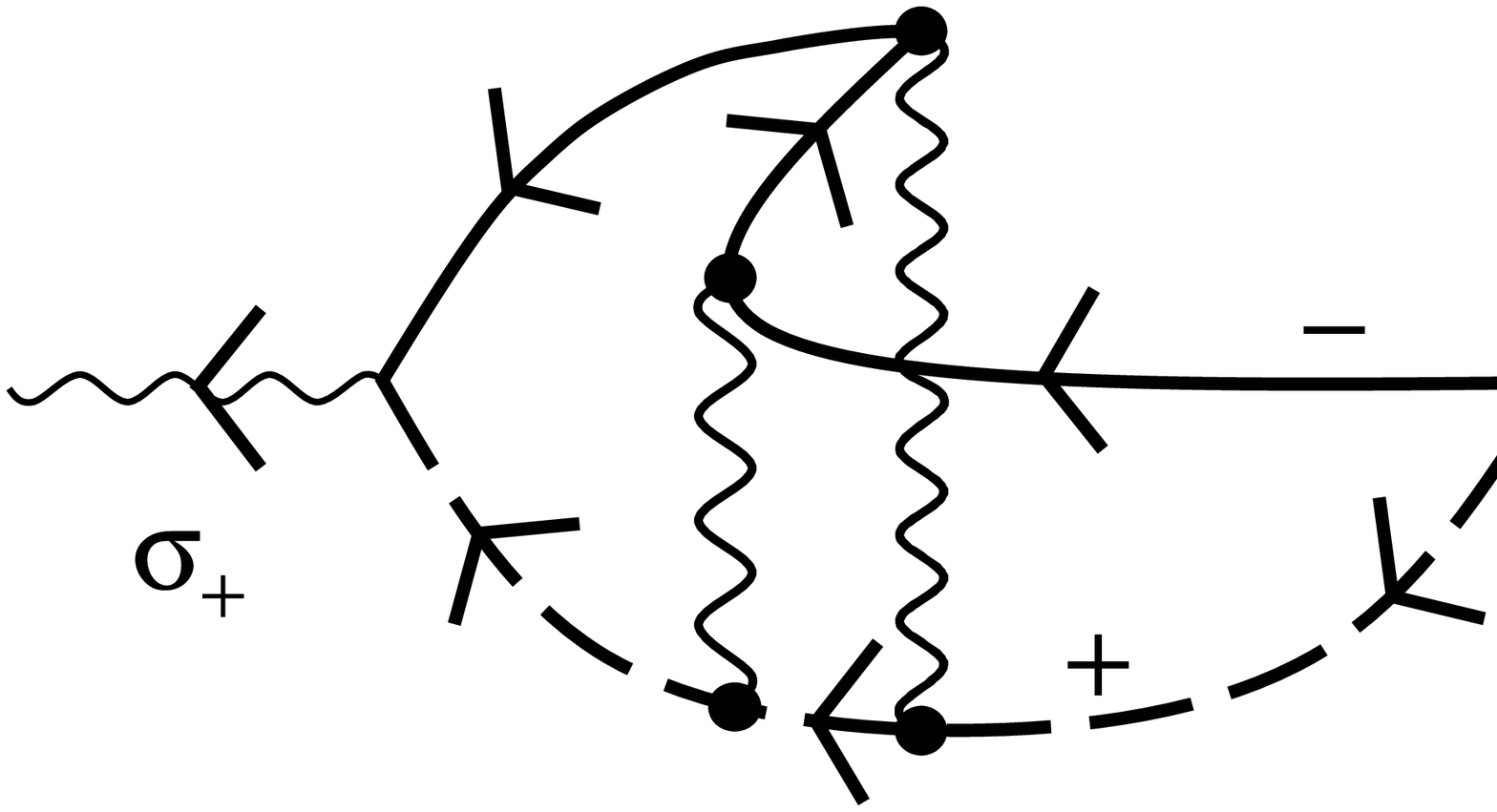}} \hspace*{0.5cm}
\scalebox{0.2}{\includegraphics{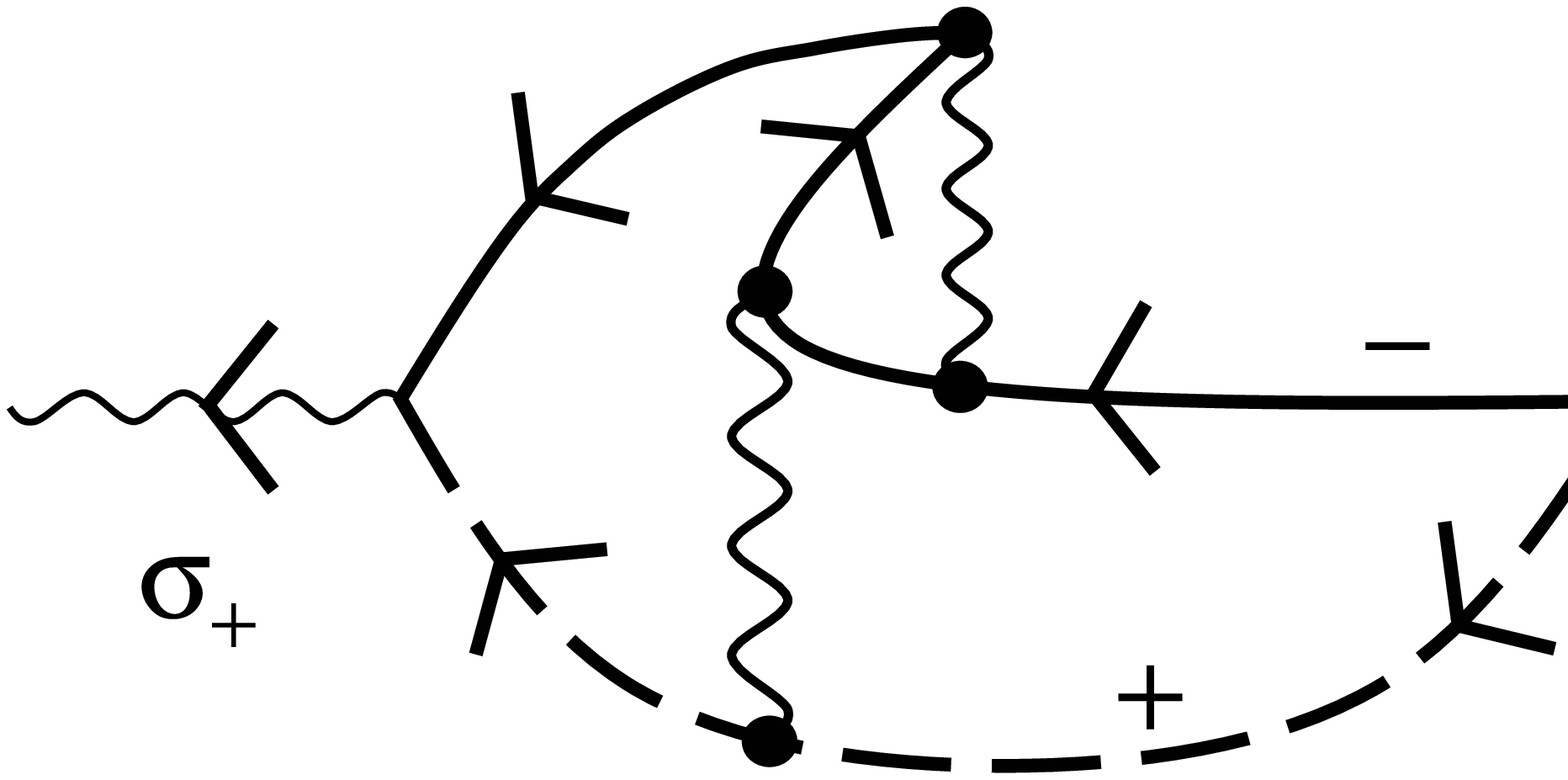}} \hspace*{0.5cm}
\scalebox{0.2}{\includegraphics{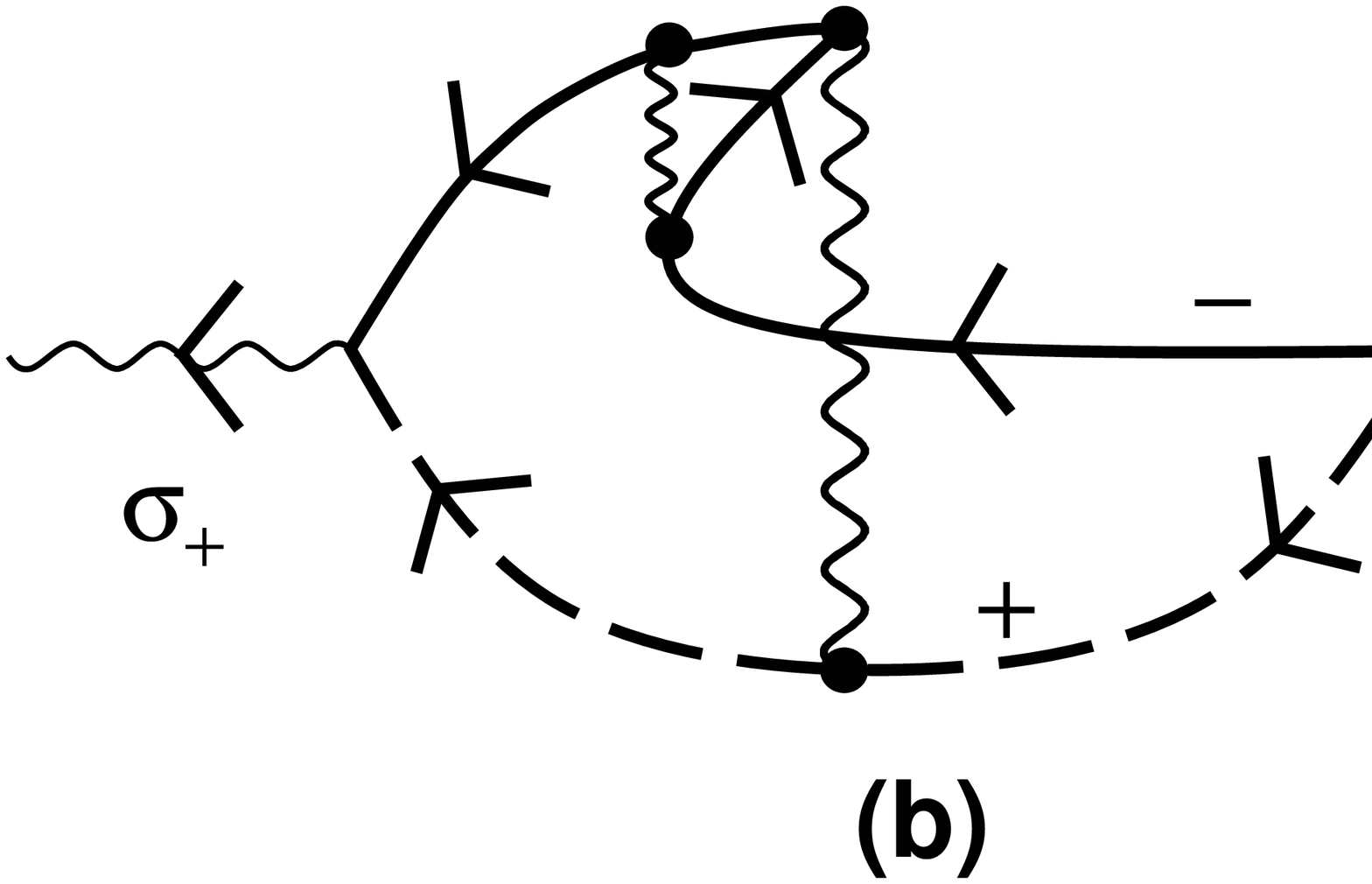}}} \caption{Absorption
of a $\sigma_+$ photon with trion formation, in the case of a
photocreated electron and an nitial electron having the same spin,
at first order (a) and second order (b) in Coulomb interaction.
The diagrams of this figure, which correspond to possible exchange
between the photocreated electron and the initial electron, have
to be added to the diagram of fig.\ (1b) and the diagram of fig.\
(1c) plus the ones of fig.\ (2), respectively.}
\end{figure}

\newpage

\begin{figure}
\centerline{ \scalebox{0.3}{\includegraphics{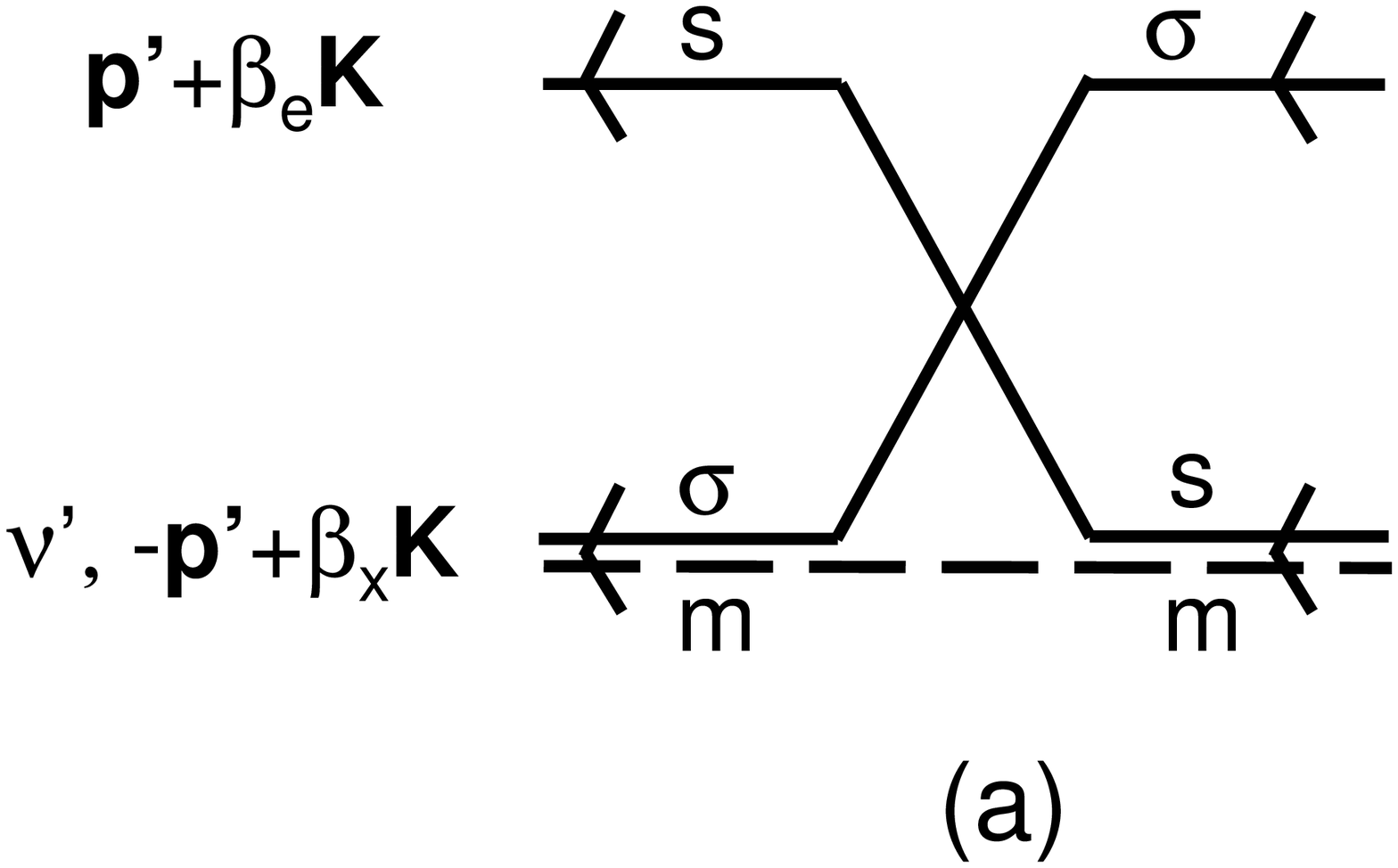}}
\hspace*{0.5cm} \scalebox{0.3}{\includegraphics{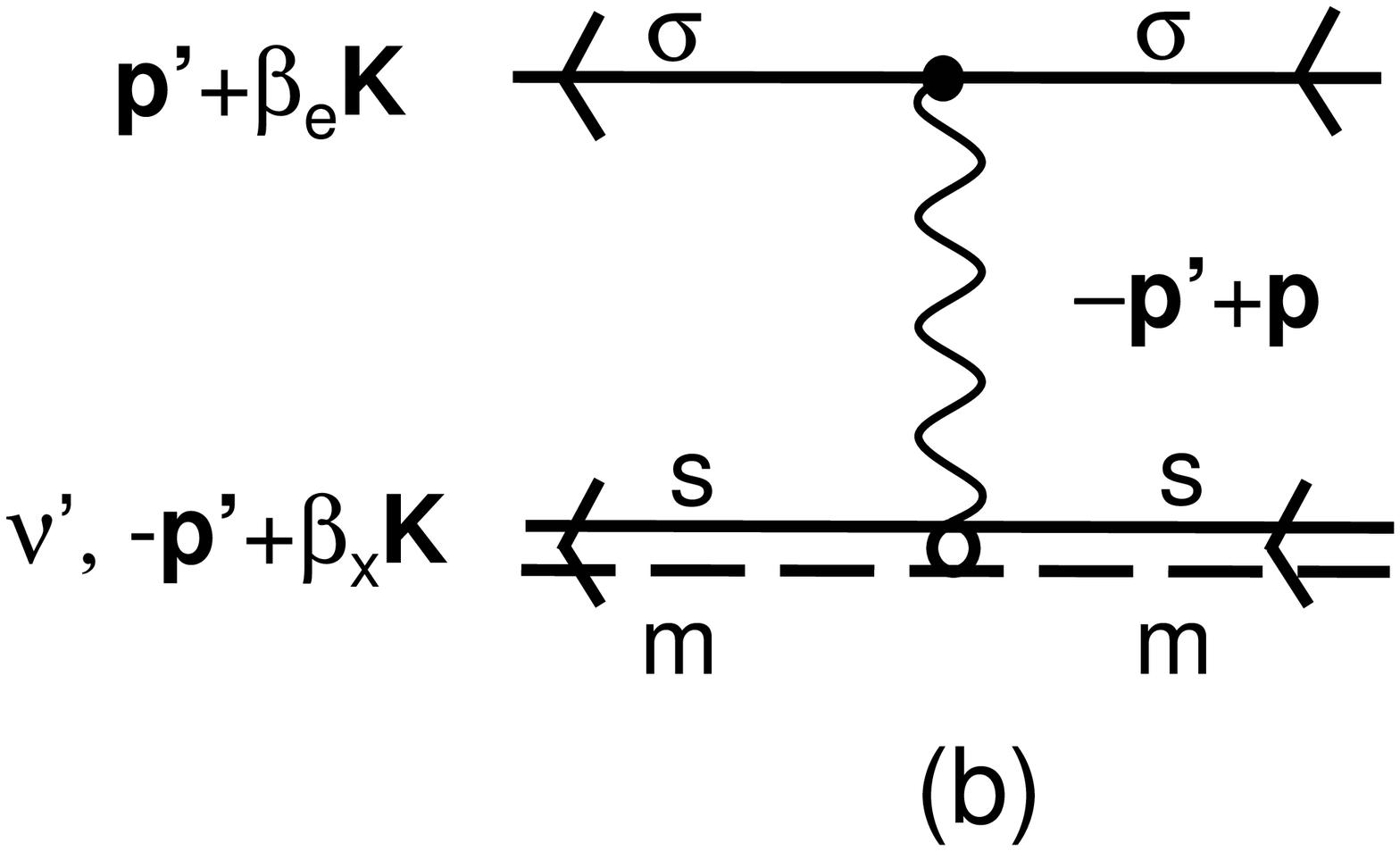}}}
\caption{(a) Exchange parameter $L_{\nu'\v p';\nu\v p}$ of the
``commutation technique''. The ``in'' exciton $\nu$ and the
``out'' exciton $\nu'$ are made with different electrons. No
Coulomb interaction takes place in this scattering. (b) Direct
coulomb scattering $C_{\nu'\v p';\nu\v p}^\mathrm{dir}$ of the
``commutation technique''. The ``in'' exciton $\nu$ and the
``out'' exciton $\nu'$ are made with the same electron.}
\end{figure}

\newpage

\begin{figure}
\begin{center}
\scalebox{0.5}{\includegraphics{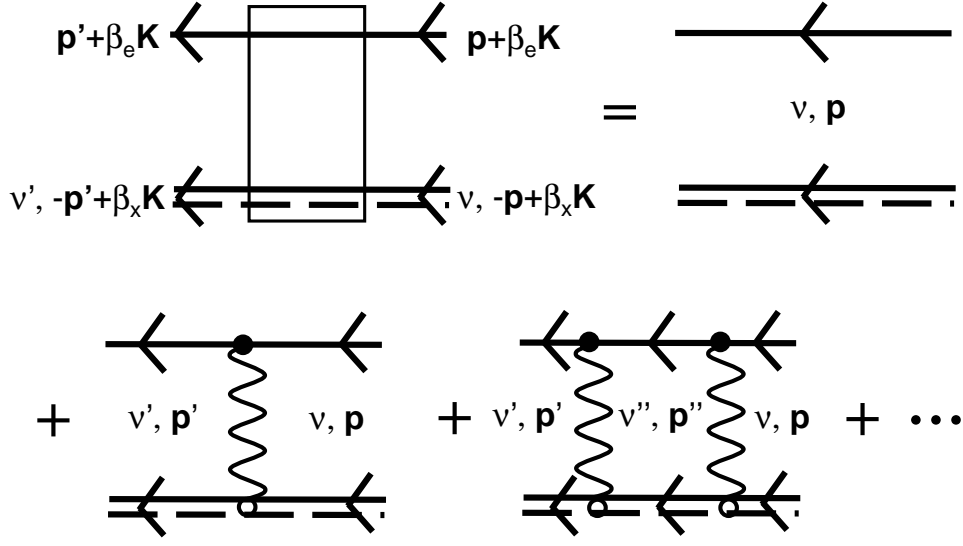}}\caption{Set of ladder
processes between one electron and one exciton. Note that, all
over, the exciton is made with the same electron and the e-X pair
has the same center of mass momentum $\v K$. The exciton
quantum number, $\nu$, and the pair relative motion momentum, $\v
p$, are the only things which change in these ladder processes. The
solid line corresponds to the electron. The double solid-dashed line
corresponds to the exciton. The wavy line corresponds to the direct
Coulomb scattering
$C_{\nu'\v p';\nu\v p}^\mathrm{dir}$.}
\end{center}
\end{figure}

\newpage

\begin{figure}
\centerline{\scalebox{0.25}{\includegraphics{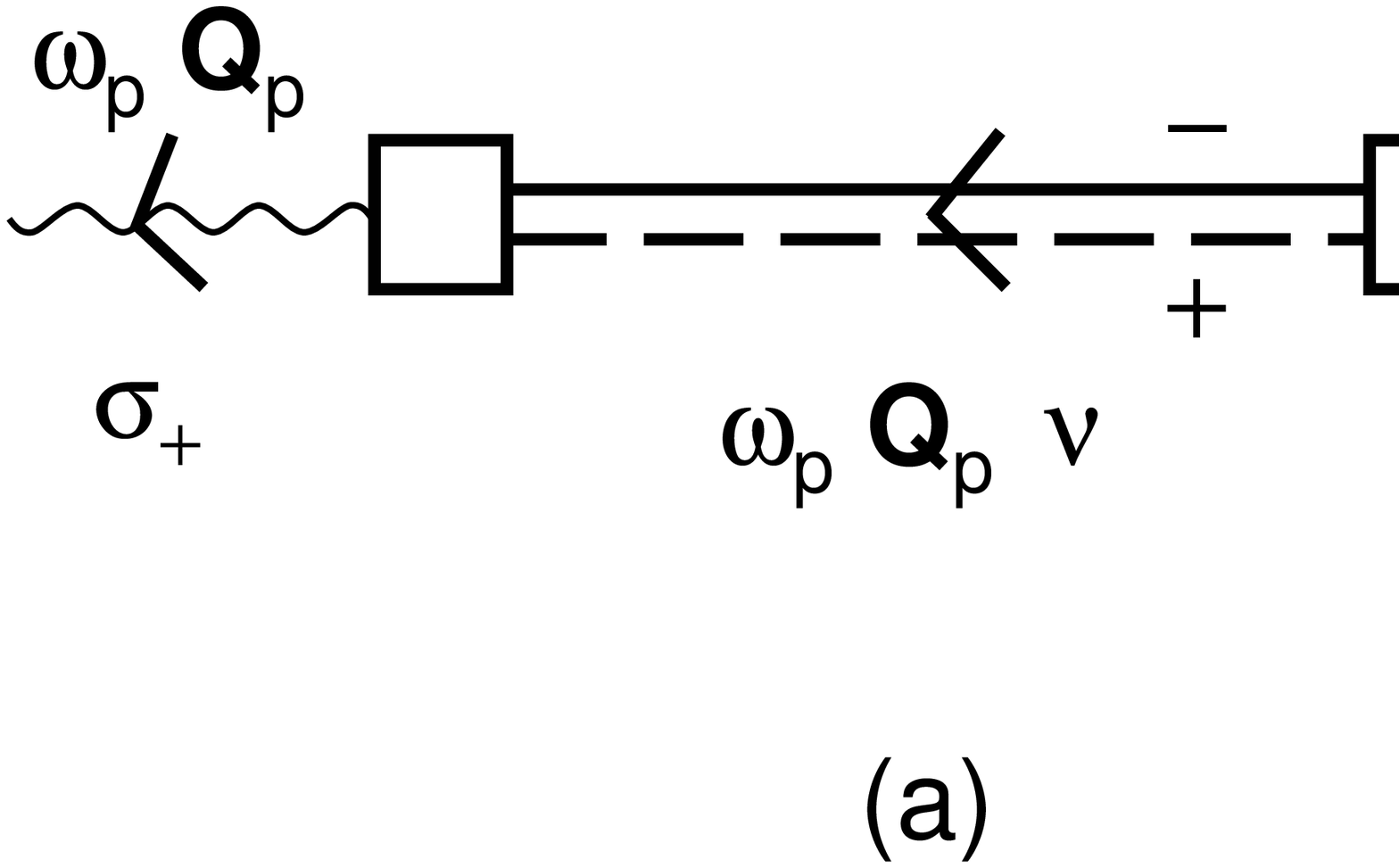}}
\hspace*{0.5cm} \scalebox{0.25}{\includegraphics{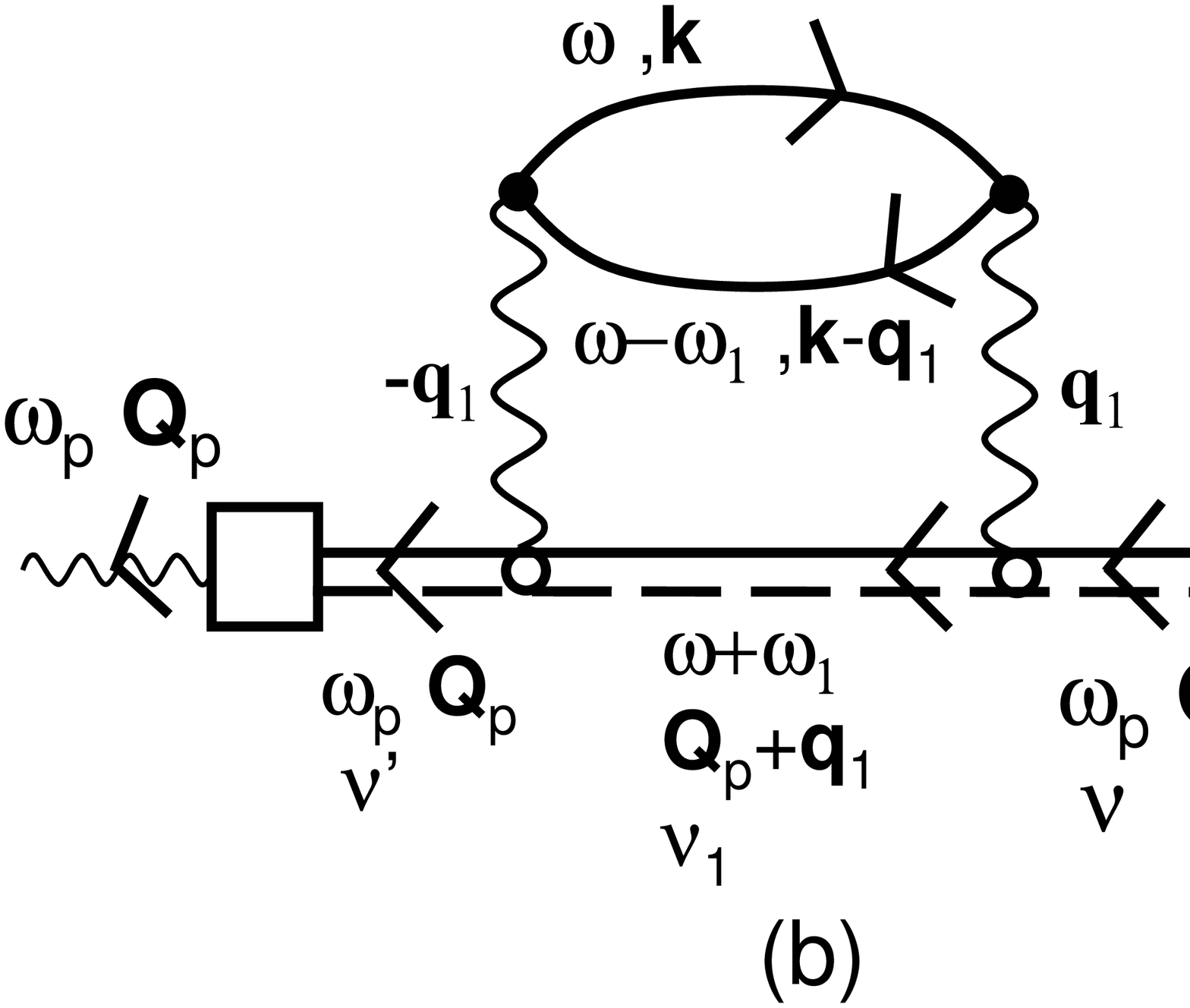}}}
\vspace*{0.5cm} \centerline{
\scalebox{0.25}{\includegraphics{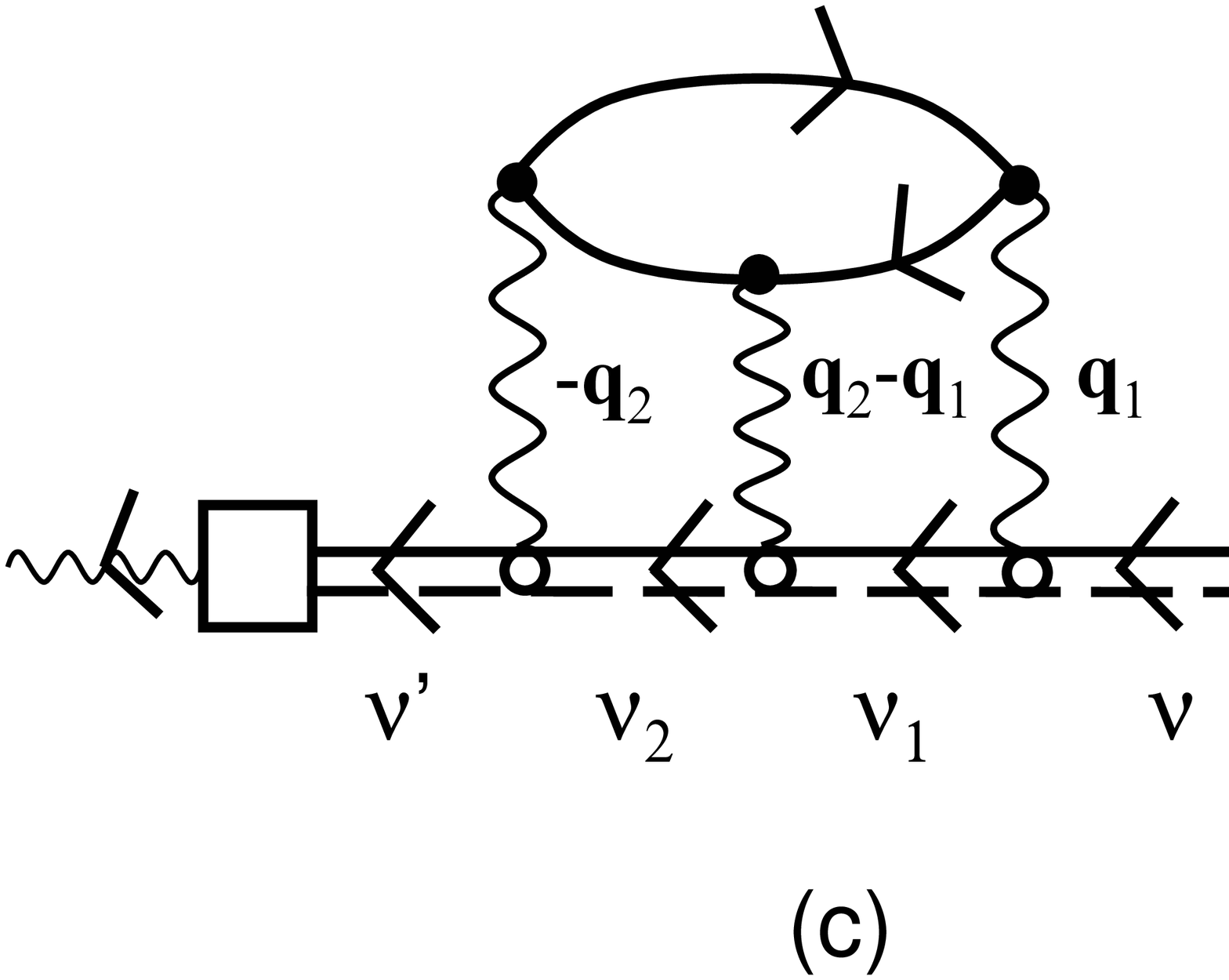}}\hspace*{0.5cm}
\scalebox{0.25}{\includegraphics{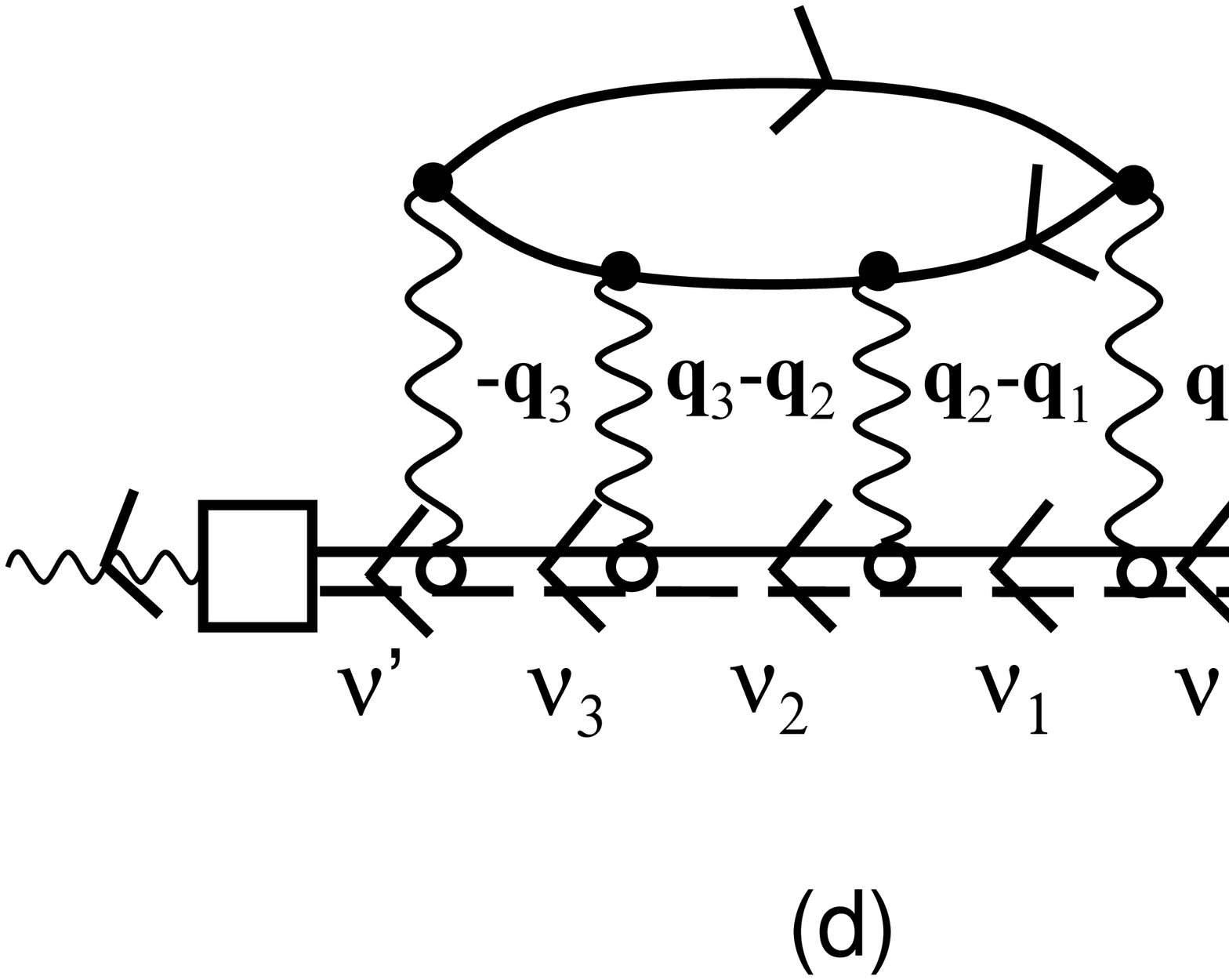}}}
\begin{center}
\scalebox{0.25}{\includegraphics{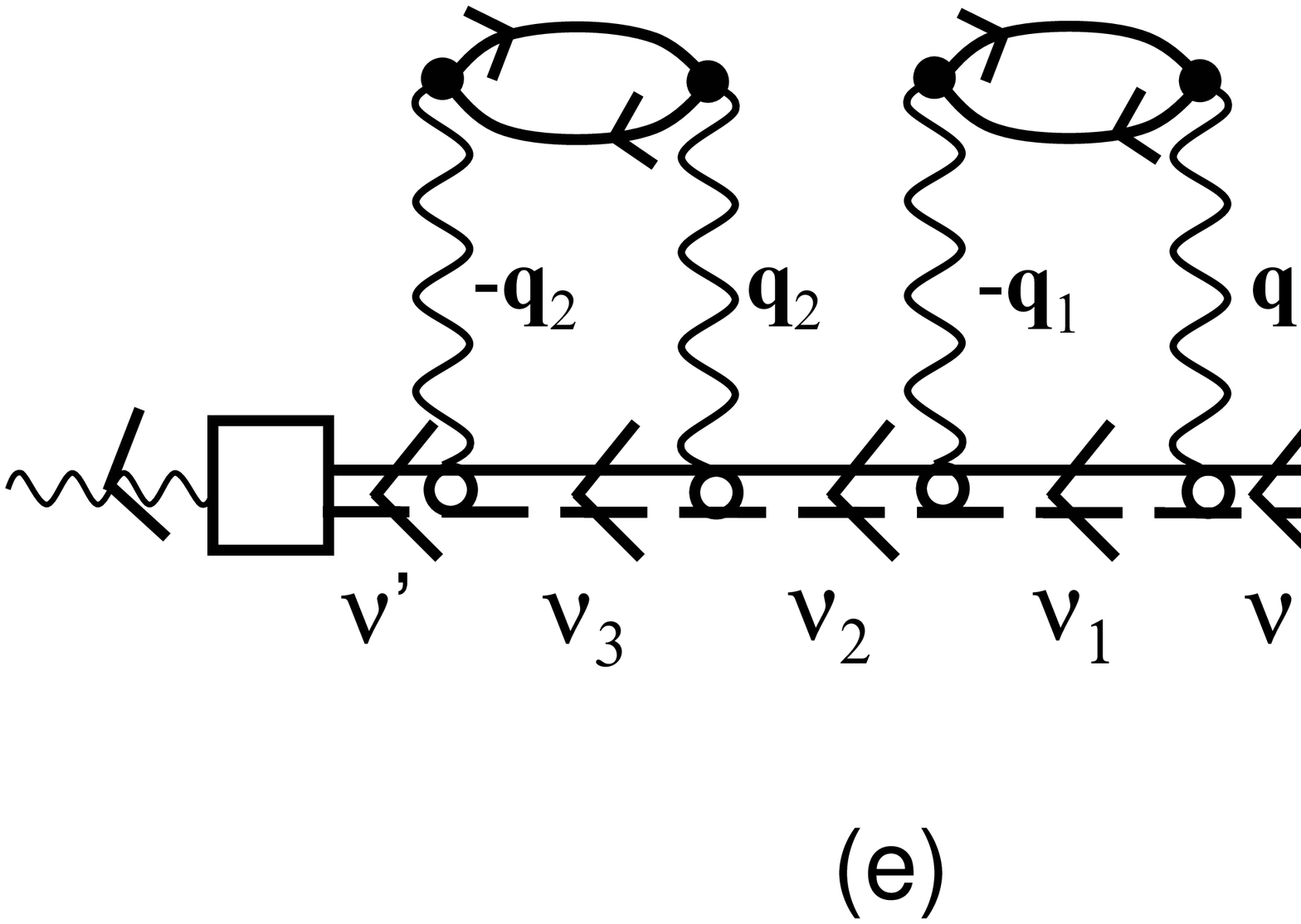}}\end{center}
\caption{(a) Absorption of a $\sigma_+$ photon with exciton
formation, using exciton diagrams. This unique diagram corresponds
to the set of electron-hole ladder diagrams of fig.\ (1). It also
corresponds to the trion absorption diagram, at zero order in
Coulomb scattering $C_{\nu'\v p';\nu\v p}^\mathrm{dir}$. (b)
Absorption of a $\sigma_+$ photon with trion formation, when the
photocreated electron and the initial electron have different
spins: second order process in (direct) Coulomb scattering between
electron and exciton $C_{\nu'\v p';\nu\v p}^\mathrm{dir}$. (c)
Same as (b), with three direct Coulomb scatterings. (d) Same as
(b), with four direct Coulomb scatterings. (e) Additional diagram
which appears in the response function at fourth order in
$C_{\nu'\v p';\nu\v p}^\mathrm{dir}$.}
\end{figure}

\newpage

\begin{figure}
\centerline{\scalebox{0.25}{\includegraphics{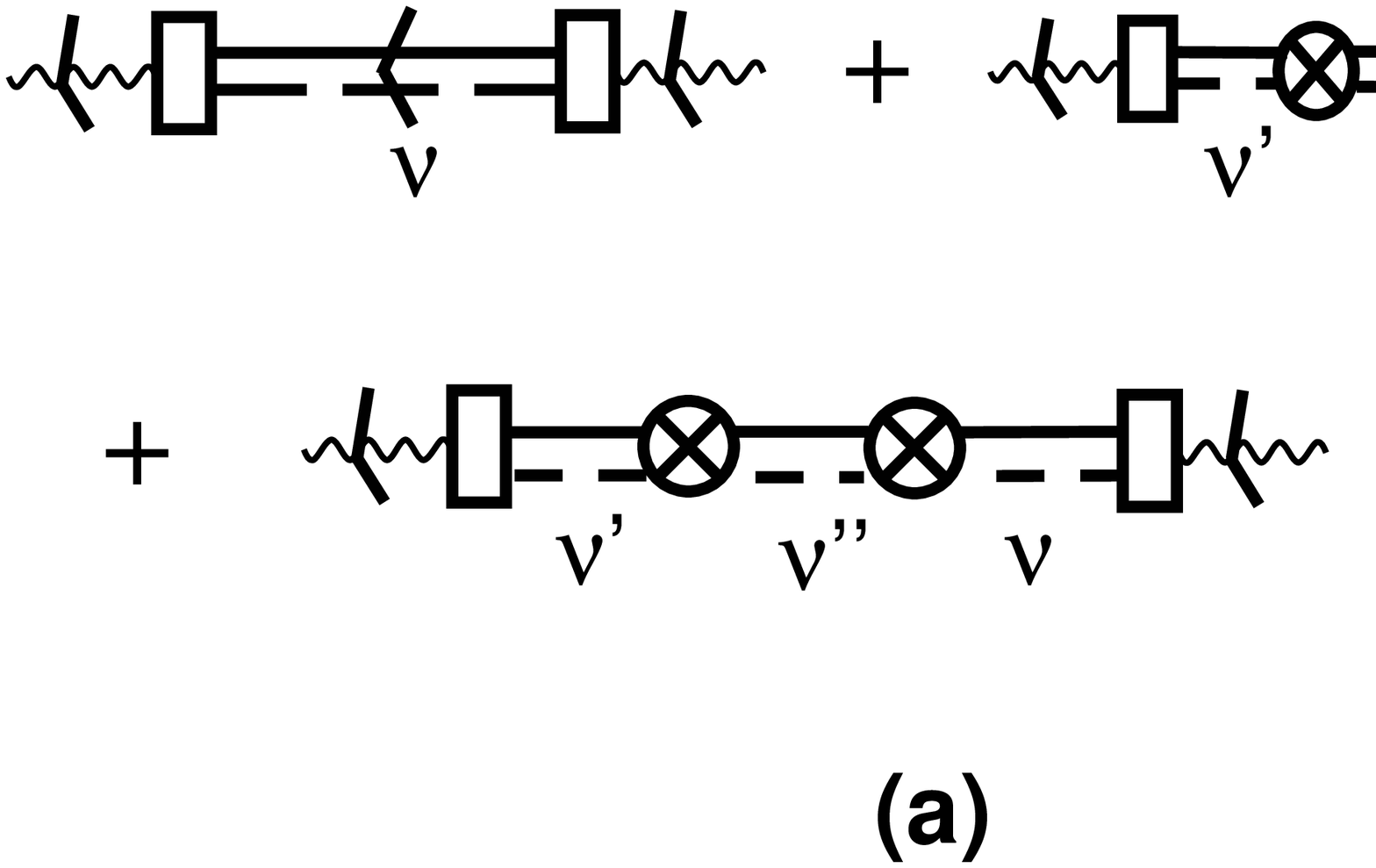}}
\hspace*{0.5cm} \scalebox{0.25}{\includegraphics{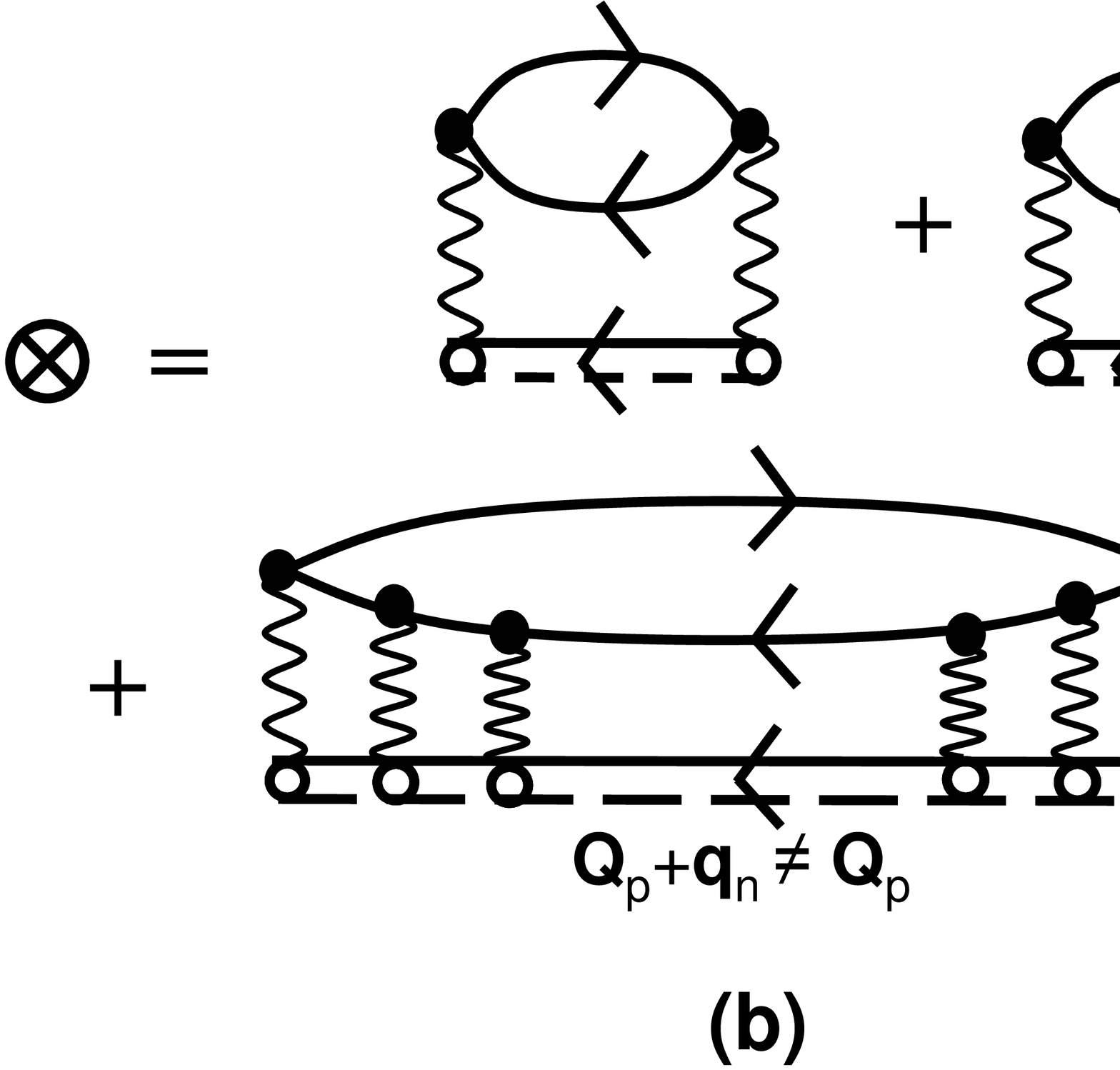}}}
\caption{(a) Integral equation (4.28) for the response function
$S_{\neq}$ when the photocreated electron and the initial electron
have different spins. The cross corresponds to all possible
diagrams shown in fig.\ (8b). (b) Processes contributing to the
exciton scattering $\Gamma_{\nu'\nu}$ appearing in the integral
equation (4.28): In these ``bubbles'', the exciton momentum always
differs from the initial momentum $\v Q_p$, by construction.}
\end{figure}

\newpage

\begin{figure}
\begin{center}
\scalebox{0.3}{\includegraphics{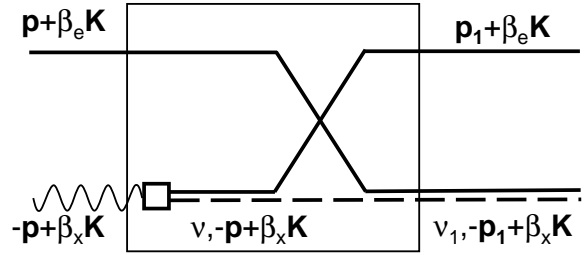}}\caption{Semiconductor-photon
interaction $\hat{\lambda}_{\v p;\nu_1\v p_1}$ dressed by the
presence of one electron $\v k=\v p+\beta_X\v K$ having the same
spin as the photocreated electron.}
\end{center}
\end{figure}

\newpage

\begin{figure}
\centerline{\scalebox{0.2}{\includegraphics{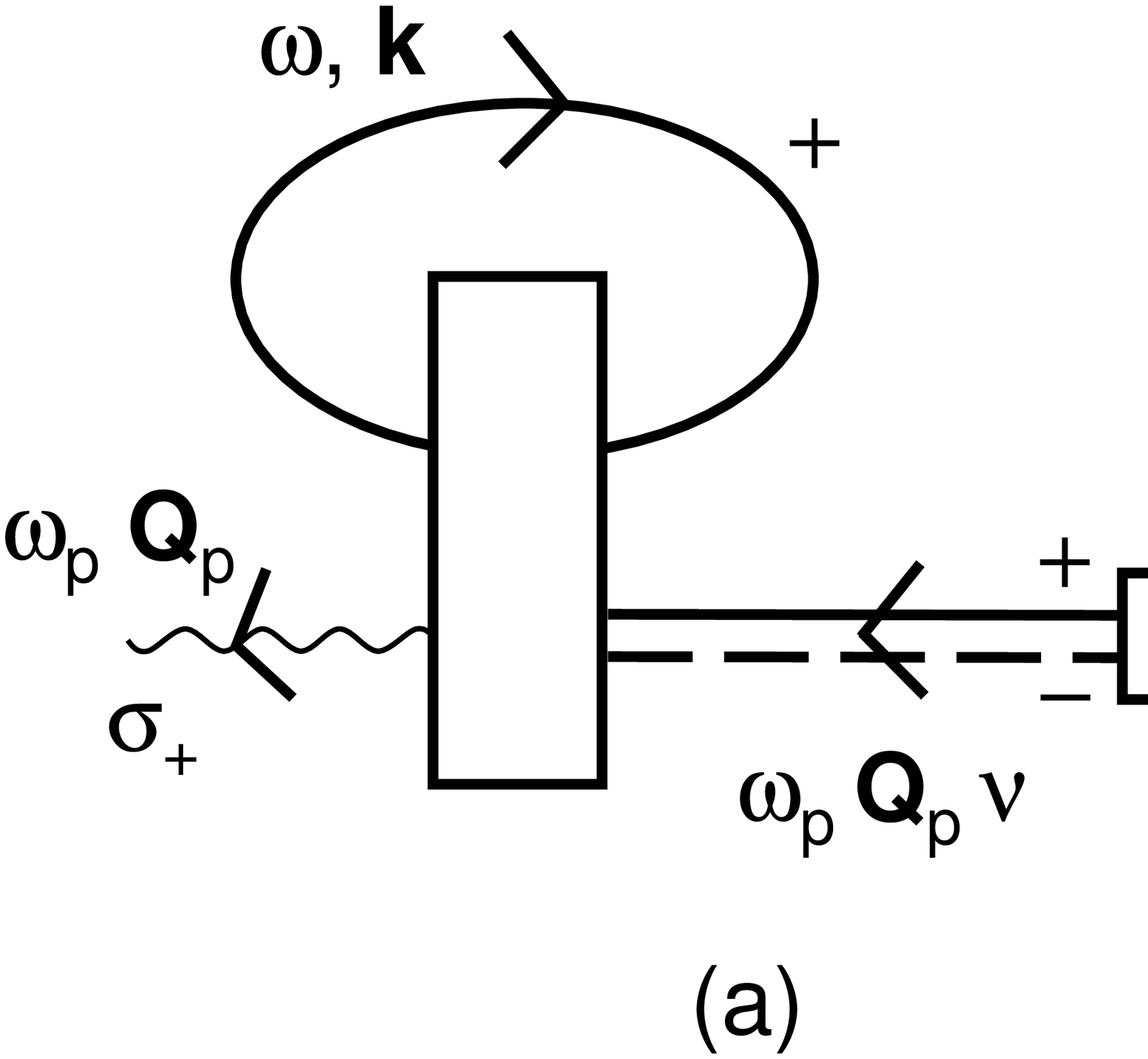}}
\hspace*{0.5cm} \scalebox{0.2}{\includegraphics{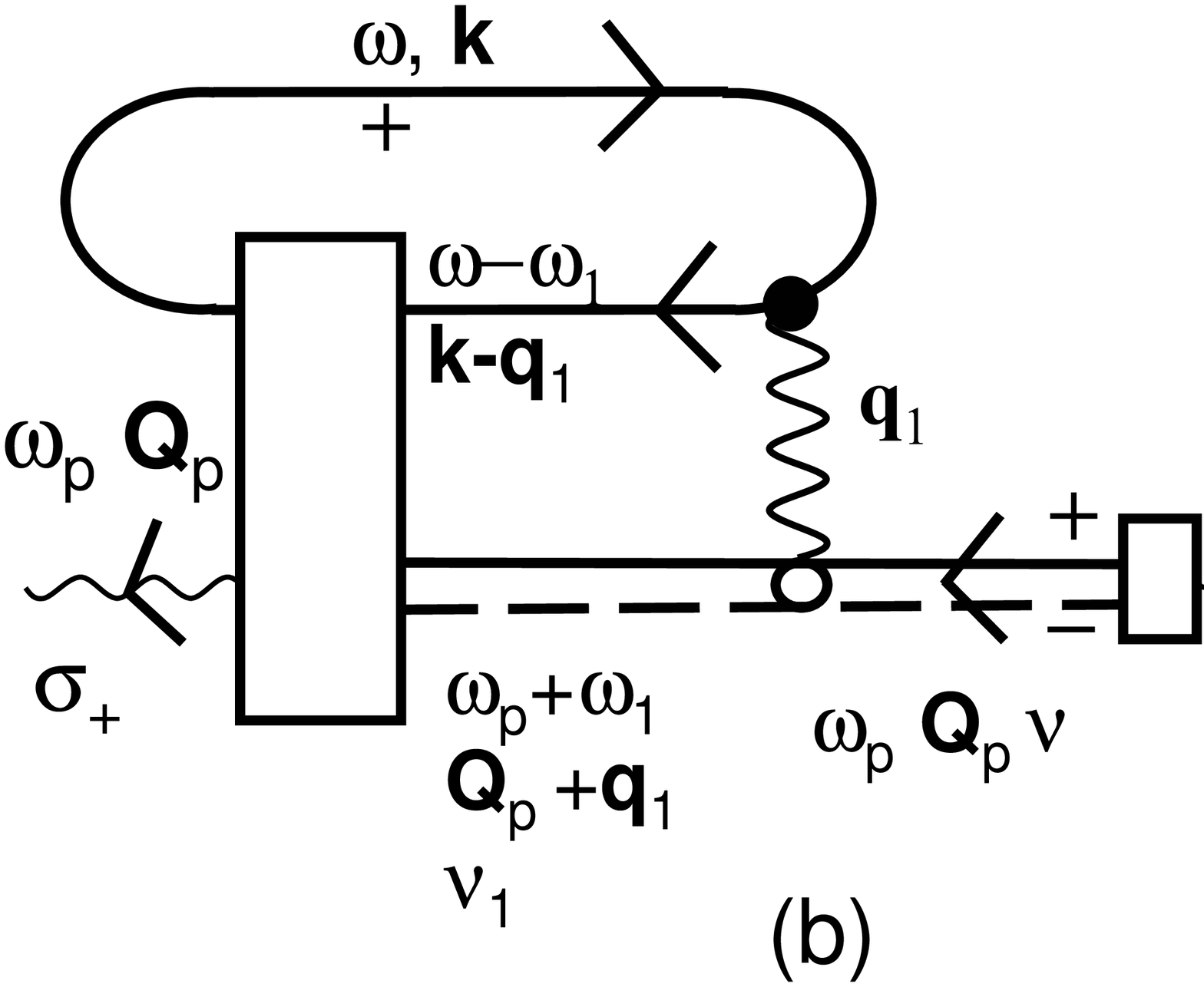}}}
\vspace*{0.5cm} \centerline{
\scalebox{0.2}{\includegraphics{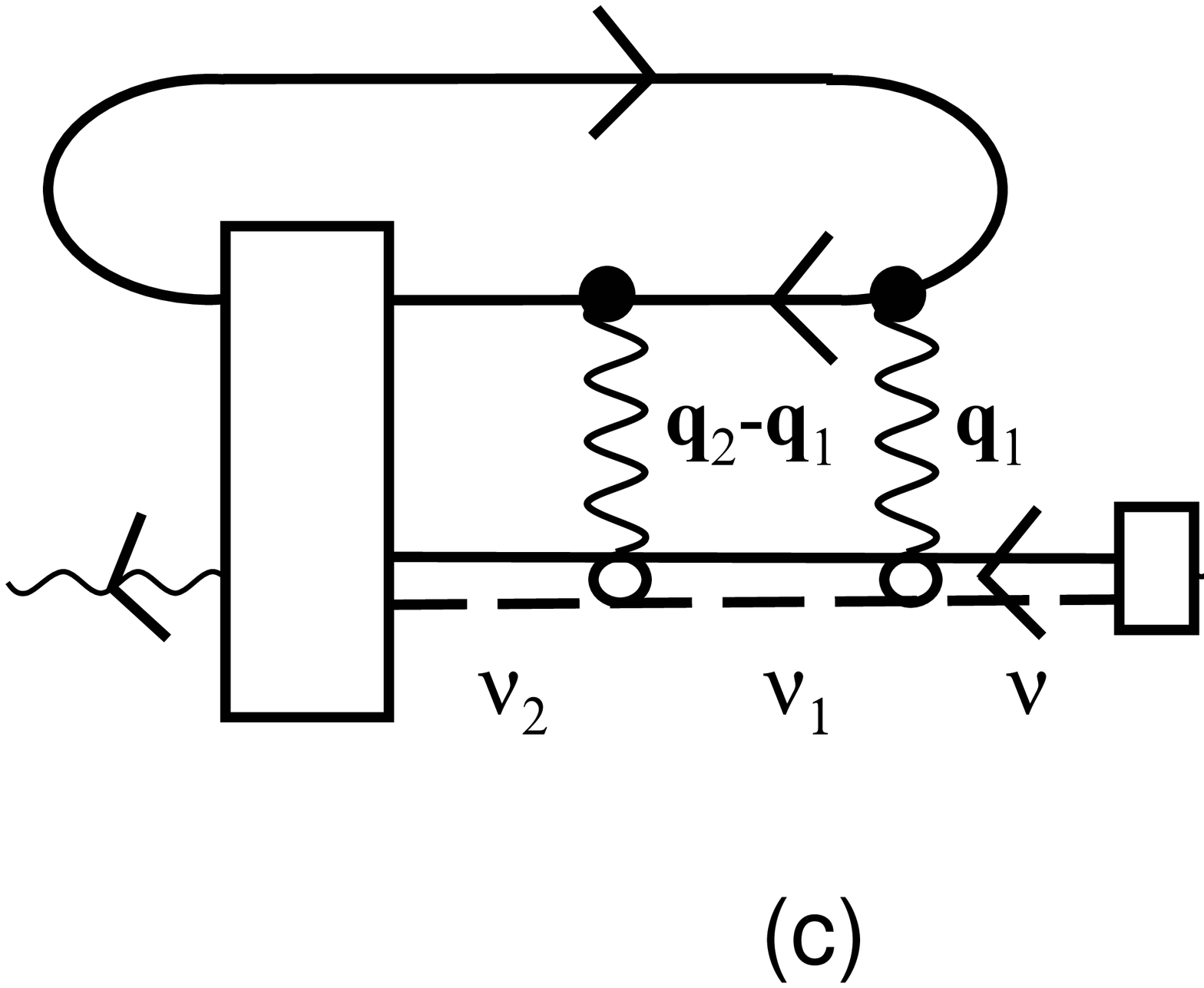}}\hspace*{0.5cm}
\scalebox{0.2}{\includegraphics{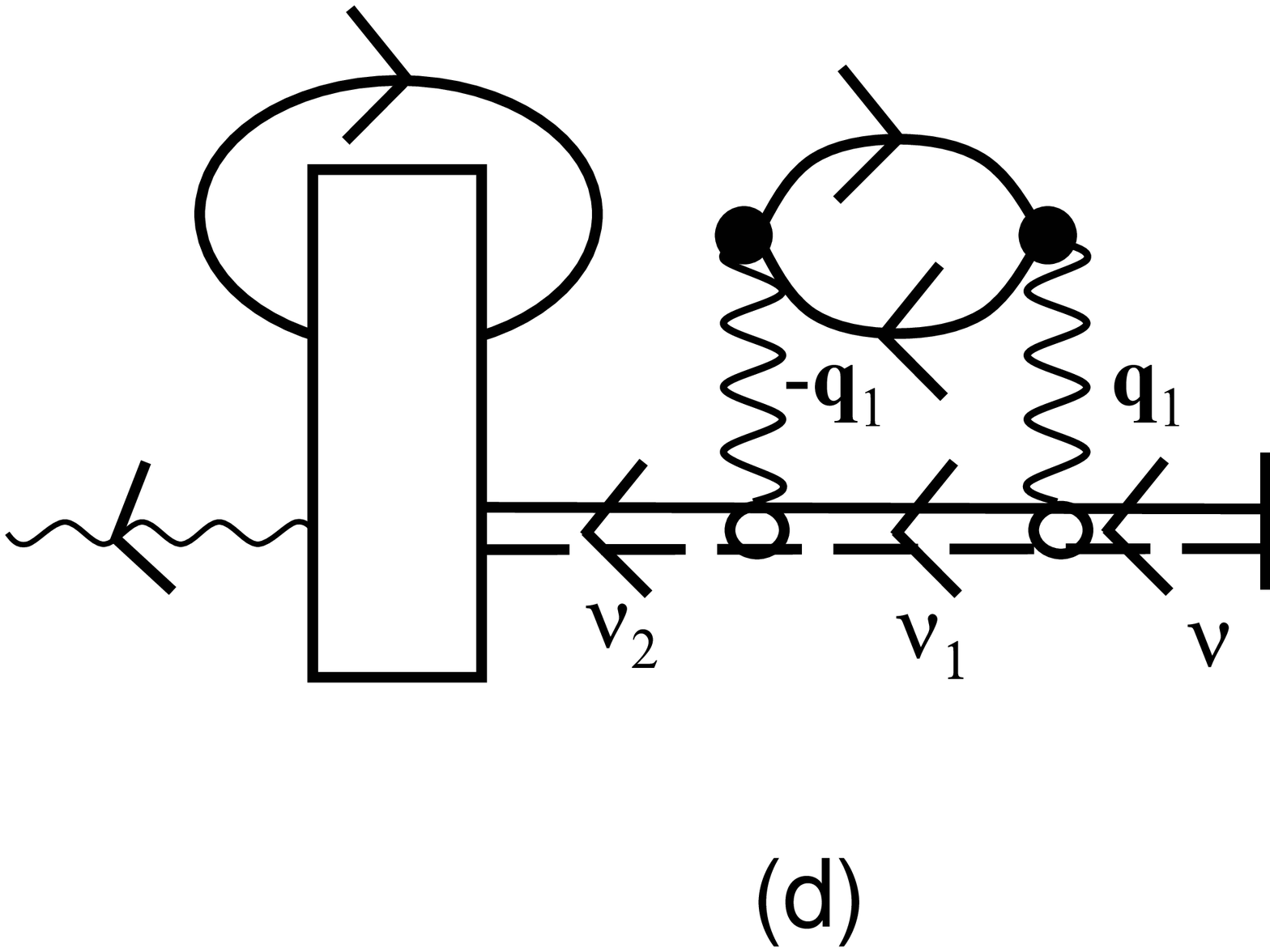}}} \vspace*{0.5cm}
\centerline{
\scalebox{0.2}{\includegraphics{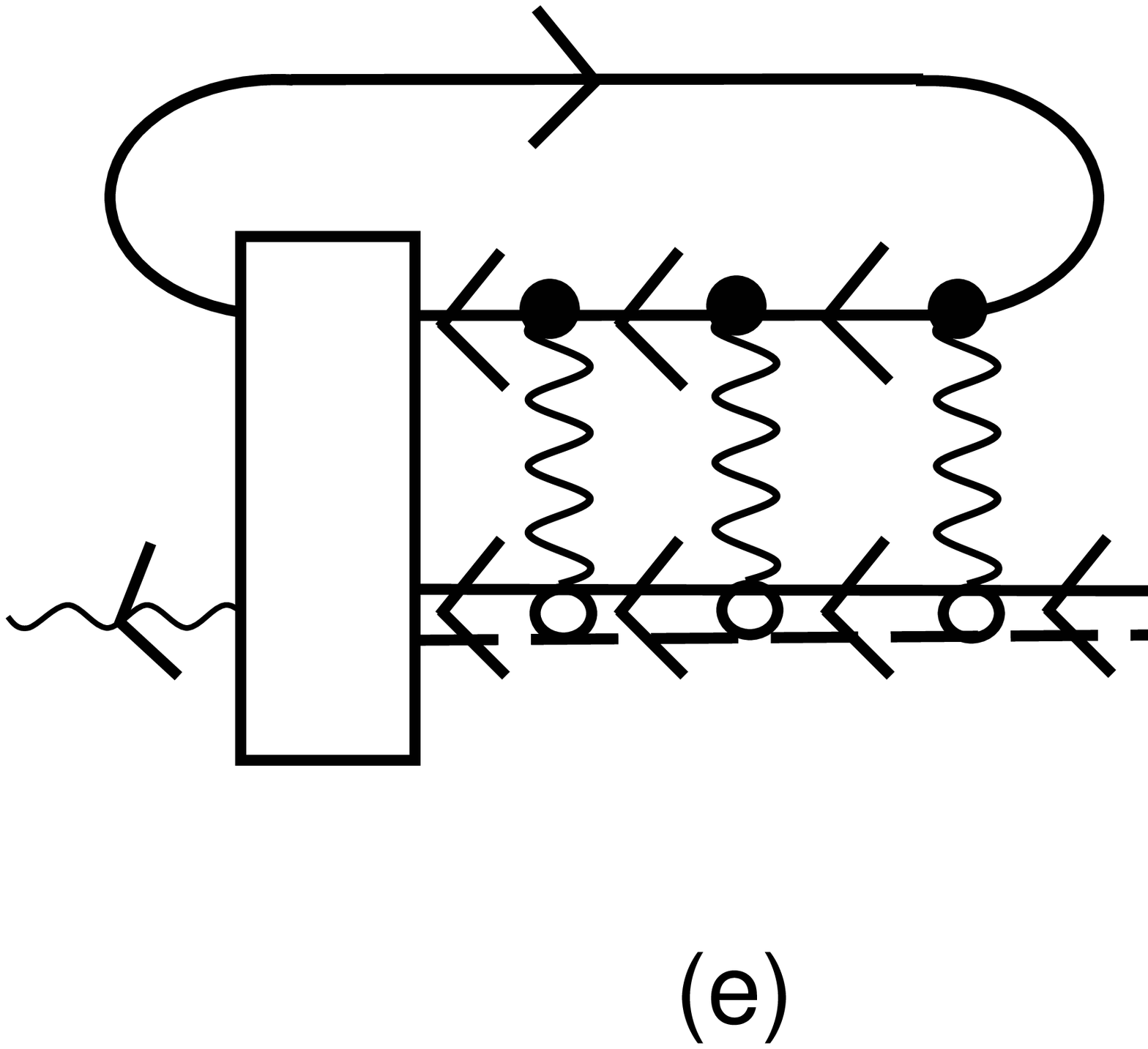}}\hspace*{0.5cm}
\scalebox{0.2}{\includegraphics{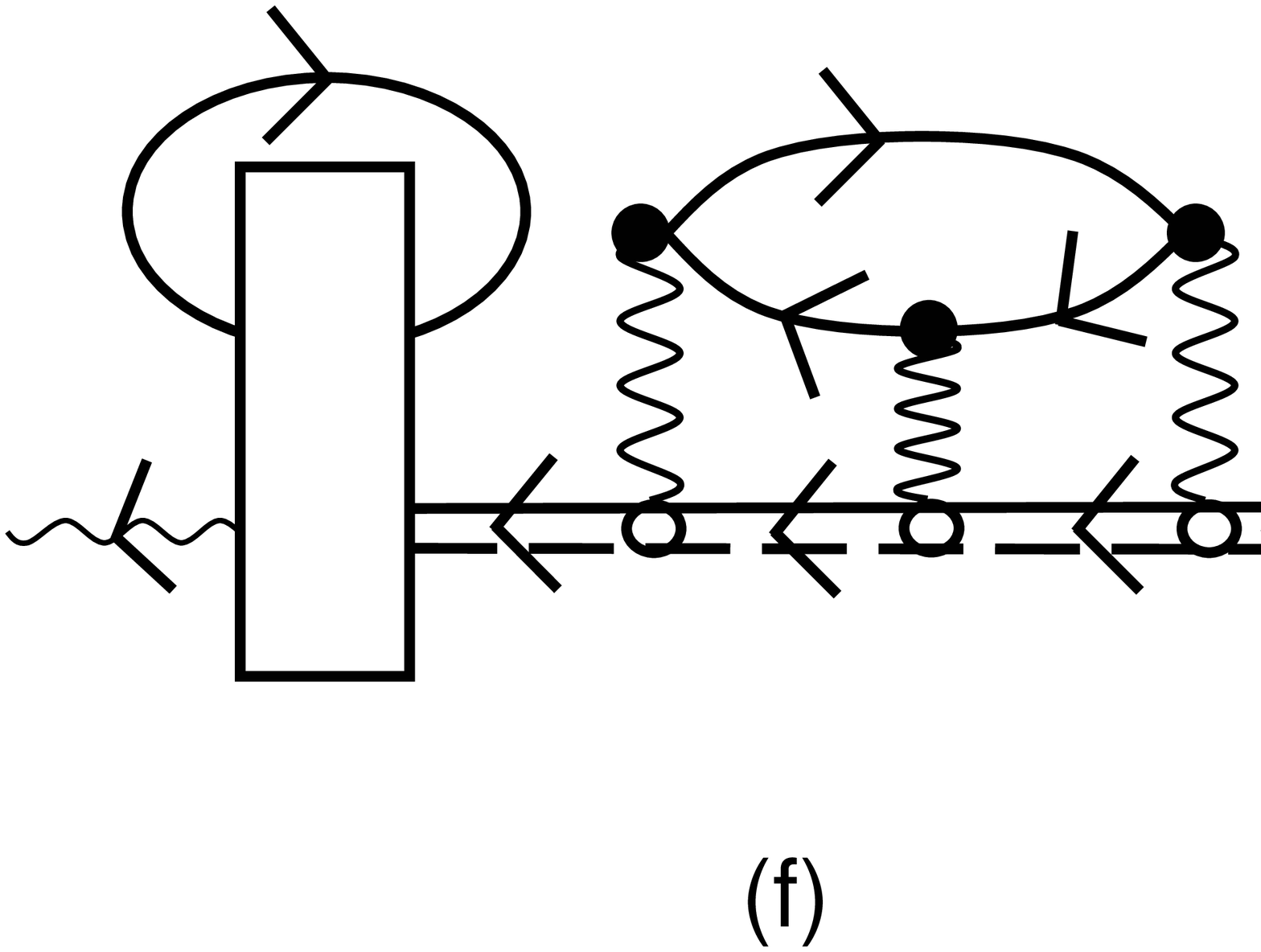}}}
\begin{center}
\scalebox{0.2}{\includegraphics{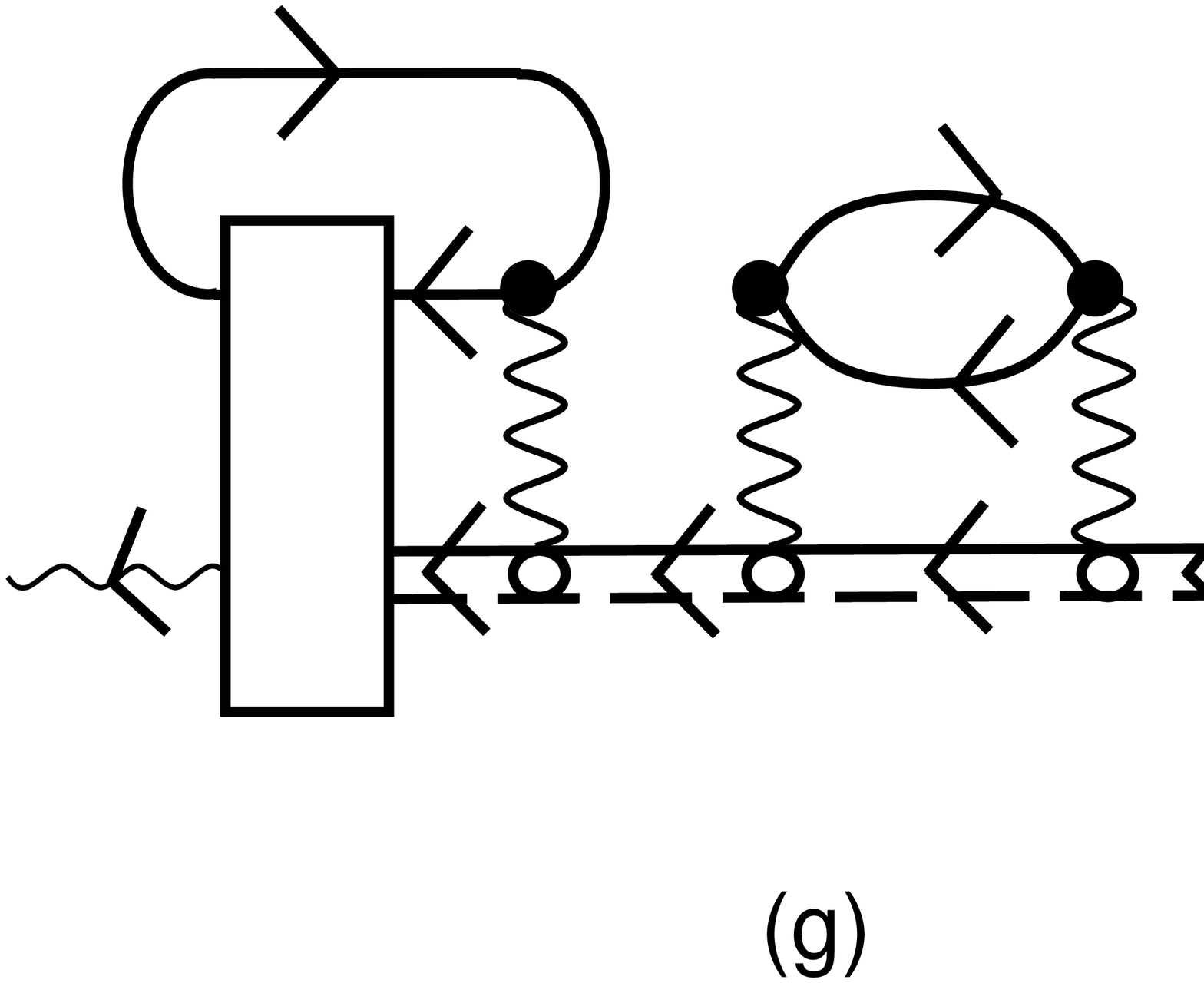}}\end{center}
\caption{Photon absorption with trion formation when the
photocreated electron and the initial electron have the same spin.
The diagrams of this figure have to be added to the diagrams of
fig.\ (7). Their main effect is to withdraw the $S=0$ trion
contributions which cannot exist when the two electron spins are
identical. (a) Zero order in direct Coulomb electron-exciton
scattering: The photocreated exciton already feels the presence of
the initial electron through exchange processes included in the
dressed semiconductor-photon interaction. (b) First order in
$C_{\nu_1\v p_1;\nu\v p}^\mathrm{dir}$. (c) and (d) Second order.
(e), (f) and (g) Third order.}
\end{figure}

\end{document}